\documentclass[onecolumn]{IEEEtran}

\ifCLASSINFOpdf
  \usepackage[pdftex]{graphicx}
\else
  \usepackage[dvips]{graphicx}
\fi

\usepackage[cmex10]{amsmath}
\usepackage{amssymb}
\usepackage{array}

\usepackage{enumitem}
\usepackage{verbatim}
\usepackage{bbold}
\usepackage[cmex10]{amsmath}
\usepackage{color}

\usepackage{amsfonts}

\usepackage{array}

\usepackage{amssymb}
\usepackage{enumitem}
\usepackage{verbatim}
\usepackage{bbold}

\newtheorem{thm}{Theorem}

\newtheorem{cor}{Corollary}
\newtheorem{lem}{Lemma}
\newtheorem{prop}{Proposition}
\newtheorem{rem}{Remark}

\newcommand{\beq}{\begin{equation}}
\newcommand{\eeq}{\end{equation}}
\newcommand{\E}{\mathbb{E}}

\DeclareMathOperator*{\argmax}{arg\,max}

\makeatletter

\newcommand{\Rmnum}[1]{\expandafter\@slowromancap\romannumeral #1@}
\makeatother

\newcommand{\bu}{\mathbf{u}}
\newcommand{\bv}{\mathbf{v}}
\newcommand{\bx}{\mathbf{x}}
\newcommand{\by}{\mathbf{y}}
\newcommand{\bz}{\mathbf{z}}
\newcommand{\bw}{\mathbf{w}}
\newcommand{\bn}{\mathbf{n}}
\newcommand{\cC}{\mathcal{C}}
\newcommand{\Var }{{\sf Var}}
\begin{document}

\title{Unequal Error Protection Querying Policies for the Noisy 20 Questions Problem }

\author{Hye Won Chung,~Brian M. Sadler,~Lizhong Zheng and Alfred O. Hero
\thanks{Hye Won Chung (hwchung@kaist.ac.kr) is with the School of Electrical Engineering at KAIST in South Korea. Brian M. Sadler (brian.m.sadler6.civ@mail.mil) is with the US Army Research Laboratory. Lizhong Zheng (lizhong@mit.edu) is with the EECS department at MIT.  Alfred O. Hero (hero@eecs.umich.edu) is with the EECS department at the University of Michigan.  This research was supported in part by ARO grants W911NF-11-1-0391 and W911NF-15-1-0479.
This research was presented in part at 2016 IEEE International Symposium on Information Theory in Barcelona, Spain~\cite{chung2016unequal}.  
}}

\maketitle

\begin{abstract}
In this paper, we propose an open-loop unequal-error-protection querying policy based on superposition coding for the noisy 20 questions problem. 
In this problem, a player wishes to successively refine an estimate of the value of a continuous random variable by posing binary queries and receiving noisy responses.
When the queries are designed non-adaptively as a single block and the noisy responses are modeled as the output of a binary symmetric channel the 20 questions problem can be mapped to an equivalent problem of channel coding with unequal error protection (UEP). 
A new non-adaptive querying strategy based on UEP superposition coding is introduced whose estimation error decreases with an exponential rate of convergence that is significantly better than that of the UEP repetition coding introduced by Variani {\it{et al.}},~\cite{jedynak2015}.
With the proposed querying strategy, the rate of exponential decrease in the number of queries matches the rate of a closed-loop adaptive scheme where queries are sequentially designed with the benefit of feedback. 
Furthermore, the achievable error exponent is significantly better than that of random block codes employing equal error protection.

\end{abstract}


\IEEEpeerreviewmaketitle

\section{Introduction}

Consider a noisy 20 questions game between a player and an oracle. The objective of the player is to estimate the value of a continuous target variable  $X\sim \text{unif}[0,1]$. The player asks binary queries to the oracle who knows the value of $X$, and receives a noisy version of the oracle's correct answers transmitted through a binary symmetric channel with flipping probability $\epsilon\in(0,1/2)$, denoted BSC($\epsilon$). 
The central question addressed here is: What is the optimal sequence of queries to estimate the value of $X$ with a minimum estimation error at a fixed number of querying? 
This general setup of noisy 20 questions game and the optimal query design problem is of broad interest, arising in various areas, including active learning~\cite{mackay1992information,settles2010active}, optimal sensing~\cite{castro2008active} and experimental design~\cite{lindley1956measure,fedorov1972theory}, with diverse applications.
For example,  a target localization problem in a sensor network~\cite{tsiligkaridis2014collaborative}  can be modeled as a noisy 20 questions game where a player (agency) aims to locate a target by receiving query responses from sensors probing the region of interest.

The problem of optimal query design for the noisy 20 questions game can be categorized into two main approaches, adaptive vs. non-adaptive designs.
In each approach, the sequence of queries is designed by a controller that may either use feedback (adaptive 20 questions) or operate open-loop (non-adaptive 20 questions) to formulate the sequence of questions.
For the adaptive case, the controller uses noisy answers to previous questions to determine the next question posed to the oracle.
For the non-adaptive case, on the other hand, the controller designs the sequence of queries ahead of time without access to future answers of the oracle.
In general, the use of feedback in the adaptive design provides an information advantage, allowing a better error rate of convergence, but at the cost of higher query design complexity and the need for a feedback channel. 

Previous studies on optimal query design for the noisy 20 questions problem often sought to design queries that acquire observations minimizing the posterior uncertainty of the target variable, where uncertainty was quantified by the Shannon entropy~\cite{luttrell1985use, mackay1992information, jedynak2012twenty, tsiligkaridis2014collaborative}. 
In these works, the utility of observation is quantified by the expected reduction of the entropy due to the observation.
This reduction is equivalent to the increase in mutual information between the target variable and the observation.
For adaptive sequential querying, greedy successive-entropy-minimization policies~\cite{jedynak2012twenty,chen2015sequential} have been extensively investigated. 

When the mutual information is used to quantify the utility of observations, any two observations that increase the mutual information by the same amount are considered to be equally valuable, regardless of how much these observations reduce the estimation error.
However, when estimation accuracy is important, queries maximizing the mutual information may not generate observations of equal importance. 
For example, when the queries are on the coefficients in the dyadic expansion of a target variable $X$ the queries on the most significant bits (MSBs) of $X$ may acquire more valuable observation than those on the least significant bits (LSBs)  in terms of reducing the estimation error.
For estimation of $X$, the important question is then how to design queries that acquire observations valuable in reducing the estimation error.

In the noisy 20 questions game, estimates on the coefficients in dyadic expansion of the target variable, which are based on the received noisy answers from the oracle, may contain errors.
Since the errors in MSBs cause a higher estimation error than do the errors in LSBs, it is desirable to provide {\it unequal error protection} (UEP) for MSBs vs. LSBs in order to minimize the estimation error with a limited number of queries.
In this paper, we provide such a non-adaptive UEP querying policy for state estimation in the context of the noisy 20 questions problem. 

To develop the UEP querying policy, we exploit a close connection between the problem of optimal query design in the noisy 20 questions problem and the problem of channel coding for the classical information-transmission problem. 
Let $\mathcal{M}=\{0,\dots, 2^{k}-1\}$ denote the set of $2^k$ possible states of the target variable $X$, determined by the first $k$ bits in its dyadic expansion.
A binary query  partitions the set $\mathcal{M}$ into two disjoint subsets, one of which contains the true state of $X$. 
For adaptive sequential querying, the partition is random, depending on the answers to the previous queries, whereas for non-adaptive querying, the partition is deterministic and determined in advance. 
By considering the true state of the target variable as a message transmitted from the oracle to the player and the oracle's binary answer bits to the sequence of queries as a codeword, the query design problem can be mapped to an equivalent problem of channel coding.
Specifically, the query design problem reduces to the channel coding with feedback for the adaptive case and to channel coding without feedback for the non-adaptive case.


The equivalence between the query design problem and the channel-coding problem allows us to apply UEP channel coding methods to design a UEP querying strategy.
Unequal-error-protection querying accounts for the fact that for estimation of a target variable, the errors in the most significant bits (MSBs) are much more costly than the errors in the least significant bits (LSBs). 

One way to provide unequal error protection is repetition coding. 
In repetition coding, each bit is repeatedly transmitted multiple times, the number of repetitions varying in accordance with the desired level of unequal error protection.
Such a UEP repetition coding approach to the noisy 20 questions problem was considered in~\cite{ jedynak2015}.
It was shown that the mean squared error (MSE) of this approach decreases exponentially in $\sqrt{N}$ where $N$ is the number of queries. The square root of $N$ rate is smaller than the linear in $N$ exponential rate of decrease achievable by the bisection-based adaptive 20 questions strategy~\cite{burnashev1974interval} that corresponds to Horstein's coding scheme for a BSC($\epsilon$) with perfect feedback~\cite{horstein1963sequential}.

The main contribution of this paper is to provide a new non-adaptive querying strategy based on superposition coding~\cite{cover1972broadcast} that can provide UEP and achieve better MSE convergence rate than that of repetition coding in~\cite{ jedynak2015}.
The proposed superposition coding strategy provides UEP for two levels of priority, i.e., a strictly better error protection for MSBs than that for LSBs. 
We show that the proposed querying strategy achieves the MSE that decreases exponentially in $N$, as contrasted to $\sqrt{N}$, matching the error rate of the adaptive 20 questions strategy~\cite{burnashev1974interval}.  Furthermore, this strategy achieves a better scale factor in the MSE exponent as compared to that of random block codes employing equal error protection.

The rest of this paper is organized as follows.
In Section~\ref{sec:problem}, we review the mathematical formulation for the noisy 20 questions problem for state estimation. We highlight the connection between the query design and the channel-coding problems both for adaptive sequential querying and for non-adaptive block querying. For query performance measures, the MSE and quantized MSE are considered. The different importances of the first $k$ bits in the dyadic expansion of the target variable are quantified for these performance measures.
In Section~\ref{sec:three_comp}, we review three well-known querying policies including the adaptive bisection policy~\cite{horstein1963sequential}, non-adaptive UEP repetition policy~\cite{ jedynak2015}, and non-adaptive block querying based on random block coding~\cite{shannon2001mathematical}. 
In Sections~\ref{sec:sub1}, we show that the bisection policy is the optimal myopic policy among successive-entropy-minimization policies in reducing the minimum MSE of the target variable (Proposition~\ref{thm:opt_bis1}).
In Sections~\ref{sec:sub2} and~\ref{sec:sub3}, two representative non-adaptive policies are presented and compared in terms of UEP property and coding gain. 
We introduce a new non-adaptive querying policy based on superposition coding in Section~\ref{sec:super}. 
We show that block querying based on superposition coding provides higher level of error protection for MSBs than for LSBs. We then establish that the proposed non-adaptive block querying strategy achieves a better quantized-MSE exponent (Theorem~\ref{thm:sup}) and better MSE exponent (Corollary~\ref{cor}) than those of random block coding.
In Section~\ref{sec:allcompare}, performance of all four policies discussed in this paper are compared by analyzing the achievable error rates of convergence for the estimation errors in the number $N$ of queries.
Finally, conclusions and future directions are discussed in Section~\ref{sec:con}. 
After presenting each result, we provide a brief discussion  but defer the technical details of the proofs to the Appendices.

\subsection{Notations}
Capital letters will represent random variables and lower case letters will represent specific realizations of those random variables.
The statistical expectation operator and the indicator operator will be denoted by $\E[]$ and $\mathbb{1}()$, respectively.
For a continuous random variable $X$ distributed as $p(x)$, $x\in\mathbb{R}$, the differential entropy $h(X)$ is defined as $h(X)=-\int p(x)\ln p(x) dx$. 
For a discrete random variable $Y$ with distribution $p(y)$, $y\in \mathcal{Y}$, the entropy $H(Y)$ is defined as $H(Y)=-\sum_{y\in\mathcal{Y}} p(y)\ln p(y)$. 
The entropy of a binary random variable $Z$ distributed as  Bernoulli$(\alpha)$, $0\leq \alpha\leq1$, is denoted $H_{\sf B}(a)=-a\ln a-(1-a)\ln(1-a)$.
The Kullback-Leibler divergence between two Bernoulli distributions Bernoulli($\alpha$) and Bernoulli($\beta$) is denoted  $D_{\sf B}(\alpha\|\beta):=\alpha\ln\frac{\alpha}{\beta}+(1-\alpha)\ln\frac{1-\alpha}{1-\beta}.$ The star $*$ operator is defined as $\alpha*\epsilon:=\alpha(1-\epsilon)+(1-\alpha)\epsilon$ for $\alpha,\epsilon\in\mathbb{R}$.

The normalized Gilbert-Varshamov distance $\gamma_{\sf GV}(R)\in[0,1/2]$ is the value $\gamma_{\sf GV}(R)$ that gives
$
D_{\sf B}(\gamma_{\sf GV}(R)\|1/2)=R.
$
The inverse of the normalized Gilbert-Varshamov distance is denoted $\gamma_{\sf GV}^{-1}(\alpha)$ for $0\leq \alpha\leq 1/2$.

Bold face $\bz$ or $z_1^N$ denotes the length-$N$ binary sequence $(z_1z_2\dots z_N)$ where $z_t$ is the $t$-th bit of $\bz$.
The Hamming weight of $\bz$ is equal to the cardinality of the set $\{t\in[1:N]: z_t=1\}$ and is denoted as $w_H(\bz)$.
The bit-wise XOR operation is symbolized by $\oplus$ and the bit-wise XOR of two binary sequences $\bx$ and $\by$ is written as $\bx\oplus \by$. The Hamming distance between two binary sequences $\bx$ and $\by$ is the cardinality of the set $\{t\in[1:N]:x_t\neq y_t\}$ and is denoted as $d_H(\bx,\by):=|\{t\in[1:N]:x_t\neq y_t\}|$. 

We will use the notation  $\doteq $, $\dot{\leq}$, and $\dot{\geq}$  as follows:
1) $ a_N\doteq e^{Nd}$ denotes $d=\liminf_{N\to\infty} \frac{\ln a_N}{N}$.
2) $a_N\dot{\leq} e^{Nd}$ denotes $d\geq \liminf_{N\to\infty} \frac{\ln a_N}{N}$.
3)  $a_N\dot{\geq} e^{Nd}$ denotes $d\leq \liminf_{N\to\infty} \frac{\ln a_N}{N}$.

\section{Problem Statement: Noisy 20 Questions for Estimation of A Target Variable }\label{sec:problem}

We consider an estimation problem for a target variable in the context of a noisy 20 questions game between a player and an oracle who communicate over a channel. 
The objective of the player is to estimate the value of a target variable, $X\sim\text{unif}[0,1]$ by posing a sequence of binary queries to the oracle and receiving noisy answers. 
To estimate $X$, the player asks the oracle whether $X$ is located within some sub-region $Q\subset [0,1]$, which may be connected or non-connected, and receives a noisy binary answer $Y\in\{0,1\}$ based on the correct answer $Z(X)=\mathbb{1}(X\in Q)$  with error probability $\epsilon\in[0,1/2)$. 
The oracle always provides a correct binary answer $Z(X)=\mathbb{1}(X\in Q)$ to the player's query. The channel through which the oracle's binary answer is transmitted to the player is modeled as a binary symmetric channel, BSC($\epsilon$).

The player asks a sequence of $N$ questions in the form of a sequence of querying regions $(Q_1,Q_2,\dots, Q_N)$. The oracle provides correct answers $(Z_1,Z_2,\dots, Z_N)$ to the queries about the target variable $X$, and the player receives a noisy version $(Y_1,Y_2,\dots Y_N)$ of the oracle's answers transmitted through $N$ uses of the BSC($\epsilon$).
Based on these answers, the player calculates an estimate $\hat{X}_N$ of $X$. For a given cost function $c(x,\hat{x}_N)$ between the true value $x$ and the estimate $\hat{x}_N$, the player's goal is to find the optimal sequence of querying regions $(Q_1,Q_2,\dots, Q_N)$ and the estimator $\hat{X}_N(Y_1,\dots,Y_N)$ that minimize the expected cost function. That is, the player aims to achieve
\beq\label{eqn:main_opt}
\min_{(Q_1,Q_2,\dots, Q_N),\hat{X}_N(\cdot)}\E[c(X,\hat{X}_N)]
\eeq
where the expectation is taken over the joint distribution of $(X,Y_1,Y_2,\dots,Y_N)$. 

Note that the joint distribution of $(X,Y_1,Y_2,\dots,Y_N)$ depends on the querying regions $(Q_1,Q_2,\dots,Q_N)$.

\begin{figure}[t]
\centerline{\includegraphics[scale=0.4]{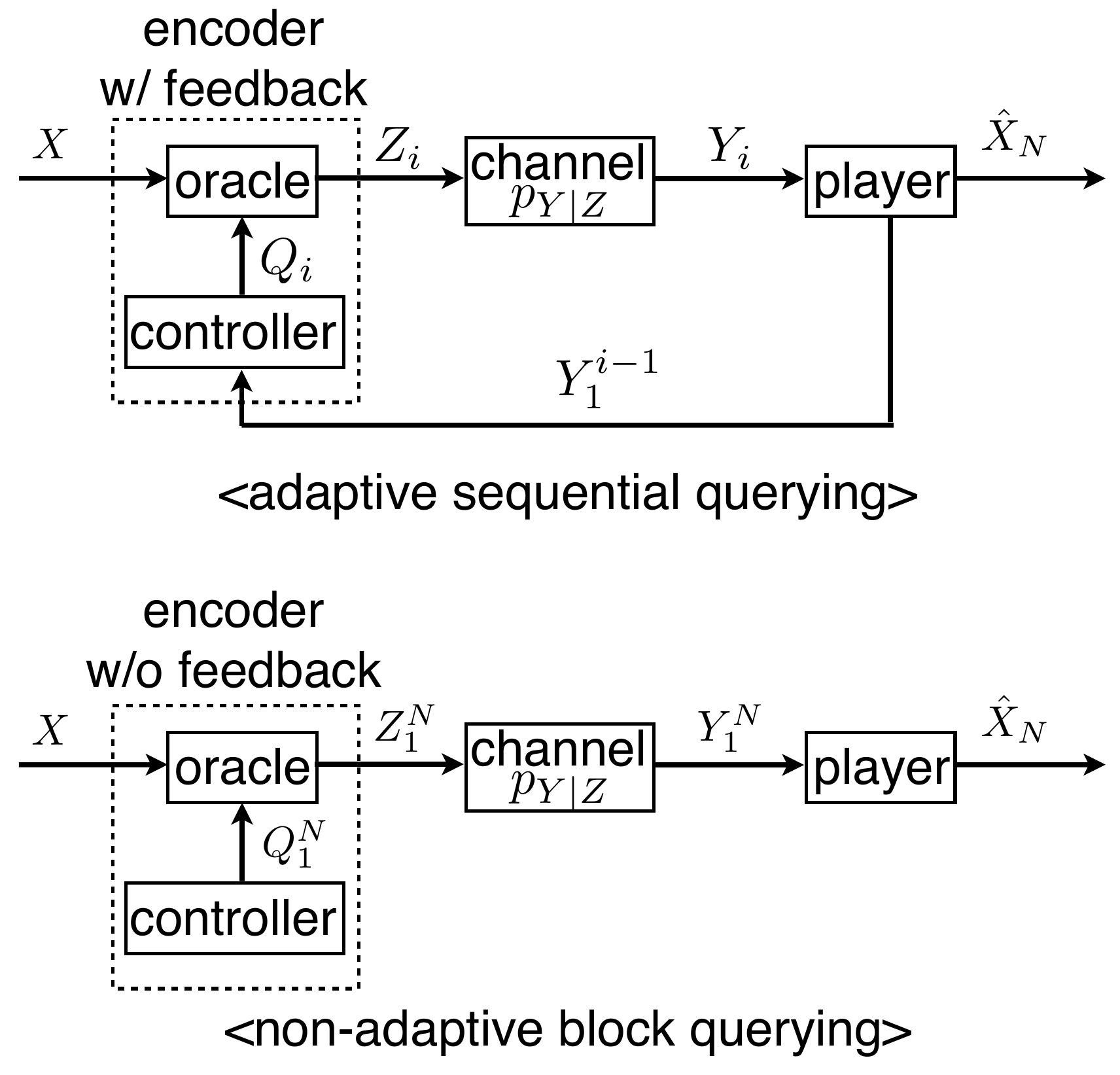}}
\caption{Noisy 20 questions problem between an oracle and a player over a BSC($\epsilon$). The controller generates questions using feedback (adaptive sequential querying) or operating open-loop (non-adaptive block querying).
For adaptive sequential querying, the controller generates queries $Q_i$ adaptively based on past answers $Y_1^{i-1}$, whereas for non-adaptive block querying the controller generates a length-$N$ block of queries $Q_1^N=(Q_1,\dots, Q_N)$ non-adaptively as a single block. The oracle gives the correct answer $Z_i$ to the query $Q_i$ about the target variable $X$. The player receives a noisy version $Y_i$ of the oracle's answer $Z_i$ transmitted through a BSC($\epsilon$), and outputs an estimate $\hat{X}_N$ based on the received answers $Y_1^N=(Y_1,\dots, Y_N)$. 
}
\label{fig:model}
\end{figure}

The sequence of questions  is designed by  a controller that may either use feedback (adaptive sequential querying) or operate open-loop (non-adaptive block querying) as depicted in Fig.~\ref{fig:model}. Depending on whether the questions are designed with or without the benefit of feedback, the optimal querying strategy and the corresponding performance can vary. 
In the next section, we highlight differences between adaptive sequential querying and non-adaptive block querying and show a connection between the noisy 20 questions problem and the channel-coding problem.

\subsection{Adaptive vs. Non-adaptive Querying Strategies and Associated Channel-Coding Problems}

In the adaptive case, the $i$-th querying region $Q_i$ can be updated based on past answers $Y_1^{i-1}:=(Y_1,\dots, Y_{i-1})$ to previous queries. 
For this case, the controller uses the updated posterior distribution  $p(x|y_1^{i-1})$ of $X$ to design the next query, i.e., the region $Q_i$. 
For example, consider the case when the $i$-th querying region $Q_i$ is designed to equalize the probabilities of $X$ belonging to $Q_i$ and of $X$ not belonging to $Q_i$, respectively, for given collected answers $Y_1^{i-1}=y_1^{i-1}$:
\beq\label{eqn:1bit_cond}
\begin{split}
&\Pr(X\in Q_i|Y_1^{i-1}=y_1^{i-1})=\Pr(X\notin Q_i|Y_1^{i-1}=y_1^{i-1})=1/2. 
\end{split}
\eeq
Since the channel input (the oracle's binary answer) $Z_i(X)$ is an indicator random variable of the event $\{X\in Q_i\}$, for the choice of $Q_i$ satisfying~\eqref{eqn:1bit_cond} the corresponding channel input $Z_i$ follows the distribution
\beq
\Pr(Z_i=0|Y_1^{i-1}=y_1^{i-1})=\Pr(Z_i=1|Y_1^{i-1}=y_1^{i-1})=1/2,
\eeq
which is an optimal input distribution for the BSC($\epsilon$) in maximizing the mutual information.
Specifically, the corresponding mutual information between the oracle's $i$-th binary answer $Z_i$ and the channel output $Y_i$ given previous answers $Y_1^{i-1}=y_1^{i-1}$ is
\beq
I(Z_i;Y_i|Y_1^{i-1}=y_1^{i-1})=C
\eeq
where 
\beq\label{eqn:forC}
C:=\ln2-(-\epsilon\ln \epsilon-(1-\epsilon)\ln(1-\epsilon)).
\eeq
To summarize, in adaptive sequential querying the channel input $Z_i(X)=\mathbb{1}(X\in Q_i)$ depends on the previous channel outputs $Y_1^{i-1}$, since the querying region $Q_i$ depends on $Y_1^{i-1}$.
As depicted in the upper figure of Fig.~\ref{fig:model},  the combined operation of the controller and the oracle
can be thought of as an encoder in a feedback communication system. Therefore, there is a one-to-one mapping between designing an adaptive sequential-querying strategy and designing a sequential channel encoder with noiseless feedback.

In the non-adaptive case the querying regions $Q_1^N:=(Q_1,\dots,Q_N)$ are specified in advance, before observing any of the answers from the oracle. 
Assume that the controller generates queries on the first $k$ bits in the dyadic expansion of $X\approx 0.B_1\dots B_k$, $B_i\in\{0,1\}$ for $i=1,\dots, k$.
The resolution parameter $k$ may depend on the number of queries $N$.
Discovering $(B_1,\dots, B_k)$ is equivalent to finding the index $M=\sum_{i=1}^k B_i 2^{k-i}\in\{0,\dots, 2^k-1\}$ of the interval $I_M:=[M2^{-k},(M+1)2^{-k})$ that contains $X$. 
Here the domain $[0,1]$ of $X$ is uniformly quantized into $2^k$ disjoint sub-intervals $\{I_0,\dots, I_{2^k-1}\}$ of length $2^{-k}$. 
If the oracle's answer  $Z_i$ to the question $Q_i$ can be transmitted to the player without noise, i.e., $\epsilon=0$, then by querying each coefficient of the dyadic expansion of $X$ from the MSB to the LSB, the player can discover the $N$ most significant bits $(B_1,\dots, B_N)$ of $X$ without error.
However, in the case of a noisy channel, the player needs to ask redundant questions in order to accurately estimate the $k$ most significant bits of $X$ for some $k<N$.   

Non-adaptive block querying can be mapped to an equivalent problem of length-$N$ block channel coding over a BSC($\epsilon$). The rate of the block code is defined as $R=(k\ln 2)/N$ (nats/channel use) for the resolution of $k$ bits of $X$.
 Designing a block of questions $(Q_1,\dots,Q_N)$ to discover the index $M$ of the sub-interval $I_M$ containing $X$ can be thought of as designing a length-$N$ and rate-$R$ block code, or, more specifically, defining an encoding map $f:\{0,\dots,2^k-1\}\to\{0,1\}^N$, to reliably transmit one of the $2^k$ messages through $N$ uses of the channel with channel coding rate $R=(k\ln 2)/N$.

A block of questions  specifies the encoding map $f:\{1,\dots,2^k\}\to\{0,1\}^N$, and vice versa.
The one-to-one mapping between the two is described as follows. 
Define sub-intervals $I_m:=[m2^{-k},(m+1)2^{-k})$ for $m\in\{0,\dots, 2^k-1\}$. 
We restrict the querying region $Q_i$ to be the union of a subset of the quantized intervals $\{I_0,\dots, I_{2^{k}-1}\}$. In other words, we fix the maximum resolution of the querying interval as $2^{-k}$.
Let $z_{i}^{(m)}$ denote the $i$-th bit of the codeword $f(m)=(z_{1}^{(m)},\dots,z_{N}^{(m)})$  for a message $m$ given an encoding map $f:\{0,\dots,2^k-1\}\to\{0,1\}^N$. Note that the bit $z_{i}^{(m)}$ is the oracle's binary answer to the query $Q_i$ indicating whether $x\in Q_i$ when $x\in I_m$. Therefore, the bit $z_i^{(m)}$ equals 1 if and only if $I_m\subset Q_i$, i.e.,
\beq
z_{i}^{(m)}=\mathbb{1}(I_{m}\subset Q_i).
\eeq
On the other hand, when the encoding map $f(\cdot)$ is specified, the associated $i$-th querying region $Q_i$ becomes the union of the sub-intervals $\{I_{m'}\}$ for message $m'$'s such that the $i$-th  answer bit $z_{i}^{(m')}$ equals 1, i.e., 
\beq
Q_i=\underset{\{m':z_{i}^{(m')}=1\}}\bigcup I_{m'}.
\eeq
Given the block of questions $(Q_1,\dots,Q_N)$, for an index $m$ such that $x\in I_m$ the oracle transmits the corresponding length-$N$ binary answer bits $f(m)$ through $N$ uses of the BSC($\epsilon$), and the player tries to decode the message $m$ given a noisy version of the codeword. 

Thus both adaptive sequential querying and non-adaptive block querying can be mapped to associated channel-coding problems in information transmission through a noisy channel, with and without feedback, respectively.
However, different from information-transmission problems where the goal is to achieve reliable communications at some positive rate $0<R\leq C$ for capacity $C$ of the channel,  the objective of the noisy 20 questions problem for state estimation is  to minimize estimation error $\E[c(X,\hat{X}_N)]$. 
In the next section, we introduce two different types of estimation errors that will be considered in this paper and discuss what kind of properties are desired for channel coding to minimize these estimation errors. 

\subsection{Estimation Errors: Mean Squared Error and Quantized Mean Squared Error }
We consider two types of estimation errors. The first is the mean squared error (MSE) 
$\E[|X-\hat{X}_N|^2]$ where $\hat{X}_N$ is the estimate of $X$ after $N$ queries.
The second is the quantized MSE $\E[c_{\sf q}(X,\hat{X}_N)]$ where the quantized cost function $c_{\sf q}(X,\hat{X}_N)$ with $2^k$ levels is a stepwise function defined as
\beq
\begin{split}\label{eqn:c_qdef}
&c_{\sf q}(X,\hat{X}_N)=(d2^{-k})^2, \;\; \text{when}\\
&d2^{-k}-\frac{2^{-k}}{2}<|X-\hat{X}_N|\leq  d2^{-k}+\frac{2^{-k}}{2},\\
&\text{for}\; d\in\{0,\dots,2^k-1\},
\end{split}
\eeq
for $X,\hat{X}_N\in[0,1]$.
We consider this cost function when the objective of the problem is to estimate the value of $X$ up to the first $k$ bits $(B_1,\dots, B_k)$ in the dyadic expansion of $X$.

\begin{figure}[t]
\centerline{\includegraphics[scale=0.55]{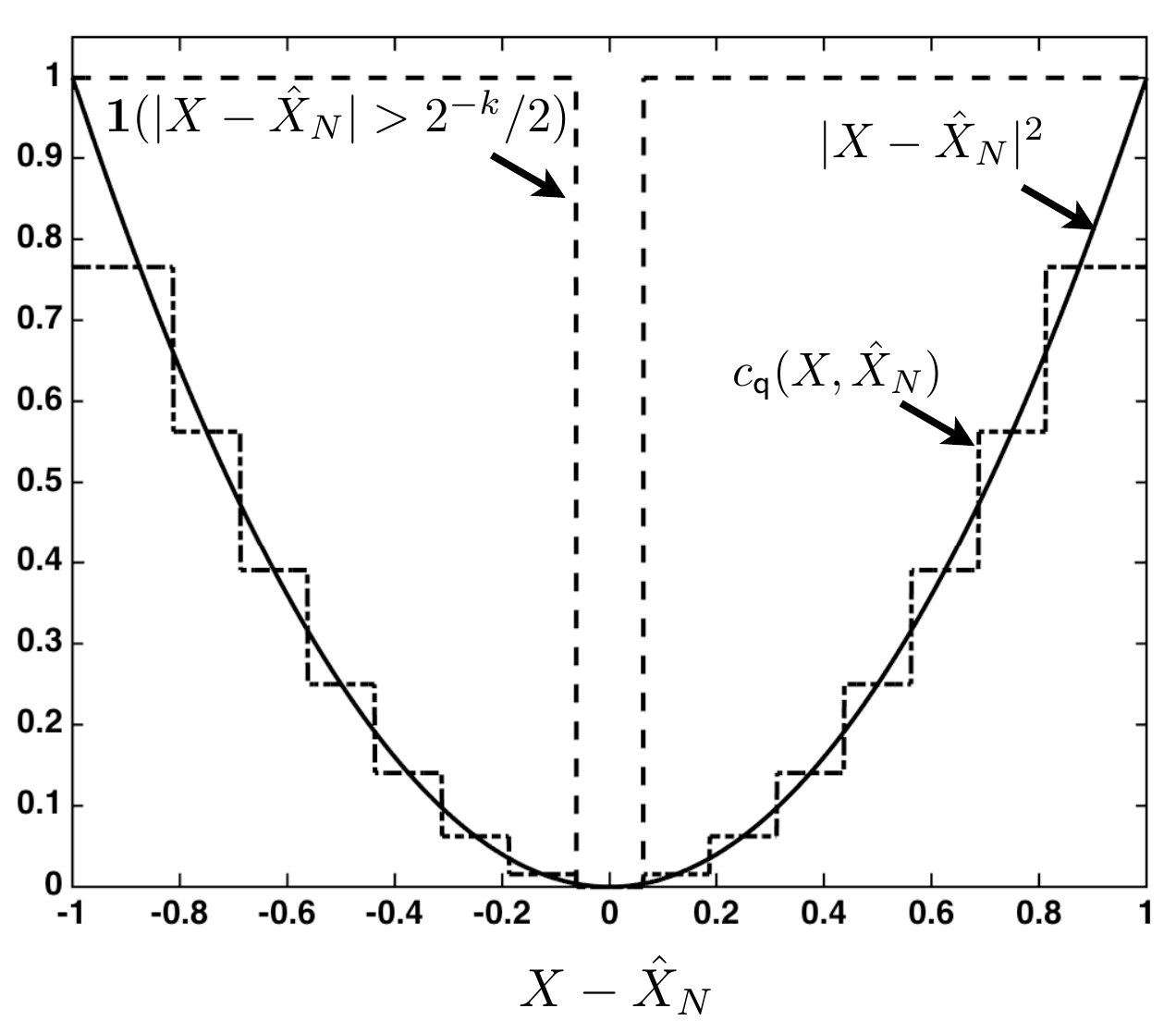}}
\caption{Plot of three different cost functions: $|X-\hat{X}_N|^2$ for the MSE, $c_{\sf q}(X,\hat{X}_N)$ for the quantized MSE and $\mathbb{1}(\hat{M}\neq M)=\mathbb{1}(|X-\hat{X}_N|>2^{-k}/2)$ for the block-decoding-error probability when the resolution parameter $k=3$.
}
\label{fig:costs}
\end{figure}

Let $(\hat{B}_1,\dots,\hat{B}_k)$ denote the estimate of $(B_1,\dots, B_k)$ and let
\beq
\hat{M}=\sum_{i=1}^k \hat{B}_i2^{k-i}
\eeq denote the estimate of the message $M=\sum_{i=1}^k B_i2^{k-i}$.
We define the decoding-error distance $d(M,\hat{M})$ between $M$ and $\hat{M}$ as
 \beq\label{eqn:dec_error_dist}
 d(M,\hat{M}):=|M-\hat{M}|=\left|\sum_{i=1}^k (B_i-\hat{B}_i)2^{k-i}\right|.
 \eeq 
By defining the finite-resolution estimator $\hat{X}_{N,\sf finite}$ as 
\beq\label{eqn:finiteresolution}
\hat{X}_{N,\sf finite}:=\hat{M}2^{-k}+2^{-k}/2,
\eeq
the quantized MSE $c_{\sf q}(X,\hat{X}_{N})$ with $\hat{X}_N=\hat{X}_{N,\sf finite}$ can be written in terms of the decoding-error distance as
 \beq\label{eqn:c_qdefFinite}
 c_{\sf q}(X,\hat{X}_{N,\sf finite})=2^{-2k}(d(M,\hat{M}))^2.
 \eeq
Note that the quantized MSE equals 0 when the player correctly decodes the message $M$. The error increases proportionally to the square of the decoding-error distance. 
On the other hand, in information-transmission problems where the cost function is $\mathbb{1}(\hat{M}\neq M)$, decoding error is claimed when $\hat{M}\neq {M}$, i.e., when $d(M,\hat{M})\neq 0$, and the cost of incorrect decoding is the same for every $\hat{M}\neq M$ regardless of the decoding-error distance $d(M,\hat{M})$. This difference in the cost functions makes the desired channel-coding strategy for state-estimation problem different from that of the information-transmission problem. Fig.~\ref{fig:costs} shows the three different cost functions, $|X-\hat{X}_N|^2$ for the MSE, $c_{\sf q}(X,\hat{X}_N)$ for the quantized MSE and $\mathbb{1}(\hat{M}\neq M)=\mathbb{1}(|X-\hat{X}_N|>2^{-k}/2)$ for the block-decoding-error probability when the resolution parameter $k=3$.

The quantized MSE $\E[c_{\sf q}(X,\hat{X}_N)]$ approximates the MSE $\E[|X-\hat{X}_N|^2]$.
In particular, with the finite-resolution estimator $\hat{X}_N=\hat{X}_{N,\sf finite}$,  the MSE $\E[|X-\hat{X}_{N,\sf finite}|^2]$ can be written as a sum of the quantized MSE $\E[c_{\sf q}(X,\hat{X}_{N,\sf finite})]$ and the error from finite resolution,
\beq\label{eqn:diff_q_MSE}
\E[|X-\hat{X}_{N,\sf finite}|^2]=\E[c_{\sf q}(X,\hat{X}_{N,\sf finite})]+c2^{-2k},
\eeq
for some constant $0<c\leq1/4$.
This can be shown by writing the difference between the two expected errors as a sum of errors conditioned on the decoding-error distance $d(M,\hat{M})=d$ for $d\in\{0,\dots, 2^k-1\},$
\beq
\begin{split}\label{eqn:conddiff_q_MSE}
&\E[|X-\hat{X}_{N,\sf finite}|^2]-\E[c_{\sf q}(X,\hat{X}_{N,\sf finite})]\\
&=\sum_{d=0}^{2^k-1}\left(\Pr(d(M,\hat{M})=d)\E\left[\left(|X-\hat{X}_{N,\sf finite}|^2-d^2 2^{-2k}\right)\Big|d(M,\hat{M})=d\right]\right).
\end{split}
\eeq
For $d=0$, the conditional expectation in~\eqref{eqn:conddiff_q_MSE} is bounded above by $2^{-2k}/4$. 
Given that $X\sim\text{unif}[0,1]$, conditioned on $d(M,\hat{M})=d$ for $d\in\{1,\dots,2^k-1\}$, $|X-\hat{X}_{N,\sf finite}|$ is uniformly distributed over $[d2^{-k}-2^{-k}/2,d2^{-k}+2^{-k}/2)$. Thus
\beq
\E\left[\left(|X-\hat{X}_{N,\sf finite}|^2-d^2 2^{-2k}\right)\Big|d(M,\hat{M})=d\right]=\frac{1}{12}2^{-2k}.
\eeq
Therefore, the difference between the MSE and quantized MSE is bounded above by a scale factor of $2^{-2k}$ as in~\eqref{eqn:diff_q_MSE}.

Consider the case when the resolution $k$ bits of the estimator $\hat{X}_{N,\sf finite}$ increases linearly in the number of queries $N$ as $k=NR/\ln 2$ for some fixed positive rate $R>0$.
Let $E^*_{\sf MSE, policy}(R)$ and $E^*_{\sf q, policy}(R)$ denote the best achievable exponentially decreasing rates of the MSE and of the quantized MSE in $N$ at a fixed rate $R$, respectively, for some policy, i.e.,
\begin{align}
E^*_{\sf MSE, policy}(R):=&\liminf_{N\to\infty}\frac{-\ln \E[|X-\hat{X}_{N,\sf finite}|^2]}{N}\label{eqn:achieve_exponetCMSE},\\
E^*_{\sf q, policy}(R):=&\liminf_{N\to\infty}\frac{-\ln \E[c_{\sf q}(X,\hat{X}_{N,\sf finite})]}{N}\label{eqn:achieve_exponetCq}.
\end{align}
Then the equality in~\eqref{eqn:diff_q_MSE} implies that for large $N$, the exponential convergence rate of the MSE $\E[|X-\hat{X}_{N,\sf finite}|^2]$ in $N$ is dominated by the minimum between the exponentially decreasing rate of the quantized MSE and 2R, i.e.,
\beq\label{eqn:diff_q_MSE_exp}
E^*_{\sf MSE, policy}(R)=\min\{E^*_{\sf q, policy}(R), 2R\}.
\eeq
For sufficiently large $R>0$ where $E^*_{\sf q, policy}(R)\leq 2R$, the MSE exponent is identical to the quantized-MSE exponent.
In this paper, we analyze performance of a querying policy by first calculating the best achievable quantized-MSE exponent $ E^*_{\sf q, policy}(R)$ at a fixed rate $R>0$ for querying resolution of $k=NR/\ln 2$ bits. 
Once the quantized-MSE exponent is calculated for every $R>0$, by using~\eqref{eqn:diff_q_MSE_exp}  we calculate the resulting MSE exponent. 

We next show how the MSE and the quantized MSE can be bounded below and above in terms of the block-decoding-error events $\{\hat{M}\neq {M}\}$ or bit-decoding-error events $\{\hat{B}_i\neq {B}_i\}$, $i\in\{1,\dots,k\}$.
The block-decoding error $\{\hat{M}\neq M\}$ occurs when any of $B_i$'s are incorrectly decoded.
For a given cost function $c(x,\hat{x}_N)$, the expected estimation error can be written in terms of block-decoding events $\{\hat{M}=M\}$ and $\{\hat{M}\neq M\}$ as
\beq
\begin{split}
&\E[c(X,\hat{X}_N)]= \Pr(\hat{M}\neq {M})\E[c(X,\hat{X}_N)\big| \hat{M}\neq {M}]+ \Pr(\hat{M}={M})\E[c(X,\hat{X}_N)\big| \hat{M}= {M}].
\end{split}
\eeq
With  the finite-resolution estimator $\hat{X}_{N,\sf finite}=\hat{M}e^{-NR}+e^{-NR}/2$, the MSE and the quantized MSE can be bounded above as
\begin{align}
&\E[|X-\hat{X}_{N,\sf finite}|^2]\leq \Pr(\hat{M}\neq M)+(e^{-NR}/2)^2\label{eqn:bd_M},\\
&\E[c_{\sf q}(X,\hat{X}_{N,\sf finite})]\leq \Pr(\hat{M}\neq M)\label{eqn:bd_M_q},
\end{align}
 by using $\E[|X-\hat{X}_{N,\sf finite}|^2|\hat{M}= M]\leq (e^{-NR}/2)^2$ and $\E[c_{\sf q}(X,\hat{X}_{N,\sf finite})|\hat{M}= M]=0$, respectively.
The achievable exponent of the MSE in~\eqref{eqn:bd_M} is determined by trade-offs between the exponentially decreasing rate of $\Pr(\hat{M}\neq M)$ at a fixed rate $R$ and the exponent $2R$.
On the other hand, in~\eqref{eqn:bd_M_q}, the achievable exponent of the quantized MSE is determined by the exponentially decreasing rate of $\Pr(\hat{M}\neq M)$ in $N$ at a fixed rate $R$.

Tighter bounds on the two estimation errors can be found by expanding the errors in terms of the   bit-decoding events. For a cost function $c(x,\hat{x}_N)$, the expected cost can be written as 
\beq\label{eqn:mmse_biterror}
\begin{split}
\E[c(X,\hat{X}_N)]&=\sum_{i=1}^{k} \left(\Pr(\hat{B}_i\neq {B}_i,\hat{B}_1^{i-1}=B_1^{i-1})\E[c(X,\hat{X}_N)\big| \hat{B}_i\neq {B}_i,\hat{B}_1^{i-1}=B_1^{i-1}]\right)\\
&\quad+ \Pr(\hat{B}_1^{k}={B}_1^{k})\E[c(X,\hat{X}_N)\big|\hat{B}_1^{k}={B}_1^{k}]
\end{split}
\eeq
where the number of information bits is $k=NR/\ln2$.
Note that conditioned on the event $\{\hat{B}_i\neq {B}_i,\hat{B}_1^{i-1}=B_1^{i-1}\}$, both the cost functions $|X-\hat{X}_{N,\sf finite}|^2$ and $c_{\sf q}(X,\hat{X}_{N,\sf finite})$ are bounded above by ${2^{-2(i-1)}}$. Thus, we have
\beq
\begin{split}
& \E[|X-\hat{X}_{N,\sf finite}|^2 \big| \hat{B}_i\neq {B}_i,\hat{B}_1^{i-1}=B_1^{i-1}]\leq {2^{-2(i-1)}},\\
& \E[c_{\sf q}(X,\hat{X}_{N,\sf finite}) \big| \hat{B}_i\neq {B}_i,\hat{B}_1^{i-1}=B_1^{i-1}]\leq{2^{-2(i-1)}}.
\end{split}
\eeq
From these bounds and~\eqref{eqn:mmse_biterror}, we can upper bound the MSE as,
\beq
\begin{split}\label{eqn:bound_bit_MSE}
\E[|X-\hat{X}_{N,\sf finite}|^2]\leq \sum_{i=1}^{k}\Pr(\hat{B}_i\neq B_i){2^{-2(i-1)}}+2^{-2k},
\end{split}
\eeq
and the quantized MSE as
\beq
\begin{split}\label{eqn:bound_bit_qMSE}
\E[c_{\sf q}(X,\hat{X}_{N,\sf finite})]\leq \sum_{i=1}^{k}\Pr(\hat{B}_i\neq B_i){2^{-2(i-1)}}.
\end{split}
\eeq

These bounds show how differently each  bit-error probability contributes to the estimation errors.
In the upper bounds on the MSE~\eqref{eqn:bound_bit_MSE} and on the quantized MSE~\eqref{eqn:bound_bit_qMSE}, we can see that the weights on the bit-error probabilities decrease exponentially in $i$ as the bit position $i$ increases corresponding to lower significance. 
In order to minimize the upper bounds on the MSE and on the quantized MSE for a fixed number of querying $N$, we need to design a querying strategy (or the associated channel coding) that can provide {unequal error protection} depending on the bit positions. 
This property differentiates a good channel-coding strategy for state estimation from that for information transmission.
In classical information-transmission problems where the cost function is $\mathbb{1}(\hat{M}\neq M)$, any bit error event $\{\hat{B}_i\neq B_i\}$ results in the same cost. Therefore, the optimal coding strategy to minimize the expected cost function $\E[\mathbb{1}(\hat{M}\neq M)]$, which equals the block-decoding-error probability $\Pr(\hat{M}\neq M)$, provides equal error protection for all the information bits. In the state-estimation problem, on the other hand, the optimal coding strategy should provide unequal error protection on information bits $\{B_1,\dots, B_k\}$ depending on the bit positions.

\section{Review of Three Different Querying Strategies}\label{sec:three_comp}
In this section, we review three well-known querying policies including the adaptive bisection policy~\cite{horstein1963sequential}, the non-adaptive UEP repetition policy~\cite{ jedynak2015}, and the non-adaptive block-querying policy based on random block coding~\cite{shannon2001mathematical} in Sections~\ref{sec:sub1},~\ref{sec:sub2}, and \ref{sec:sub3}, respectively. 
The performance of these policies are analyzed by the best achievable quantized-MSE exponent defined in~\eqref{eqn:achieve_exponetCq} with the finite-resolution estimator $\hat{X}_{N,\sf finite}$~\eqref{eqn:finiteresolution}.

\subsection{Adaptive Bisection Policy}\label{sec:sub1}
For adaptive sequential querying, greedy successive entropy minimization of a target variable is often proposed as a way to design a querying strategy for estimation of the target variable~\cite{tsiligkaridis2014collaborative,jedynak2012twenty}. 
Successive-entropy-minimization strategies select a binary query that maximally reduces the remaining uncertainty of the target variable at each round. This can be accomplished by choosing a querying region $Q_i$ that balances the probability of the event $\{X\in Q_i\}$ and the probability of the event $\{X\notin Q_i\}$, given past answers $y_1^{i-1}$, i.e.,
\beq\label{eqn:1bit_cond_5}
\begin{split}
&\Pr(X\in Q_i|Y_1^{i-1}=y_1^{i-1})=\Pr(X\notin Q_i|Y_1^{i-1}=y_1^{i-1})=1/2. 
\end{split}
\eeq
The uncertainty of the target variable is quantified by the differential entropy $h(X):=-\int p(x)\ln p(x) dx$ where $X\sim p(x)$, and the expected reduction of the uncertainty by the $i$-th querying equals
\beq\label{eqn:Cred}
h(X|Y_1^{i-1}=y_1^{i-1})-h(X|Y_i,Y_1^{i-1}=y_1^{i-1})
\eeq
where $Y_i$ is the noisy observation of the oracle's answer $Z_i=\mathbb{1}(X\in Q_i)$ transmitted through a BSC($\epsilon$).
For $Q_i$ satisfying~\eqref{eqn:1bit_cond_5},  the expected reduction of the uncertainty in~\eqref{eqn:Cred} is equal to $C:=\max_{Y_i}I(X;Y_i|Y_1^{i-1}=y_1^{i-1})= H_{\sf B}(1/2)- H_{\sf B}(\epsilon)$ where $I(X;Y_i|Y_1^{i-1}=y_1^{i-1})$ is the conditional mutual information between $X$ and $Y_i$ given $Y_1^{i-1}=y_1^{i-1}$.
After $N$ rounds of querying,  successive-entropy-minimization strategies reduce the entropy of $X$ by $NC$. From the elementary bound
\beq\label{eqn:max_ent_red2}
h(X)-h(X|Y_1^N)\leq \max_{Y^N}I(X;Y^N)=NC,
\eeq
we can see that the successive-entropy-minimization policy achieves the maximum possible entropy reduction of $X$ for $N$ uses of the BSC($\epsilon$).

\begin{figure}[t]
\centerline{\includegraphics[scale=0.4]{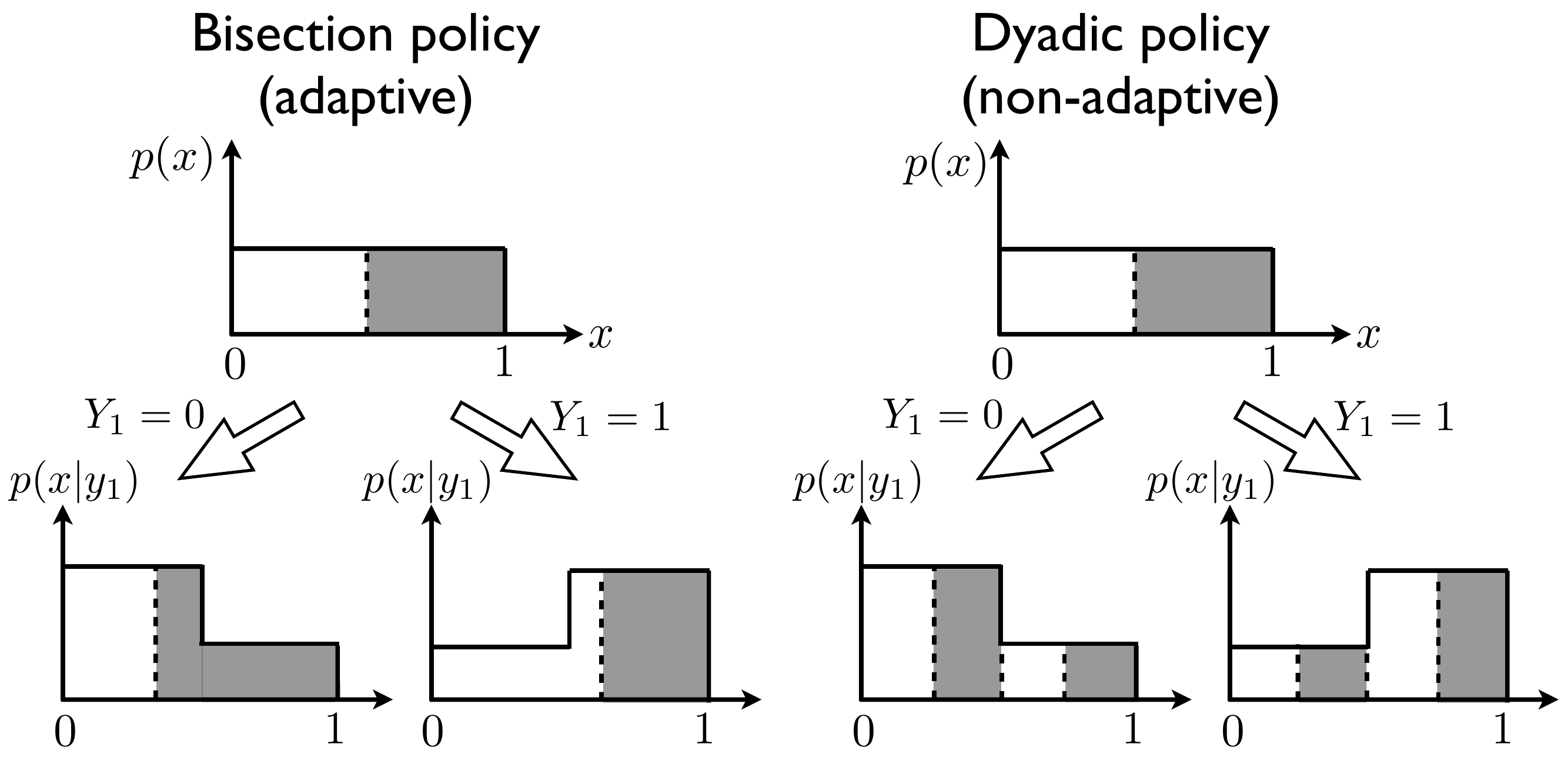}}
\caption{Illustration of two successive-entropy-minimization policies, adaptive bisection policy and non-adaptive dyadic policy, for the first two rounds. The shaded regions correspond to the posterior distribution over the querying region $Q_i$ at the $i$-th round. The bisection policy designs $Q_i$ to be right region of the median of the posterior distribution, while the dyadic policy assigns $Q_i$ to be the region that corresponds to the $i$-th bit $B_i$ of the binary expansion of $X$ being equal to 1.  The querying region $Q_i$ of the bisection policy changes depending on the received answers $Y_1^{i-1}$ of the previous queries, while that of the dyadic policy does not change on $Y_1^{i-1}$. For both the querying strategies, the shaded areas take 1/2 of the posterior distribution.}
\label{fig:bis_dya1}
\end{figure}
The bisection policy, which is also called Horstein's coding scheme~\cite{horstein1963sequential}, is one example of the successive-entropy-minimization policies. 
This policy asks whether $X$ lies to the left or right of the median of the updated posterior distribution at each round. The left figure of Fig.~\ref{fig:bis_dya1} illustrates the bisection policy for the first two rounds of querying. At the first round, the value of $X$ is uniformly distributed over $[0,1]$ and the median of the prior distribution equals 1/2. Thus, the player asks whether $X$ belongs to the right half of the region of interest $[0,1]$ by choosing $Q_1=[1/2,1]$, i.e., the player tries to extract the most significant bit of $X$.
Given the observed answer $Y_1\in\{0,1\}$, the player updates the posterior distribution $p(x|y_1)$ of $X$, and then chooses $Q_2\subset[0,1]$ that bisects the posterior distribution such that $\Pr(X\in Q_2|Y_1=y_1)=1/2$, i.e., it queries whether $X$ lies to the right of the median of the posterior distribution $p(x|y_1)$. Depending on the answer $Y_1$ to the previous query, the updated posterior distribution $p(x|y_1)$ and the median of the distribution change, so that the second querying region $Q_2$ changes as a function of the answer to the previous query. At each round,  the player keeps updating the posterior distribution $p(x|y_1^{i-1})$ of the target variable given collected answers and designs the querying region $Q_i$ to be right of the median of the updated $p(x|y_1^{i-1})$. 

The bisection policy is known to work well in practice, but there are few available theoretical guarantees for the performance of this policy. 
Here we  demonstrate that among successive-entropy-minimization policies satisfying~\eqref{eqn:1bit_cond_5} for every $i\in\{1,\dots, N\}$ the bisection policy maximally reduces the conditional variance of $X$ at each round. 
More specially, we show that the bisection policy chooses $Q_i$ that maximizes the predicted variance reduction at the $i$-th round given the answers $y_1^{i-1}$ of the previous rounds, i.e.,
\beq\label{eqn:var_gap}
\max_{Q_i}\left({\sf Var}(X|Y_1^{i-1}=y_1^{i-1})-\E[{\sf Var}(X|Y_i, Y_1^{i-1}=y_1^{i-1})]\right).
\eeq
The predicted variance reduction depends on the choice of the querying region $Q_i$ since the posterior distribution $p(x|y_i,y_1^{i-1})$ is a function of $Q_i$.
The minimum-mean-square-error  (MMSE)  estimator $\hat{X}_{N,\sf MMSE}=\E[X|Y_1^{N}=y_1^{N}]$ minimizes the MSE and makes it equal to the conditional variance of $X$ given $Y_1^N$,
\beq\label{eqn:MMSE}
 \min_{\hat{X}_N}\E[|X-\hat{X}_N|^2]=\E[(X-\hat{X}_{N,\sf MMSE})^2]=\E[\Var(X|Y_1^N)].
 \eeq
Therefore, the bisection policy, which maximizes the predicted one-step variance reduction in~\eqref{eqn:var_gap}, is the optimal myopic (greedy) policy in reducing the MSE at each round given the previous answers. But this does not necessarily mean that the bisection policy is the globally optimal policy in minimizing the MSE for a fixed number of querying. 
\begin{prop}\label{thm:opt_bis1}{\it
Among successive-entropy-minimization policies, which choose the $i$-th querying region $Q_i$ satisfying $\Pr(X\in Q_i|Y_1^{i-1}=y_1^{i-1})=\Pr(X\notin Q_i|Y_1^{i-1}=y_1^{i-1})=1/2$ given previous answers $y_1^{i-1}$, the bisection policy maximizes the predicted one-step variance reduction~\eqref{eqn:var_gap} at each round.
}
\end{prop}
\begin{IEEEproof}
Appendix~\ref{sec:adaptive}
\end{IEEEproof}

\vspace{0.1in}
\begin{rem}
In the proof of Proposition~\ref{thm:opt_bis1} in Appendix~\ref{sec:adaptive}, we show that the predicted variance reduction of $X$ due to the $i$-th query $Q_i$ is a function of not only $\{\Pr(X\in Q_i|Y_1^{i-1}=y_1^{i-1}),\Pr(X\notin Q_i|Y_1^{i-1}=y_1^{i-1})\}$ but also $\{\E[X|X\in Q_i, Y_1^{i-1}=y_1^{i-1}],\E[X|X\notin Q_i, Y_1^{i-1}=y_1^{i-1}]\}$. Successive-entropy-minimization policies select $Q_i$ that makes $\Pr(X\in Q_i|Y_1^{i-1}=y_1^{i-1})=1/2$ but do not care about the corresponding conditional expectations $\{\E[X|X\in Q_i, Y_1^{i-1}=y_1^{i-1}],\E[X|X\notin Q_i, Y_1^{i-1}=y_1^{i-1}]\}$, which also governs the conditional variance of $X$. What we show in Proposition~\ref{thm:opt_bis1} is that the choice of the querying region $Q_i$ from the bisection policy results in the selection of $\{\E[X|X\in Q_i, Y_1^{i-1}=y_1^{i-1}],\E[X|X\notin Q_i, Y_1^{i-1}=y_1^{i-1}]\}$ that maximizes the predicted variance reduction among all the possible querying regions $Q_i$ satisfying $\Pr(X\in Q_i|Y_1^{i-1}=y_1^{i-1})=1/2$. More discussions on successive-entropy-minimization policies and the proof of Proposition~\ref{thm:opt_bis1} are provided in Appendix~\ref{sec:adaptive}.
\end{rem}
\vspace{0.1in}

In this paper, we are particularly interested in the error rates of convergence achievable with the bisection policy.
Even though the error rate for the bisection policy is very hard to analyze and not known in general, a slight modification of the bisection policy proposed by Burnashev and Zigangirov  in~\cite{burnashev1974interval}, and called the BZ algorithm, is analyzable. 
The BZ algorithm works very similarly to the bisection policy, except that the boundary of the querying regions is not equal to the median of the posterior distribution. Rather, the BZ boundary is chosen among a set of uniformly quantized thresholds $\mathcal{T}=\{0,2^{-k}, 2 (2^{-k}),\dots, 2^k ( 2^{-k})\}$ with resolution $2^{-k}$. More specifically, the threshold is chosen by sampling between the two points in the set $\mathcal{T}$ that are closest to the median of the posterior distribution. 
Let $M$ denote the true index of the interval $I_M=[M2^{-k},(M+1)2^{-k})$ where the target variable $X$ belongs.
After $N$ rounds of querying with the BZ algorithm, the controller finds the sub-interval $I_{\hat{M}}=[\hat{M}2^{-k},(\hat{M}+1)2^{-k})$  where the posterior probability of $\{X\in I_{\hat{M}}\}$ is maximized and defines such a $\hat{M}$ as the estimate of $M$. 
In~\cite{burnashev1974interval,castro2008active}, it is shown that the probability of the error event $\{\hat{M}\neq M\}$ with the BZ algorithm decreases exponentially  in $N$ as 
\beq
\Pr(\hat{M}\neq M)\leq 2^k e^{-N\left(-\ln\left(1/2+\sqrt{\epsilon(1-\epsilon)}\right)\right)}
\eeq
for a fixed $k$. 
When we consider a sequence of BZ algorithms with different resolutions $\{2^{-k}\}$ where $k$ scales as $k=RN/\ln2$ for a fixed rate $R>0$, the probability of decoding error $\{\hat{M}\neq M\}$ is bounded above by
\beq\label{eqn:blockdecoding_BZ}
\Pr(\hat{M}\neq M)\leq e^{-N\left(-\ln\left(1/2+\sqrt{\epsilon(1-\epsilon)}\right)-R\right)}.
\eeq
Since the quantized MSE can be bounded above by the block-decoding-error probability as shown in~\eqref{eqn:bd_M_q}, the quantized-MSE exponent defined in~\eqref{eqn:achieve_exponetCq} is bounded below by the exponent on the right hand side of~\eqref{eqn:blockdecoding_BZ}.
\begin{lem}[Quantized-MSE exponent with BZ algorithm]{\it
The best achievable quantized-MSE exponent with the BZ algorithm, denoted $E^*_{\sf q,BZ}(R)$, is bounded below as
\beq
E^*_{\sf q, BZ}(R)\geq E_{\sf q, BZ}(R):=-\ln\left(1/2+\sqrt{\epsilon(1-\epsilon)}\right)-R
\eeq
when the resolution of the querying region scales as $k=NR/\ln 2$ bits for a fixed rate $R>0$.}
\end{lem}\label{lem:BZ_ref}


\subsection{Non-Adaptive Unequal-Error-Protection Repetition Querying }\label{sec:sub2}
Different from the adaptive policy where the updated posterior distribution $p(x|y_1^{i-1})$ is available to the controller for the design of the $i$-th querying region $Q_i$,  
for the non-adaptive policy a block of queries is determined independently of previous answers from the oracle. Our objective is to design a block of queries to estimate $X$ up to the first $k$ bits in the binary expansion of $X\approx 0.B_1B_2\dots B_k$ with the minimum estimation error $\E[c(X,\hat{X}_N)]$ for a given cost function $c(X,\hat{X}_N)$. 

We first point out that even for the non-adaptive case, there exists a block of queries $(Q_1,\dots, Q_N)$ that does not depend on $Y_1^N$ but still meets the condition~\eqref{eqn:1bit_cond_5} of successive-entropy-minimization policies for every $Y_1^N\in\{0,1\}^N$. Such a policy is the dyadic policy~\cite{jedynak2012twenty} and works as follows: The dyadic policy queries the coefficients in the dyadic expansion of $X\approx 0.B_1B_2\dots B_N$ from $B_1$ to $B_N$ one at a time over $N$ rounds of querying.  The right figure of Fig.~\ref{fig:bis_dya1} illustrates the procedure of the dyadic policy. At the first round, as does the bisection policy, the dyadic policy queries the MSB $B_1$. At the second round, regardless of $Y_1\in\{0,1\}$ it queries the second MSB $B_2$ by choosing $Q_2$ to be $Q_2=[1/4,2/4]\cup[3/4,1]$, which is the region of $X$ where $B_2=1$. The player continues the procedure of asking about $B_i$ at the $i$-th round for $i=1,\dots, N$. 
Since the prior distribution of $X$ is uniform over $[0,1]$, the quantized bits $\{B_i\}$ are i.i.d. with Bernoulli(1/2).
Moreover,  since the channel outputs $Y_1^{i-1}\in\{0,1\}^{i-1}$ contain  information only about $B_1^{i-1}$ but not about $B_i$, the events $\{B_{i}=1\}$ and $\{B_{i}=0\}$ are independent of $Y_1^{i-1}$. Therefore, the dyadic policy satisfies the condition~\eqref{eqn:1bit_cond_5} for every $y_1^{i-1}\in\{0,1\}^{i-1}$ and achieves the maximum reduction~\eqref{eqn:max_ent_red2} of the conditional entropy.

Even though the dyadic policy maximally reduces the uncertainty of $X$ measured by the entropy, this policy fails to make the estimation error converge to 0 even when $N\to\infty$.
This is because, in the BSC($\epsilon$), with $\epsilon\in(0,1/2)$ probability the player receives an incorrect value for the information bit $B_i$.
Since each bit $B_i$ is queried only once by the dyadic policy, if the player receives an incorrect answer for some bit $B_i$ there is no way to recover from this error. 
Therefore, the estimation error of the dyadic policy does not converge to 0. 

To correctly estimate $(B_1,\dots,B_k)$ through $N$ uses of the noisy BSC($\epsilon$), the player needs to design a block of queries $(Q_1,\dots, Q_N)$ with some redundancy, or equivalently design a block code with encoding map $f:\{0,\dots,2^k-1\}\to \{0,1\}^N$ to guarantee a reliable transmission of the information bits $(B_1,\dots,B_k)$. As pointed out earlier, the decoding error of each $B_i$ has different effect on the estimation error. The different importances of $B_i$'s can be quantified by the different weights on the bit error probabilities $\Pr(\hat{B}_i\neq B_i)$ in the upper bounds~\eqref{eqn:bound_bit_MSE} and~\eqref{eqn:bound_bit_qMSE} on the MSE and on the quantized MSE, respectively.
For non-adaptive block querying, in order to minimize the estimation error with a limited number $N$ of queries it is desirable to provide different levels of error protection. 

One way to provide unequal error protection is to repeat the query on  the information bits multiple times, the number of repetitions varying in accordance with the desired level of error protection.
Such a UEP repetition coding approach  was considered in~\cite{ jedynak2015}.
For this policy, the controller queries each information bit $B_i$ in the dyadic expansion of $X\approx 0.B_1\dots B_k$ repeatedly $N_i$ times and the oracle sends the uncoded bit $B_i$ repeatedly by $N_i$ uses of the BSC($\epsilon$). 
The total number of channel uses is restricted to $\sum_{i=1}^k N_i=N$ where $k$ is the resolution of  the quantification of $X$.

Note that this repetition-coding policy cannot achieve the maximum entropy reduction~\eqref{eqn:max_ent_red2} of the target variable $X$, which is achievable only when the player keeps asking the most informative query at each round. 
The repeated queries on $B_i$ successively reduce the uncertainty of $B_i$, and the bit error probability of $B_i$ decreases exponentially in the number of repeated queries, $N_i$.
The minimum bit-error probability of $B_i$ is achievable with a simple majority-voting algorithm, which claims the estimate $\hat{B}_i$ to be the more frequently received binary value at the $N_i$ channel outputs. This simple algorithm is equivalent to  maximum-likelihood (ML) decoding for $B_i$. 
\begin{lem}\label{lem:rep_biterror}{\it
When the oracle sends  a binary bit $B_i\sim\text{Bernoulli}(1/2)$ repeatedly $N_i(\geq 1)$ times through a BSC($\epsilon$), the best achievable bit-error probability with the majority-voting algorithm decreases exponentially in $N_i$ as 
\beq\label{eqn:rep_biterror}
\frac{e^{-1/(3N_i)}}{\sqrt{{2\pi N_i}}}e^{-N_i D_{\sf B}\left(1/2\|\epsilon\right)}\leq \Pr(\hat{B}_i\neq B_i)\leq e^{-N_i D_{\sf B}\left(1/2\|\epsilon\right)}.
\eeq
}
\end{lem}
\begin{IEEEproof}
Appendix~\ref{app:lem:rep_biterror}.
\end{IEEEproof}
By assigning different numbers of repetitions $(N_1,N_2,\dots N_k)$ for each information bit $B_i$ we can provide unequal error protection for the information bits.
The remaining  issue is the optimal solution for the number of repetitions $(N_1,N_2,\dots, N_k)$ that minimize the estimation error where $k$ is the total number of queried information bits.
These should be selected to minimize the upper bound on the MSE $\E[|X-\hat{X}_{N,\sf finite}|^2]$ in~\eqref{eqn:bound_bit_MSE} or the upper bound on the quantized MSE $\E[c_{\sf q}(X,\hat{X}_N)]$ in~\eqref{eqn:bound_bit_qMSE}. 
Since the weight $2^{-2(i-1)}$ on $\Pr(\hat{B}_i\neq B_i)$ decreases exponentially in $i$  and $\Pr(\hat{B}_i\neq B_i)$ decreases exponentially in $N_i$ as shown in~\eqref{eqn:rep_biterror}, the optimal $N_i^*$ that minimizes the upper bounds should decrease linearly in $i$ from MSB to LSB. This condition then implies that $N=\sum_{i=1}^k N_i^* = O(k^2)$. Therefore, the number of information bits that are queried by the optimal UEP repetition coding increases in $N$ on the order of $k=O(\sqrt{N})$, and  the corresponding rate $R=k/N$ goes to 0 as $N\to\infty$.
The resulting MSE and quantized MSE decrease exponentially only as $\sqrt{N}$.

By using the similar arguments, in~\cite{ jedynak2015} it was shown that with the UEP repetition coding, the MSE minimized over all choices of $(N_1,\dots, N_k)$ and $k$ decreases exponentially in $\sqrt{N}$ but not faster than that
\beq
 c_1 e^{-c_2\sqrt{N}}\leq \min_{(N_1,\dots, N_k),k}\E[|X-\hat{X}_{N,\sf finite}|^2]\leq c_3 e^{-c_4\sqrt{N}},
\eeq
for some positive constants $c_1,c_2,c_3,c_4>0$. 
Therefore, compared to the adaptive bisection-based policy, whose estimation error decreases exponentially in $N$, the UEP repetition coding achieves a quadratically worse exponential rate of converenge.
Moreover, the UEP repetition coding gives a MSE exponent~\eqref{eqn:achieve_exponetCMSE} and quantized-MSE exponent~\eqref{eqn:achieve_exponetCq} that is equal to zero at any positive rate $R>0$ where $k=NR/\ln2 $ bits.
\begin{lem} {\it With the UEP repetition coding, the best achievable MSE exponent and the quantized-MSE exponent are
\beq
E^*_{\sf MSE, repetition}(R)=E^*_{\sf q,repetition}(R)=0
\eeq
at any positive rate $R>0$.}
\end{lem}

For non-adaptive block querying, in order to improve the error rates of convergence we need to use more sophisticated codes that can efficiently encode $k=O(N)$ information bits in a length-$N$ codeword while guaranteeing reliable transmission of those $k=O(N)$ bits. For this purpose, we consider a non-adaptive block querying based on random block coding in the following section.

\subsection{Non-Adaptive Block Querying Based on Random Block Coding}\label{sec:sub3}
\begin{figure}[t]
\centerline{\includegraphics[scale=0.4]{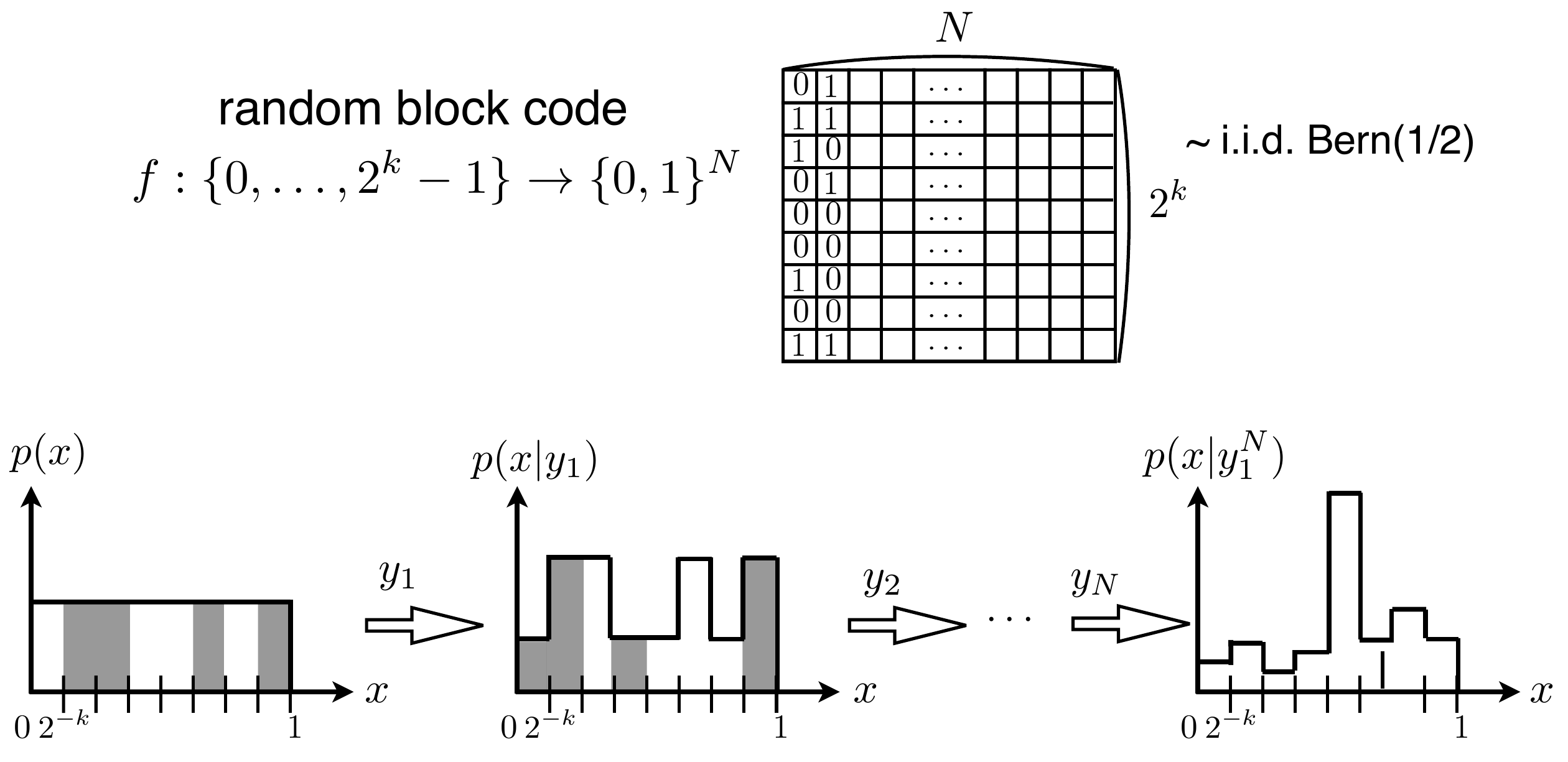}}
\caption{Non-adaptive block querying based on random block coding with encoder $f:\{0,\dots,2^{k}-1\}\to\{0,1\}^N$. The $i$-th querying region (shade region) is the union of the sub-intervals $I_{m'}=[m'2^{-k},(m'+1)2^{-k})$ for  messages $m'$ for which the associated codeword has bit 1 at the $i$-th position. Since every symbol of every codeword are i.i.d. with Bernoulli(1/2), at each querying about a half of the sub-intervals belong to the querying region. As the querying progresses, if the posterior probability of the event $\{x\in I_m\}$ for the correct message $m$ becomes higher than those of the other incorrect messages, the player can correctly decode the index $m$ of the sub-interval where the value $x$ of the target variable belongs. }
\label{fig:map_enc}
\end{figure}

In this section, we introduce a non-adaptive block-querying strategy based on random block coding~\cite{shannon2001mathematical}.
The encoding map $f:\{0,\dots,e^{NR}-1\}\to\{0,1\}^N$ of the random block codes of rate $R$ independently generates length-$N$ codewords $\bz^{(m)}=(z_1^{(m)},\dots,z_N^{(m)}):=f(m)$ each of which is composed of i.i.d. symbols of  Bernoulli(1/2) distribution. 
The player and the oracle agree on the encoding map, which in turn specifies a block of queries $(Q_1,\dots, Q_N)$.
Fig.~\ref{fig:map_enc} illustrates the one-to-one mapping between the codebook  and the block of queries. 
For a given block code with codewords $\{\bz^{(m)}\}$, $m\in\{0,\dots, e^{NR}-1\}$, where the querying resolution is $k=NR/\ln2 $ bits, the corresponding $i$-th querying region $Q_i$ becomes the union of the intervals $I_{m'}=[m'2^{-k},(m'+1)2^{-k})$ of $m'$'s such that the $i$-th answer bit $z_i^{(m')}=1$.

When the value of the target variable $X$ belongs to the sub-interval $ I_m$, the oracle transmits the length-$N$ answer bits $\bz^{(m)}=(z_1^{(m)},\dots,z_N^{(m)})$ to the block of queries $(Q_1,\dots, Q_N)$ by $N$ uses of the BSC($\epsilon$).
The length-$N$ channel-output sequence that the player receives is denoted by $\by=\bz^{(m)}\oplus \bn$ where $\bn$ is the noise sequence composed of i.i.d. symbols with Bernoulli($\epsilon$) distribution. Given the channel-output sequence $\by$, the player finds an estimate $\hat{m}$ of $m$ that maximizes the likelihood (ML decoder)
\beq
\hat{m}=\argmax_m p^N(\by|\bz^{(m)})
\eeq
where $p^N(\by|\bz)=\prod_{i=1}^N p_{Y|Z}(y_i|z_i)$ and $p_{Y|Z}(y|z)$ is the transition probability of the BSC($\epsilon$).
 Define the set of $\by$'s that are mapped to the message $m'$ by the ML decoder as $\mathcal{Y}_{m'}$ for $m'\in\{0,\dots, e^{NR}-1\}$.
Since the message $M$ is uniformly distributed over $\{0,\dots,e^{NR}-1\}$ for $X\sim \text{unif}[0,1]$, the average decoding-error probability is
\beq
\Pr(\hat{M}\neq M)=\sum_{m=0}^{e^{NR}-1} e^{-NR}\sum_{\by\notin \mathcal{Y}_m}p^N(\by|\bz^{(m)}).
\eeq
We review previous results on analyzing the exponentially decreasing rate of $\Pr(\hat{M}\neq M)$ for random block codes with the ML decoding, and use it to analyze the best achievable quantized-MSE exponent~\eqref{eqn:achieve_exponetCq} with the random block codes.


For the random block codes of rate $R$, define the best achievable error exponent for the block-decoding-error probability $\Pr(\hat{M}\neq M)$ as
\beq
E_{\sf r}(R):=\liminf_{N\to\infty}\frac{-\ln \Pr(\hat{M}\neq M)}{N}.
\eeq
For a BSC($\epsilon$) with the optimal input distribution Bernoulli(1/2), Forney's analysis~\cite{forney2001exponential} provides a  closed form solution for $E_{\sf r}(R)$,
\beq\label{eqn:randombck1}
E_{\sf r}(R)=\begin{cases}
E_0(1/2,\epsilon)-R, &  0\leq R< R_{\sf crit}(\epsilon),\\
D_{\sf B}(\gamma_{\sf GV}(R)\|\epsilon), & R_{\sf crit}(\epsilon)\leq R\leq C,
\end{cases}
\eeq
where  $E_0(a,b)=-\ln(1-2a(1-a)(\sqrt{b}-\sqrt{1-b})^2)$ and thus $E_0(1/2,\epsilon)=-\ln(1/2+\sqrt{\epsilon(1-\epsilon)})$, $R_{\sf crit}(\epsilon)=D_{\sf B}(\gamma_{\sf crit}(\epsilon)\|1/2)$ with $\gamma_{\sf crit}(\epsilon)=\frac{\sqrt{\epsilon}}{\sqrt{\epsilon}+\sqrt{1-\epsilon}}$, $C=H_{\sf B}(1/2)-H_{\sf B}(\epsilon)$, and  $\gamma_{\sf GV}(R)$ is the normalized Gilbert-Varshamov distance, defined such that $
D_{\sf B}(\gamma_{\sf GV}(R)\|1/2)=R
$. 
The exponent $E_{\sf r}(R)$ is a decreasing function of the rate $R$.
As shown in~\cite{gallager1968information} (pp. 147-149), for a very noisy channel ($\epsilon\approx 0.5$) the error exponent in~\eqref{eqn:randombck1} can be approximated as
\beq\label{eqn:randombck_approx}
E_{\sf r}(R)\approx\begin{cases}
\frac{C}{2}-R, & 0\leq  R< \frac{C}{4},\\
(\sqrt{C}-\sqrt{R})^2, & \frac{C}{4}\leq R\leq C.
\end{cases}
\eeq

From the upper bound in~\eqref{eqn:bd_M_q} we obtain the bound on the quantized MSE:
\beq\label{eqn:thm_rc2}
\E[c_{\sf q}(X,\hat{X}_{N,\sf finite})]\dot{\leq}e^{-NE_{\sf r}(R)}.
\eeq
Therefore, $E_{\sf r}(R)$ is the achievable quantized-MSE exponent. 
Moreover, we can also show that the exponent $E_{\sf r}(R)$ is not just an achievable quantized-MSE exponent but the best achievable quantized-MSE exponent with the random block coding. 
In Lemma~\ref{prop:RC_exp}, we prove this result by using fact that the random block codes provide equal error protection for every information bit, which makes the exponent of every bit-decoding-error probability equal to the exponent of the block-decoding-error probability, i.e.,
\beq
\begin{split}
&\Pr(B_i\neq B_i)\doteq \Pr(\hat{M}\neq M), \forall i\in\{1,\dots,k=NR/\ln2\}.
\end{split}
\eeq

\begin{lem}\label{prop:RC_exp}{\it
The best achievable quantized-MSE exponent $E^*_{\sf q, rc}(R)$ with the non-adaptive block-querying strategy  based on random block codes of rate $R$ is equal to
\beq\label{eqn:fact42}
E^*_{\sf q, rc}(R)=E_{\sf r}(R)
\eeq
for the random-coding exponent $E_{\sf r}(R)$ defined in~\eqref{eqn:randombck1}.
 }
\end{lem}
\begin{IEEEproof}
Appendix~\ref{app:prop:RC_exp}.
\end{IEEEproof}

Compared to the UEP repetition coding that achieves MSE and the quantized MSE decreasing exponentially only in $\sqrt{N}$, the block querying based on random block coding achieves the estimation errors exponentially decreasing  in $N$, matching the error rates of the adaptive bisection policy. 
However, the random block coding is not a MSE-optimal non-adaptive policy since it does not take into account the different contributions of decoding error of each information bit to the MSE. 
In the next section, we introduce a new non-adaptive block querying strategy based on superposition coding, which employs both coding gain and unequal error protection. 

\section{Non-Adaptive Block Querying Based on Superposition Coding}\label{sec:super}

Superposition coding~\cite{cover1972broadcast}  was originally developed as a channel-coding scheme for communications over a degraded broadcast channel where one receiver is {\it statistically stronger} than the other so that the stronger receiver can always recover the weaker receiver's message as well as its own message. The weaker receiver's message is thus treated as a public message and the stronger receiver's message as a private message. Since the public message should be decodable not only to the stronger receiver but also to the weaker receiver, a better error protection is required for the public message than for the private message. 
Superposition-coding scheme provides a higher level of error protection for the public message than for the private message.

In this section, we use this superposition-coding principles to develop a non-adaptive block-querying strategy that provides better error protection for MSBs than for LSBs in the dyadic expansion of the target variable $X\approx 0.B_1B_2\dots B_k$. Not only does the proposed strategy provide unequal error protection for MSBs vs. LSBs, but it also achieves reliable communications for $k=NR/\ln 2$ information bits at any fixed rate $0<R\leq C$ where $C$ is the capacity of a given channel. 
By unequally distributing a fixed amount of querying resource to the MSBs and LSBs of the target variable, the UEP querying strategy achieves better MSE convergence rates than that of the querying strategy based on random block coding, which distributes the querying resource equally to all the queried information bits. 

We first partition the information bits $(B_1,\dots, B_k)$ into two sub-groups, a group containing the first $k_1<k$ bits of $X$ $(B_1,\dots, B_{k_1})$  and the other group containing the remaining $k_2:=k-k_1$ bits of $X$ $(B_{k_1+1},\dots,B_{k_1+k_2})$.  The group of MSBs $(B_1,\dots, B_{k_1})$ determines the more important partial message $M_1\in\{0,\dots, 2^{k_1}-1\}$, while the group of LSBs $(B_{k_1+1},\dots,B_{k_1+k_2})$ determines the less important partial message $M_2\in\{0,\dots, 2^{k_2}-1\}$. Denote the rates of $M_1$ (MSBs) and of $M_2$ (LSBs) by $R_1=(k_1\ln2)/N$ and $R_2=(k_2\ln2)/N$, respectively.

Upon transmission of $M=(M_1,M_2)$ of total rate $R=R_1+R_2$, the quantized MSE $\E[c_{\sf q}(X,\hat{X}_N)]$  can be expressed in terms of the decoding events of the two partial messages $(M_1,M_2)$ as
\beq
\begin{split}
&\E[c_{\sf q}(X,\hat{X}_N)]\\
&=\Pr(\hat{M}_1\neq M_1)\E[c_{\sf q}(X,\hat{X}_N)|\hat{M}_1\neq M_1]\\
&\quad+\left(\Pr(\hat{M}_1= M_1,\hat{M}_2\neq M_2) \E[c_{\sf q}(X,\hat{X}_N)|\hat{M}_1= M_1,\hat{M}_2\neq M_2]\right)\\
&\quad+\left(\Pr(\hat{M}_1= M_1,\hat{M}_2= M_2) \E[c_{\sf q}(X,\hat{X}_N)|\hat{M}_1= M_1,\hat{M}_2= M_2]\right).
\end{split}
\eeq
When the partial message $M_1$, which is composed of the $NR_1$-most significant bits of $X$, can be correctly decoded, the quantized MSE associated with the finite-resolution estimator $\hat{X}_{N}=\hat{X}_{N,\sf finite}$ in~\eqref{eqn:finiteresolution} can be bounded above by $e^{-2NR_1}$. By using this bound and the fact that $\E[c_{\sf q}(X,\hat{X}_N)|\hat{M}_1= M_1,\hat{M}_2= M_2]=0$, the quantized MSE can be bounded above as
\beq\label{eqn:first_second_error2}
\begin{split}
&\E[c_{\sf q}(X,\hat{X}_{N,\sf finite})]\leq\Pr(\hat{M}_1\neq M_1)+\Pr(\hat{M}_2\neq M_2|\hat{M}_1=M_1)e^{-2NR_1}
\end{split}
\eeq
for $R_1<R$.
By the weight $e^{-2NR_1}$ on $\Pr(\hat{M}_2\neq M_2|\hat{M}_1=M_1)$, the decoding error of the partial message $M_2$ (LSBs), conditioned on the correctly decoded $M_1$ (MSBs), contributes less to the estimation error, than does the decoding error of $M_1$ (MSBs).

When we use random block coding, which provides equal error protection for every information bit of the message $M$, the best achievable decoding-error probabilities for the partial message $M_1$ (MSBs) and for $M_2$ (LSBs) conditioned on the correct estimate $\hat{M}_1= M_1$ are 
\beq
\begin{split}\label{eqn:partial_exp_rc}
&\Pr(\hat{M}_1\neq M_1)\doteq  e^{-N E_{\sf r}(R_1+R_2)},\\
&\Pr(\hat{M}_2\neq M_2|\hat{M}_1=M_1)\doteq  e^{-N E_{\sf r}(R_2)}
\end{split}
\eeq
where $E_{\sf r}(R)$ is the error exponent of the random block coding at rate $R$, defined in~\eqref{eqn:randombck1}.
Since $E_{\sf r}(R)$ is a decreasing function in the rate $R$ and thus $E_{\sf r}(R_1+R_2)<E_{\sf r}(R_2)$ for $R_1>0$, the decoding-error probability $\Pr(\hat{M}_1\neq M_1)$ of the partial message $M_1$ decreases in a slower rate than does the conditional decoding-error probability $\Pr(\hat{M}_2\neq M_2|\hat{M}_1=M_1)$ of the partial message $M_2$. 
Therefore, the exponentially decreasing rate of the quantized MSE in~\eqref{eqn:first_second_error2} is dominated by the exponentially decreasing rate of $\Pr(\hat{M}_1\neq M_1)$, and as demonstrated in Lemma~\ref{prop:RC_exp}, the best achievable quantized-MSE exponent $E^*_{\sf q, rc}(R)$ with the random block coding is equal to $E_{\sf r}(R)$ for $R=R_1+R_2$.

To improve the quantized-MSE exponent compared to that of random block coding, we need to design a UEP coding scheme that can provide higher level of error protection for $M_1$ (MSBs) to achieve a better exponentially decreasing rate of $\Pr(\hat{M}_1\neq M_1)$ than that of the random block coding in~\eqref{eqn:partial_exp_rc}. In this section, we provide such a UEP coding scheme based on superposition-coding principles. By using the improved convergence rates  of $\Pr(\hat{M}_1\neq M_1)$, we demonstrate that the proposed UEP coding scheme achieves a strictly positive gain in the exponentially decreasing rate of the quantized MSE $\E[c_{\sf q}(X,\hat{X}_N)]$ for high rate regimes of $R>0$.

\subsection{ Encoding of Superposition Codes and the Associated Non-Adaptive Block Querying}\label{sec:sup_A}
\begin{figure}[t]
\centerline{\includegraphics[scale=0.4]{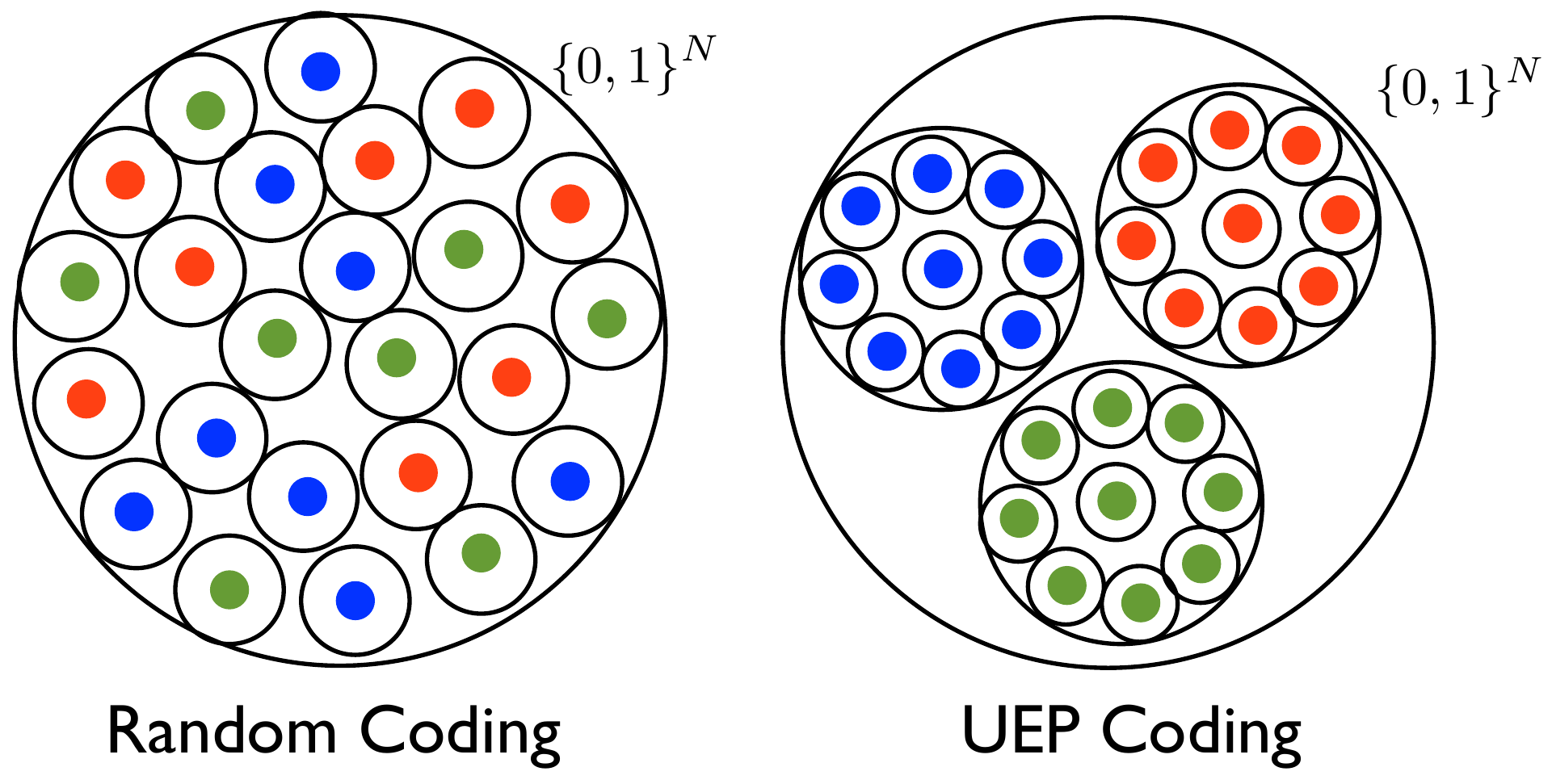}}
\caption{The distributions of codewords (each color dot) in the output space $\{0,1\}^N$ for random block coding and for UEP coding with two levels of error protection.  To better protect the color information of the codewords, which represents the MSBs of the message of the codewords, the same color codewords should be clustered together. However, this clustering makes it harder to decode the correct codeword among the same color codewords, i.e., harder to decode the LSBs of the message. }
\label{fig:UEP-2}
\end{figure}

In Fig.~\ref{fig:UEP-2}, we illustrate the codeword distributions of random block coding and of desired UEP coding with two levels of error protection, where the MSBs of the message are protected with a higher priority than are the LSBs of the message.
Each color dot is a codeword, and the shell around it is the decoding region for $M=(M_1,M_2)$ in the output space $\{0,1\}^N$. Here the partial message $M_1$ (MSBs) is represented by the color  of the codeword.
Codewords with the same color  have the same partial message $M_1$ (MSBs), while their $M_2$'s (LSBs) are different. 
For the random block coding, the same color codewords are uniformly distributed in $\{0,1\}^N$. When a noise vector corrupts the transmitted codeword beyond the correct decoding region, the decoded codeword may not have the same color as that of the transmitted codeword, since the codewords are uniformly distributed regardless of their colors and there are $e^{NR_1}$ different colors of the codewords.
On the other hand, if the same color codewords are concentrated together as shown in the right figure, even if the channel noise corrupts the transmitted codeword,  the color information will have higher probability of being correctly decoded. 
However, the probability of $M_2$ being correctly decoded given a correct estimate for $\hat{M}_1=M_1$ will be lower for the UEP coding, since the codewords of the same color are closer to each other and thus harder to be distinguished. 
We next construct codes that satisfy such a geometric property to provide two levels of unequal error protection by using superposition-coding principles.

\begin{figure}[t]
\centerline{\includegraphics[scale=0.4]{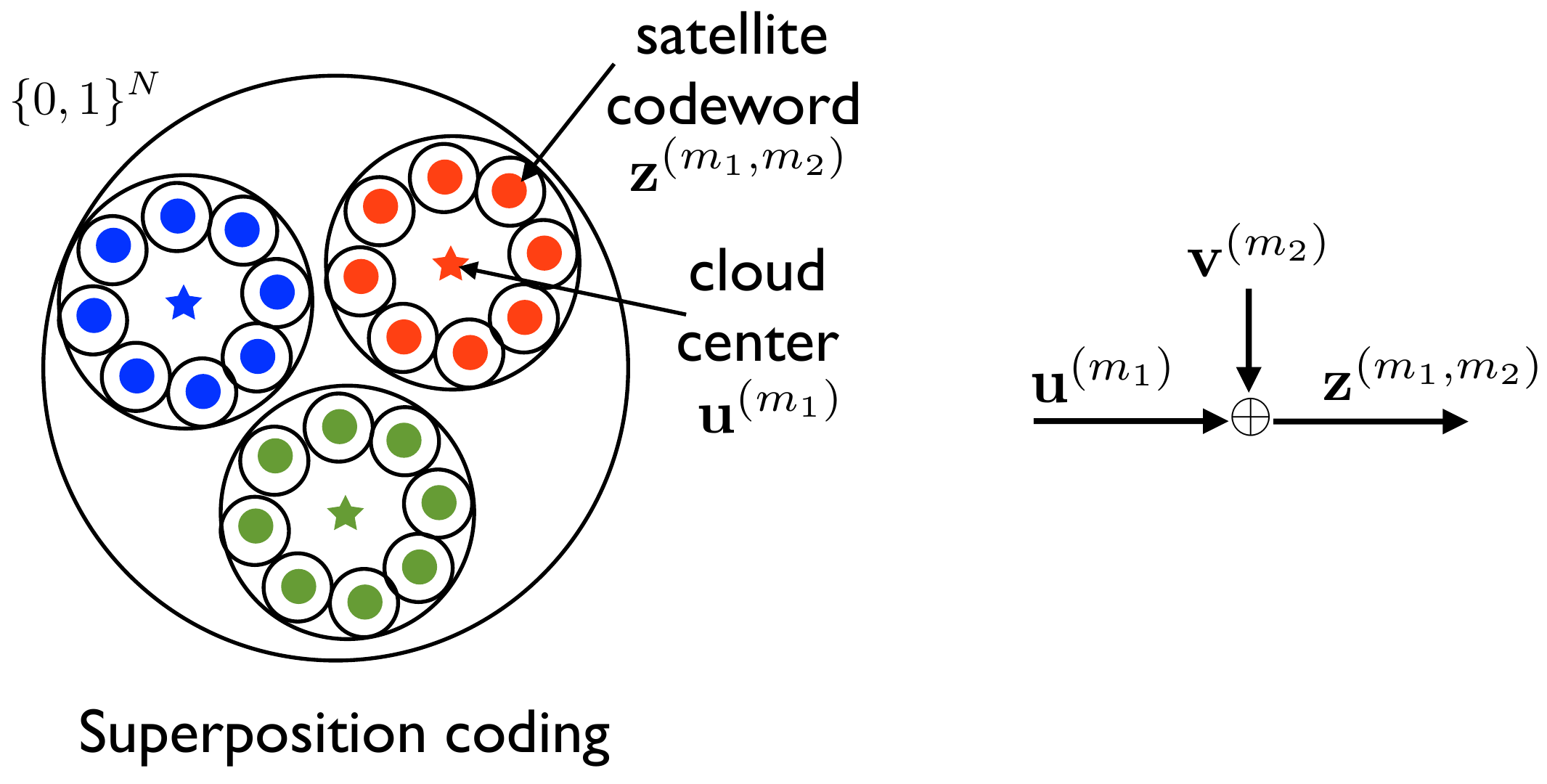}}
\caption{Superposition coding with two levels of priority, where the first partial codeword $\bu^{(m_1)}$ encodes the MSBs of the message (color information of the codewords) and the second partial codeword $\bv^{(m_2)}$ encodes the LSBs of the message. }
\label{fig:satellite}
\end{figure}

Superposition codes are constructed by superimposing two types of random block codes generated by different  distributions.
The first type of random block codes of length $N$ and rate $R_1$ is composed of $e^{NR_1}$ binary length-$N$ codewords, $\{\mathbf{u}^{(m_1)}\}$,  $m_1\in\{0,\dots, e^{NR_1}-1\}$, which encode the more important partial message $m_1$ (MSBs). The symbols of every codeword are chosen independently at random with Bernoulli(1/2) distribution.
We call these partial codewords ``cloud centers'' in the output space $\{0,1\}^N$. 
The second type of random block codes  of length $N$ and rate $R_2$ is composed of codewords $\{\mathbf{v}^{(m_2)}\}$,  $m_2\in\{0,\dots, e^{NR_2}-1\}$, and it encodes the less important partial message $m_2$ (LSBs). Every symbol of every codeword in $\{\mathbf{v}^{(m_2)}\}$ is independent and identically distributed with Bernoulli($\alpha$) distribution for a fixed $\alpha\in(0,1/2)$. This parameter $\alpha$ determines the distribution of codewords in superposition coding.
The codeword $\bz^{(m_1,m_2)}$ for the total message $(m_1,m_2)$ is designed by the bit-wise XOR of the two partial codewords $\bu^{(m_1)}$ and $\bv^{(m_2)}$. 
The superposition codes $\cC_s$ of rate $R=R_1+R_2$ are thus composed of $\{\mathbf{z}^{(m_1,m_2)}\}$ for messages $(m_1,m_2)\in\{0,\dots, e^{NR_1}-1\}\times\{0,\dots, e^{NR_2}-1\}$,  where $\mathbf{z}^{(m_1,m_2)}=\mathbf{u}^{(m_1)}\oplus \mathbf{v}^{(m_2)}$. The set of codewords $\{\mathbf{z}^{(m_1,m_2)}\}$ for a fixed $m_1$ is called ``satellite codewords'' for the respective cloud center $\bu^{(m_1)}$. There are $e^{NR_2}$ satellite codewords around each cloud center $\bu^{(m_1)}$.
Fig.~\ref{fig:satellite} illustrates the distribution of codewords with superposition coding. 

Note that when $\alpha=1/2$ the distribution of the codewords in the superposition codes $\mathcal{C}_s$ becomes the same as that of random block codes that are composed of $e^{N(R_1+R_2)}$ independent and identically distributed codewords where every symbol of every codeword is chosen independently at random with Bernoulli(1/2) distribution. Therefore, the random block codes with independent codewords of i.i.d. symbols of Beronoulli(1/2) distribution  are a special case of the superposition codes. In contrast to the case of $\alpha=1/2$, where every codeword is independent, for  superposition codes with $\alpha\in(0,1/2)$ the satellite codewords  $\{\bz^{(m_1,m_2)} \}$, $m_2\in\{0,\dots, e^{NR_2}-1\}$, for a fixed $m_1$ (the same color codewords), are mutually dependent.
Since the typical Hamming weight of $\bv^{(m_2)}$ is $N\alpha$, the typical distance between a satellite codeword $\bz^{(m_1,m_2)}=\mathbf{u}^{(m_1)}\oplus \mathbf{v}^{(m_2)}$ and its cloud center $\mathbf{u}^{(m_1)}$ is $N\alpha$. As $\alpha$ decreases from 1/2 to 0, the satellite codewords become more and more concentrated around its cloud center. 
Therefore, the superposition codes satisfy the desired geometric property for unequal error protection with two levels of error protection. The parameter $\alpha\in(0,1/2)$ determines how much the satellite codewords are concentrated around its cloud center, which determines the trade-offs between decoding-error probabilities of $M_1$ and of $M_2$. 

There exists a one-to-one mapping between the length-$N$ superposition codewords $\{\bz^{(m_1,m_2)}\}$ and the corresponding block of querying regions $(Q_1,\dots, Q_N)$.
The block of querying regions $(Q_1,\dots,Q_N)$ associated with the superposition codewords $\{\bz^{(m_1,m_2)}=\bu^{(m_1)}\oplus \bv^{(m_2)}\}$, $m_1\in\{0,\dots,2^{k_1}-1\}$, $m_2\in\{0,\dots, 2^{k_2}-1\}$,  can be represented in terms of the sub-intervals $I_{m_1}:=[m_1 2^{-k_1}, (m_1+1)2^{-k_1})$ of length $2^{-k_1}$ and another set of sub-intervals $I_{m_1, m_2}:=[m_1 2^{-k_1}+m_2 2^{-(k_1+k_2)}, m_12^{-k_1}+(m_2+1)2^{-(k_1+k_2)})$ of length $2^{-(k_1+k_2)}$ as:
\beq
\begin{split}
Q_i=&\underset{(m_1,m_2):z_{i}^{(m_1,m_2)}=1} \bigcup I_{m_1,m_2}\\
=&\left(\underset{m_1:u_i^{(m_1)}=1}\bigcup\left(I_{m_1}\cap \left(\underset{m_2:v_i^{(m_2)}=0}\bigcup I_{m_1,m_2}\right)\right) \right)\cup \left(\underset{m_1:u_i^{(m_1)}=0}\bigcup\left(I_{m_1}\cap \left(\underset{m_2:v_i^{(m_2)}=1}\cup I_{m_1,m_2}\right)\right) \right)
\end{split}
\eeq
where $u_i^{(m_1)}$ and $v_i^{(m_2)}$ are the $i$-th bit of the partial codeword $\bu^{(m_1)}$ and that of the partial codeword $\bv^{(m_1)}$, respectively.  For a partial message $m_1$ whose $i$-th bit $u_i^{(m_1)}$ of the codeword $\bu^{(m_1)}$ equals 1, about $(1-\alpha)$-fraction of the sub-intervals $\{I_{m_1,m_2}\}$ within the $I_{m_1}$ are included  in $Q_i$ since  $v_i^{(m_2)}$  is i.i.d. with Bernoulli($\alpha$), $\alpha\in(0,1/2)$. On the other hand, if   $u_i^{(m_1)}=0$ for some $m_1$, about  $\alpha$-fraction of the sub-intervals $\{I_{m_1,m_2}\}$ within the $I_{m_1}$ are included in $Q_i$.
Therefore, different from  block querying based on random block coding, where each sub-interval of $\{I_{m_1,m_2}: m_1\in\{0,\dots, 2^{k_1}-1\}, m_2\in\{0,\dots, 2^{k_2}-1\}\}$ is independently included in $Q_i$ with probability 1/2, for block querying based on superposition coding the events $\{I_{m_1,m_2}\subset Q_i\}$ ($0\leq m_2\leq 2^{k_2}-1$) for a fixed $m_1$ depend on each other. 

\begin{figure}[t]
\centerline{\includegraphics[scale=0.4]{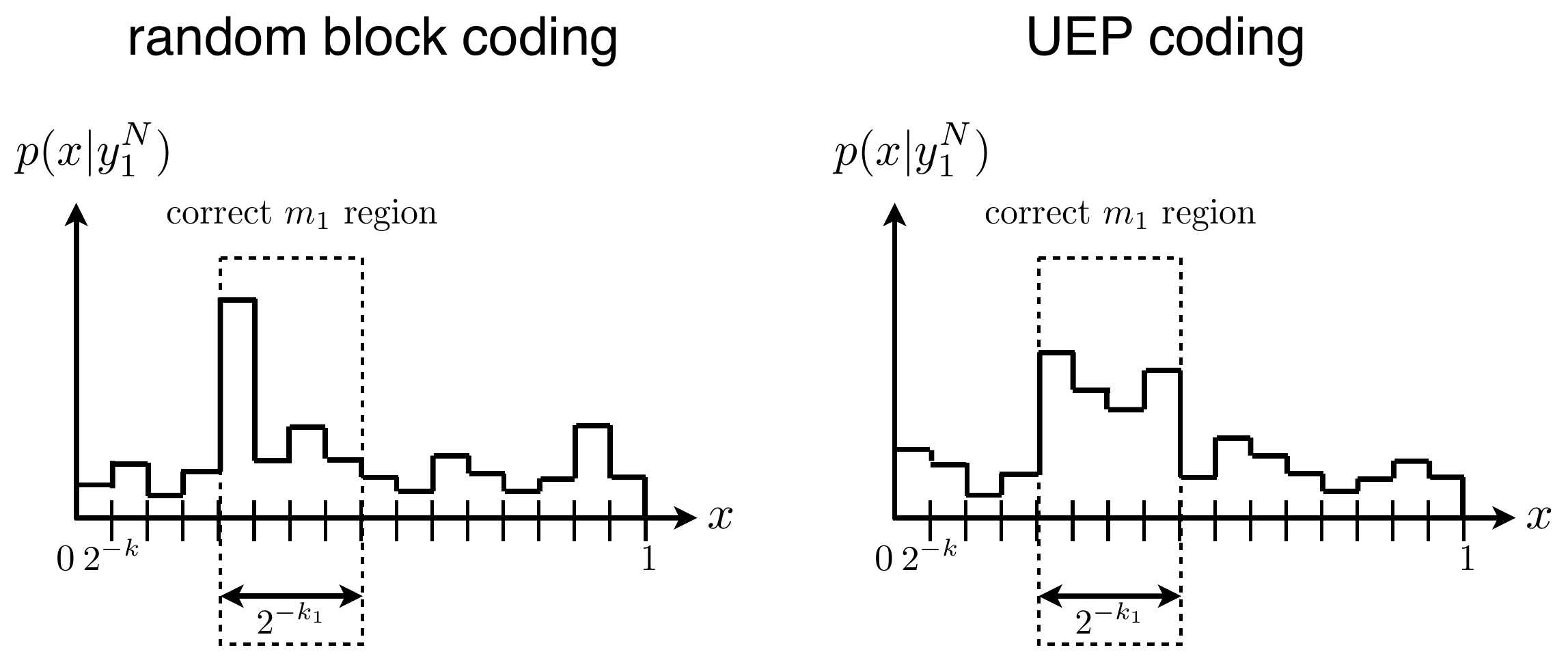}}
\caption{Illustration of the typical posterior distributions of $X$ after $N$ rounds of querying for random block coding (left) and for UEP coding with two levels of error protection (right). Consider the case where the posterior probability of $X$ in the correct $m_1$ region is higher for the UEP coding than it is for the random block coding but the peaks within the correct $m_1$ region are smoother for the UEP coding. For such a case, the UEP coding provides better error protection for $m_1$ but worse error protection for $m_2$ given the correct $\hat{m}_1=m_1$ than does the random block coding.   }
\label{fig:post_comp}
\end{figure}

Fig.~\ref{fig:post_comp} illustrates how the posterior distribution of $X$ after $N$ rounds of querying with the UEP coding (right figure) will appear as compared to that of the random block coding (left figure). 
Given the channel outputs $y_1^N$, if the posterior probability over $X\in[m_1'2^{-k_1},(m_1'+1)2^{-k_1})$ is largest for the correct $m_1'=m_1$ among all  $m_1'\in\{0,\dots, 2^{k_1}-1\}$, then the partial message $m_1$ can be correctly decoded by the optimal ML decoding for the partial message $m_1$.
If the posterior probability associated with the correct $m_1$ region is larger for the UEP coding than it is for the random block coding, one obtains an improvement in the decoding-error probability $\Pr(\hat{M}_1\neq M_1)$ of the partial message $M_1$. However, this improvement might come at the cost of degraded $\Pr(\hat{M}_2\neq M_2|\hat{M}_1=M_1)$ due to the geometric structure of codewords with the UEP coding. 


\subsection{Decoding of Superposition Codes and the Analysis of Error Exponents for Decoding-Error Probabilities}
In this section, we show that the non-adaptive block querying based on the superposition coding achieves an improved error exponent for the decoding-error probability $\Pr(\hat{M}_1\neq M_1)$  of the more important partial message $M_1$ (MSBs) as compared to that of non-adaptive block querying based on random block coding. This improvement occurs when $R_1$, the rate of $M_1$ (MSBs), is sufficiently small and $R_2<C_2(\alpha):=H_{\sf B}(\alpha*\epsilon)-H_{\sf B}(\epsilon)$, the rate of $M_2$ (LSBs), is sufficiently larger, where $\alpha*\epsilon=\alpha(1-\epsilon)+(1-\alpha)\epsilon$ and $\alpha$ is the parameter that determines the distribution of the superposition codes as explained in Section~\ref{sec:sup_A}.

Denote the maximum achievable error exponents of $\Pr(\hat{M}_1\neq M_1)$ and of $\Pr(\hat{M}_2\neq M_2|\hat{M}_1=M_1)$ with the superposition coding of rates $(R_1,R_2)$ by
\beq 
\begin{split}\label{eqn:EMSBs} 
&E^*_{\sf MSBs}(R_1,R_2,\alpha)=\liminf_{N\to\infty}\frac{-\ln\Pr(\hat{M}_1\neq M_1)}{N},\\
&E^*_{\sf LSBs}(R_2,\alpha)=\liminf_{N\to\infty}\frac{-\ln\Pr(\hat{M}_2\neq M_2|\hat{M}_1=M_1)}{N}.
\end{split}
\eeq
We analyze these exponents and compare those to the best achievable decoding-error exponents~\eqref{eqn:partial_exp_rc} of the random block codes.

There have been many previous works~\cite{kaspi2011error,korner1980universally,gallager1974capacity} to analyze the error exponents $E^*_{\sf MSBs}(R_1,R_2,\alpha)$ and $E^*_{\sf LSBs}(R_2,\alpha)$ of superposition codes.
A lower bound on $E^*_{\sf LSBs}(R_2,\alpha)$ can be calculated by directly applying the Gallager's error-exponent analysis for a discrete memoryless channel with random block codes, where codewords are independent and composed of  i.i.d. symbols having Bernoulli($\alpha$) distribution~\cite{gallager1974capacity}.
On the other hand, the analysis of $E^*_{\sf MSBs}(R_1,R_2,\alpha)$ is much more complicated, since in order to find the most probable $M_1$ (MSBs, or the color of the transmitted codeword) it involves comparisons between sums of likelihoods of exponentially many satellite codewords in Fig.~\ref{fig:satellite}, which are mutually dependent. 
The optimal maximum-likelihood (ML) decoding for the partial message $M_1$ finds $\hat{m}_1$ such that
\beq
\hat{m}_1=\argmax_{m_1}\left(\sum_{m_2} p^N\left(\by|\bz^{(m_1,m_2)}\right)\right)
\eeq
where $\by$ is the length-$N$ channel-output sequence for the input codeword $\bz^{(m_1,m_2)}$, $p^N(\by|\bz)=\prod_{i=1}^N p_{Y|Z}(y_i|z_i)$ and $p_{Y|Z}(y|z)$ is the transition probability of the BSC($\epsilon$). 
Even though there exist a few lower bounds on  $E^*_{\sf MSBs}(R_1,R_2,\alpha)$ and some bounds are shown to be numerically tighter than the others, there has been no simple closed form solution for $E^*_{\sf MSBs}(R_1,R_2,\alpha)$. Since our goal is not to exactly calculate the error exponent $E^*_{\sf MSBs}(R_1,R_2,\alpha)$ of $\Pr(\hat{M}_1\neq M_1)$ but to prove gains in this error exponent from the UEP superposition coding, we consider two well-known sub-optimal decoding rules that provide lower bounds on $E^*_{\sf MSBs}(R_1,R_2,\alpha)$. We show that these lower bounds  are  already greater than the optimal error exponent $E_{\sf r}(R_1+R_2)$ of the random block codes.

The first sub-optimal decoding rule we consider is joint-maximum-likelihood (JML) decoding for $m=(m_1,m_2)$. 
Given the received word $\by= \bz^{(m_1,m_2)}\oplus \bn$, which is a noisy version of the transmitted codeword $\bz^{(m_1,m_2)}$ added by a length-$N$ noise word $\bn$ composed of i.i.d. symbols of Bernoulli($\epsilon)$ distribution, this decoding rule finds the most probable $(\hat{m}_1,\hat{m}_2)$ such that
\beq
(\hat{m}_1,\hat{m}_2)=\argmax_{(m_1,m_2)} p^N\left(\by|\bz^{(m_1,m_2)}\right).
\eeq
Note that this decoding rule minimizes the probability of block-decoding error $(\hat{M}_1,\hat{M}_2)\neq (M_1, M_2)$ but not the probability of the partial-decoding error $\hat{M}_1\neq M_1$, so that this is a sub-optimal decoding rule for $M_1$.
The decoding error of $M_1$ happens only when $\hat{M}_1\neq M_1$, regardless of whether or not $\hat{M}_2= M_2$. 
Let $E_{\sf MSBs, JML}(R_1,R_2,\alpha)$ denote the best achievable error exponent of $\Pr(\hat{M}_1\neq M_1)$ with the JML decoding rule. 
In Lemma~\ref{lem:sup_exp_general}, we show that $E_{\sf MSBs, JML}(R_1,R_2,\alpha)\geq E_{\sf r}(R_1+R_2)$ for every $(R_1,R_2)$, regardless of the choice of $\alpha\in(0,1/2)$. 
This implies that the superposition codes provide a better, or at least as good, error protection for the partial message $M_1$ than does the random block codes for every $(R_1,R_2)$, independent of the choice of $\alpha\in(0,1/2)$.

The second sub-optimal decoding rule we consider is successive-cancellation (SC) decoding. To decode $\hat{m}_1$, this decoding rule focuses only on the geometry of the partial codewords $\left\{\bu^{(m_1)}\right\}$, $m_1\in\{0,\dots, e^{NR_1}-1\}$, (cloud centers in Fig.~\ref{fig:satellite}) while ignoring the true structure of the overall codewords $\left\{\bz^{(m_1,m_2)}\right\}$. More specifically, this decoding rule behaves as if one of  $\{\bu^{(m_1)}\}$ is transmitted and the received word $\by$ is corrupted by a noise word $\bv^{(m_2)}\oplus \bn$.
Note that $\bu^{(m_1)}$, $\bv^{(m_2)}$, and $\bn$ are independent of each other, and every symbol of the partial codeword $\bv^{(m_2)}$ is i.i.d. with Bernoulli($\alpha$) and every symbol of the noise word $\bn$ is i.i.d. with Bernoulli($\epsilon$). Therefore, the new noise word $\bv^{(m_2)}\oplus \bn$ is modeled as a sequence of i.i.d. symbols following  Bernoulli($\alpha*\epsilon$) distribution where $\alpha*\epsilon=\alpha(1-\epsilon)+(1-\alpha)\epsilon$. Denoting by $q_{Y|U}(y|u)$ the transition probability of the BSC($\alpha*\epsilon$) and defining $q^N(\by|\bu)=\prod_{i=1}^N q_{Y|U}(y_i|u_i)$, this sub-optimal decoding rule produces an estimate $\hat{m}_1$ of $m_1$ such that
\beq
\hat{m}_1=\argmax_{m_1}q^N\left(\by|\bu^{(m_1)}\right)
\eeq
for a given channel output sequence $\by$. 
After decoding $m_1$ and having the estimate $\hat{m}_1$, the SC decoding rule subtracts $\bu^{(\hat{m}_1)}$ from $\by$ and finds the estimate $\hat{m}_2$ for the partial message $m_2$ (LSBs) that maximizes the likelihood of $p^N\left(\by\oplus \bu^{(\hat{m}_1)}|\bv^{(m_2)}\right)$
\beq
\hat{m}_2=\argmax_{m_2} p^N\left(\by\oplus \bu^{(\hat{m}_1)}|\bv^{(m_2)}\right).
\eeq
Let $E_{\sf MSBs, SC}(R_1,\alpha)$ and $E_{\sf LSBs, SC}(R_2,\alpha)$ denote the best achievable error exponents of $\Pr(\hat{M}_1\neq M_1)$ and of  $\Pr(\hat{M}_2\neq M_2|\hat{M}_1= M_1)$, respectively, with the SC decoding rule.
Forney's analysis~\cite{forney2001exponential} yields the exponentially-tight error exponent $E_{\sf MSBs, SC}(R_1,R_2,\alpha)$ for $\Pr(\hat{M}_1\neq M_1)$ with the SC decoding rule: 
\beq\label{eqn:ofstrcit2}
\begin{split}
E_{\sf MSBs, \sf SC}(R_1, \alpha)
=
\begin{cases}
E_0(1/2,\alpha*\epsilon)-R_1, &0\leq R_1\leq R_{\sf crit}(\alpha*\epsilon),\\
D_{\sf B}(\gamma_{\sf GV}(R_1)\|\alpha*\epsilon), &R_{\sf crit}(\alpha*\epsilon)<R_1\leq C-C_2(\alpha).
\end{cases}
\end{split}
\eeq
Here $\gamma_{\sf GV}\in[0,1/2]$ is the Gilbert-Varshamov distance satisfying $D_{\sf B}(\gamma_{\sf GV}(R)\|1/2)=R$, $E_0(a,b)=-\ln(1-2a(1-a)(\sqrt{b}-\sqrt{1-b})^2)$ and thus $E_0(1/2,\alpha*\epsilon)=-\ln(1/2+\sqrt{(\alpha*\epsilon)(1-(\alpha*\epsilon))})$, $C=H_{\sf B}(1/2)-H_{\sf B}(\epsilon)$, $C_2(\alpha)=H_{\sf B}(\alpha*\epsilon)-H_{\sf B}(\epsilon)$, $R_{\sf crit}(\alpha*\epsilon)=D_{\sf B}(\gamma_{\sf crit}(\alpha*\epsilon)\|1/2)$ for $\gamma_{\sf crit}(\alpha*\epsilon)=\frac{\sqrt{\alpha*\epsilon}}{\sqrt{\alpha*\epsilon}+\sqrt{1-\alpha*\epsilon}}$.
For a given $\alpha\in(0,1/2)$, the error exponent $E_{\sf LSBs,SC}(R_2,\alpha)$ of $\Pr(\hat{M}_2\neq M_2|\hat{M}_1= M_1)$ can be shown to be positive for every $0\leq R_2< C_2(\alpha)$.

The following lemma summarizes two lower bounds on the error exponent $E^*_{\sf MSBs}(R_1,R_2,\alpha)$~\eqref{eqn:EMSBs}  achievable with the sub-optimal JML decoding and with the sub-optimal SC decoding, respectively. 
\begin{lem}\label{lem:sup_exp_general}
{\it
Superposition coding provides a better, or at least as good, error protection for the partial message $M_1$ (MSBs of the message) than does the random block coding for every pair of rates $(R_1,R_2)$ of the partial messages $(M_1,M_2)$, regardless of the choice of the parameter $\alpha\in(0,1/2)$ of the superposition coding.  With joint-maximum-likelihood (JML) decoding for superposition codes, the error exponent $E_{\sf MSBs, JML}(R_1,R_2)$ of $\Pr(\hat{M}_1\neq M_1)$, which is greater than or equal to $E_{\sf r}(R)$, is achievable, i.e.,
\beq\label{eqn:Em1Er_general}
E^*_{\sf MSBs}(R_1,R_2,\alpha)\geq E_{\sf MSBs, JML}(R_1,R_2)\geq E_{\sf r}(R_1+R_2).
\eeq
Moreover, for a sufficiently small $R_1>0$ and sufficiently large $R_2<C_2(\alpha)=H_{\sf B}(\alpha*\epsilon)-H_{\sf B}(\epsilon)$, a strictly positive gain in the error exponent can be achieved using successive-cancellation (SC) decoding rule, i.e.,
\beq\label{eqn:Em1Er_general_strict1}
E^*_{\sf MSBs}(R_1,R_2,\alpha)\geq E_{\sf MSBs, SC}(R_1,\alpha)> E_{\sf r}(R_1+R_2)
\eeq
where $E_{\sf MSBs, SC}(R_1,\alpha)$ is the best achievable decoding-error exponent for $M_1$ using the SC decoding rule.
}
\end{lem}
\begin{IEEEproof}
Appendix~~\ref{sec:pf_sup_exp}.
\end{IEEEproof}

For a very noisy BSC($\epsilon$), we can further demonstrate that, when we choose the rate $R_2$ of the partial message $M_2$ (LSBs) equal to the maximum possible rate $C_2(\alpha)$ to guarantee $\Pr(\hat{M}_2\neq M_2|\hat{M}_1=M_1)\to0$ as $N\to\infty$,
the superposition coding provides a strictly positive gain in the error exponent of $\Pr(\hat{M}_1\neq M_1)$ as compared to that of the random block coding, for the entire regime of $R_1\in[0,C-C_2(\alpha))$ where $E^*_{\sf MSBs}(R_1,R_2,\alpha)$ is positive. 
\begin{lem}\label{lem:strict1}
{\it
For a very noisy BSC($\epsilon$) where $\epsilon= 0.5-\delta$ for a sufficiently small $\delta>0$, assume a fixed $\alpha\in(0,1/2)$ and the rate $R_2=C_2(\alpha)$. Then the best achievable error exponent $E^*_{\sf MSBs}(R_1,R_2,\alpha)$ of $\Pr(\hat{M}_1\neq M_1)$ for superposition coding is strictly larger than the best achievable error exponent $E_{\sf r}(R_1+R_2)$ of random block coding  for every $R_1\in[0,C-C_2(\alpha))$. In particular, with successive cancellation (SC) decoding rule we can achieve an error exponent $E_{\sf MSBs,SC}(R_1,\alpha)$ that is strictly larger than $E_{\sf r}(R_1+R_2)$,
\beq\label{eqn:ofstrcit1}
E^*_{\sf MSBs}(R_1,R_2,\alpha)\geq E_{\sf MSBs,SC}(R_1,\alpha)>E_{\sf r}(R_1+R_2),
\eeq
for every $R_1\in[0,C-C_2(\alpha))$.
}
\end{lem}
\begin{IEEEproof}
Appendix~\ref{app:cor:strict}.
\end{IEEEproof}
\begin{figure}[t]
\centerline{\includegraphics[scale=0.6]{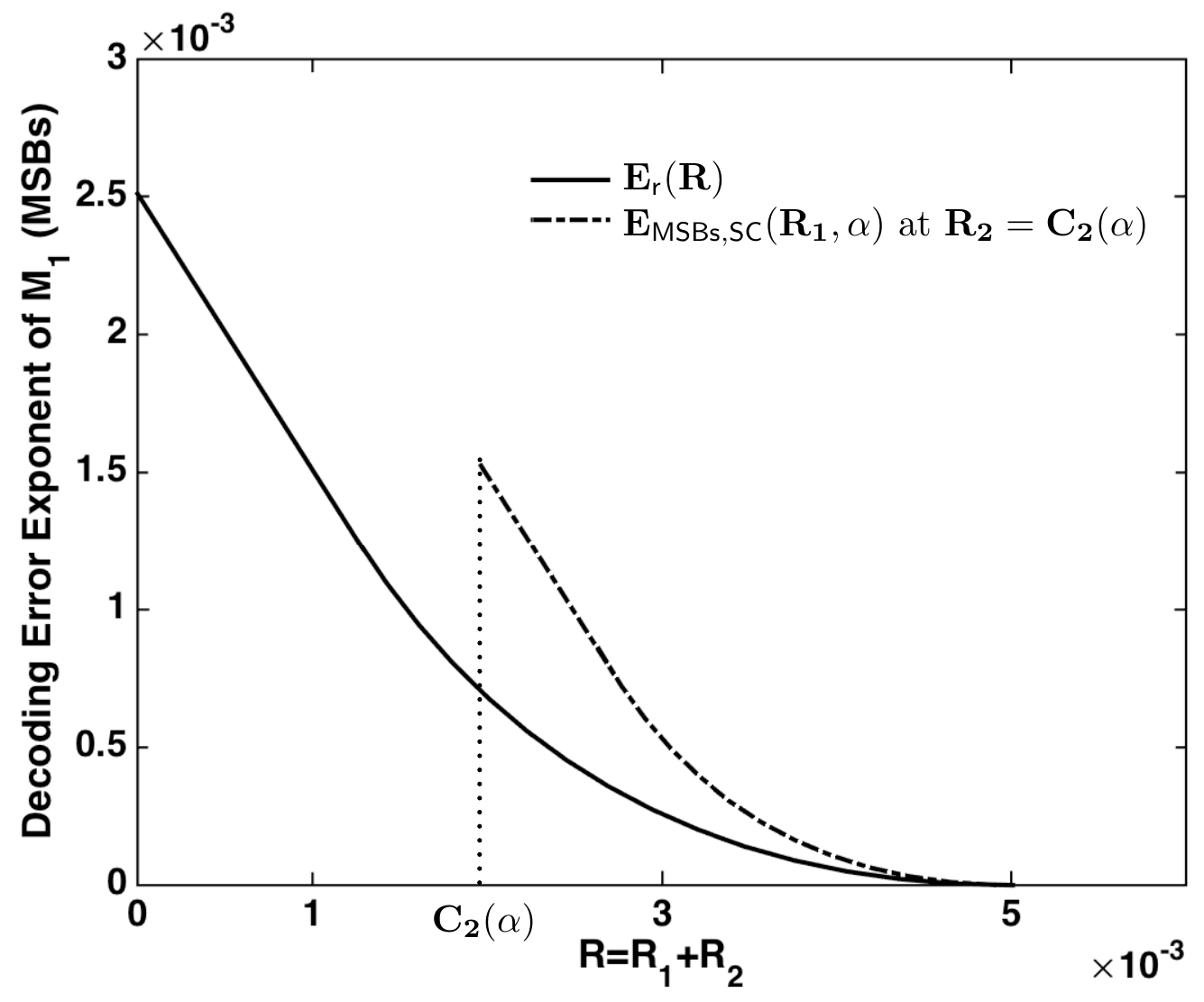}}
\caption{A plot of decoding-error exponents of the more important partial message $M_1$ (MSBs) for random block coding, $E_{\sf r}(R_1+R_2)$ (solid line), and for superposition coding with successive-cancellation decoding, $E_{\sf MSBs,\sf SC}(R_1,\alpha)$ (dash-dot line), where $\epsilon=0.45$ and $\alpha=0.11$. When the rate $R_2$ of the less important partial message $M_2$ (LSBs) equals $C_2(\alpha)$, which is the maximum rate to guarantee $\Pr(\hat{M}_2\neq M_2|\hat{M}_1=M_1)\to 0$ as $N\to\infty$, for every $R_1\in[0,C-C_2(\alpha))$ the error exponent $E_{\sf MSBs, SC}(R_1,\alpha)$ of superposition coding is larger than the error exponent $E_{\sf r}(R_1+R_2)$ of random block coding.
}
\label{fig:Em1SCgain}
\end{figure}

In Fig~\ref{fig:Em1SCgain}, we provide a plot of the error exponent $E_{\sf r}(R_1+R_2)$ of random block coding (solid line) and the error exponent $E_{\sf MSBs,\sf SC}(R_1,\alpha)$ of superposition coding with successive-cancellation decoding (dash-dot line) over $R=R_1+R_2$ for a BSC(0.45) with a fixed $\alpha=0.11$ and $R_2=C_2(\alpha)$.
The plot for $E_{\sf MSBs,\sf SC}(R_1,\alpha)$  starts from $R=C_2(\alpha)$ at which $R_1=0$.
It is shown that $E_{\sf MSBs,\sf SC}(R_1,\alpha)$ is larger than $E_{\sf r}(R_1+R_2)$  for every $R_1\in[0,C-C_2(\alpha))$.

Lemma~\ref{lem:sup_exp_general} and~\ref{lem:strict1} demonstrate that even with the sub-optimal decoding rules (either JML or SC rule) the superposing coding can provide a better error protection for the MSBs of the transmitted message than does the random block coding. 
In the next section, we use this result to show that the superposition coding achieves strictly positive gains in the exponentially decreasing rates of quantized MSE and MSE. 

\subsection{Gains in the quantized-MSE exponent and MSE Exponent from Superposition Coding}

By using the improvement in the decoding-error exponent of  $M_1$ (MSBs) from superposition coding, we next demonstrate a gain in the exponentially decreasing rate of the quantized MSE $\E[c_{\sf q}(X,\hat{X}_{N,\sf finite})]$ for the quantized cost function  $c_{\sf q}(X,\hat{X}_{N,\sf finite})$~\eqref{eqn:c_qdefFinite} of resolution $k=NR/\ln 2$ bits for a fixed rate $R>0$. %
Define $E^*_{\sf q, spc }(R)$ the best achievable exponentially  decreasing rate of  $\E[c_{\sf q}(X,\hat{X}_{N,\sf finite})]$ with the non-adaptive block querying based on the superposition coding (SPC) of rate $R$:
\beq
E^*_{\sf q, spc }(R)=\liminf_{N\to\infty} \frac{-\ln \E[c_{\sf q}(X,\hat{X}_{N,\sf finite})]}{N}.
\eeq

As shown in~\eqref{eqn:first_second_error2}, the quantized MSE is bounded above by
\beq\label{eqn:first_second_error222}
\begin{split}
&\E[c_{\sf q}(X,\hat{X}_{N,\sf finite})]\leq\Pr(\hat{M}_1\neq M_1)+\Pr(\hat{M}_2\neq M_2|\hat{M}_1=M_1)e^{-2NR_1},
\end{split}
\eeq
and the exponentially decreasing rate of the quantized MSE is dominated either by  the error exponent of $\Pr(\hat{M}_1\neq M_1)$ or by the error exponent of $\Pr(\hat{M}_2\neq M_2|\hat{M}_1=M_1)$ plus $2R_1$. For random block coding, the exponentially decreasing rate $E_{\sf r}(R)$ of $\Pr(\hat{M}_1\neq M_1)$ is smaller than that of $\Pr(\hat{M}_2\neq M_2|\hat{M}_1=M_1)$, so that the random block coding achieves the quantized-MSE exponent   $E^*_{\sf q, rc}(R)$ equal to $E_{\sf r}(R)$, as demonstrated in Lemma~\ref{prop:RC_exp}.

In Lemma~\ref{lem:sup_exp_general} and~\ref{lem:strict1}, we showed that the superposition coding achieves a better error exponent of  $\Pr(\hat{M}_1\neq M_1)$ by providing higher-level  error protection for $M_1$ (MSBs) than does the random block coding. But this improvement comes with  degraded error protection for $M_2$ (LSBs). Therefore, to analyze the best achievable quantized-MSE exponent $E^*_{\sf q, spc}(R)$ with the superposition coding, we need to consider the trade-offs in the levels of error protection for $M_1$ (MSBs) and for $M_2$ (LSBs), which can be controlled by the choice of the respective rates $(R_1,R_2)$ of the two partial messages, under the constraint of the total rate $R_1+R_2=R$, and the choice of the distribution parameter $\alpha\in(0,1/2)$ of the superposition coding. From~\eqref{eqn:first_second_error222}, the best achievable quantized-MSE exponent $E^*_{\sf q, spc}(R)$ is bounded below by
\beq\label{eqn:error_exp_trade_off}
\begin{split}
&E^*_{\sf q, spc}(R)\geq\max_{\substack{\{(R_1,R_2,\alpha):\\ R_1+R_2=R, \\ \alpha\in(0,1/2)\}}}\min\{E^*_{\sf MSBs}(R_1,R_2,\alpha),E^*_{\sf LSBs}(R_2,\alpha)+2R_1\},
\end{split}
\eeq
where $E^*_{\sf MSBs}(R_1,R_2,\alpha)$ and $E^*_{\sf LSBs}(R_2,\alpha)$ are the best achievable error exponents of $\Pr(\hat{M}_1\neq M_1)$ and of $\Pr(\hat{M}_2\neq M_2|\hat{M}_1= M_1)$, respectively, with the superposition coding, as defined in~\eqref{eqn:EMSBs}. 

For a given $\alpha\in(0,1/2)$, when we choose the rate $R_2$ of the partial message $M_2$ (LSBs) equal to $C_2(\alpha)=H_{\sf B}(\alpha*\epsilon)-H_{\sf B}(\epsilon)$, which is the maximum possible rate of $M_2$ to guarantee $\Pr(\hat{M}_2\neq M_2|\hat{M}_1= M_1)\to 0$ as $N\to\infty$, the resulting error exponent $E^*_{\sf LSBs}(R_2,\alpha)$ equals 0. This particular choice of $R_2=C_2(\alpha)$ provides a lower bound on $E^*_{\sf q,spc}(R)$ such that
\beq\label{eqn:lower_Eqspc_new}
\begin{split}
&E^*_{\sf q, spc}(R)\geq \max_{\{\alpha: \alpha\in(0,1/2)\}}\min\{E^*_{\sf MSBs}(R-C_2(\alpha),C_2(\alpha),\alpha), 2(R-C_2(\alpha))\}.
\end{split}
\eeq
The optimization in the right-hand side is about finding the optimal value of the distribution parameter $\alpha\in(0,1/2)$ of the superposition coding.

In the theorem below, we prove that for a very noisy BSC($\epsilon$) the quantized-MSE exponent $E^*_{\sf q,spc}(R)$ is strictly larger than that of random block coding, i.e.,
\beq
E^*_{\sf q,spc}(R)>E^*_{\sf q, rc}(R)= E_{\sf r}(R),
\eeq
at high-rate regimes of $R\in(E_0(1/2,\epsilon)/3,C)$, by solving the optimization in the right-hand side of~\eqref{eqn:lower_Eqspc_new} and proving that this lower bound is greater than the best achievable quantized-MSE exponent $E^*_{\sf q, rc}(R)$ of random block coding. To prove this theorem, we use Lemma~\ref{lem:strict1} where we showed that for a very noisy BSC($\epsilon$) successive-cancellation decoding for superposition coding provides a strictly positive gain in the error exponent of $M_1$ (MSBs) at every rate $R_1\in(0,C-C_2(\alpha))$ of $M_1$ (MSBs) when the rate $R_2$ of $M_2$ (LSBs) is fixed as $R_2=C_2(\alpha)$.

\begin{thm}\label{thm:sup} {\it  For a very noisy BSC($\epsilon$) with $\epsilon= 0.5-\delta$ for a sufficiently small $\delta>0$, the best achievable quantized-MSE exponent  $E^*_{\sf q,spc}(R)$ of superposition coding  is strictly larger than the best achievable quantized-MSE exponent $E^*_{\sf q, rc}(R)$ of random block coding for every rate $R\in(E_0(1/2,\epsilon)/3,C)$ where $E_0(1/2,\epsilon)=-\ln(1/2+\sqrt{\epsilon(1-\epsilon)})\approx C/6$ and $C=H_{\sf B}(1/2)-H_{\sf B}(\epsilon)$. In particular, successive-cancellation (SC) decoding for superposition coding achieves the quantized-MSE exponent $E_{\sf q, spc}(R)$ that is strictly larger than $E^*_{\sf q,rc}(R)$ for $R\in(E_0(1/2,\epsilon)/3,C)$, i.e.,
\beq\label{eqn:thmeqan}
E^*_{\sf q, spc}(R)\geq E_{\sf q,spc}(R)>E^*_{\sf q,rc}(R)=E_{\sf r}(R),
\eeq
where 
\beq\label{eqn:fact62}
E_{\sf q, spc}(R)=E_{\sf MSBs, SC}(R-C_2(\alpha^*),\alpha^*)
\eeq
for $E_{\sf MSBs, SC}(R_1,\alpha)$ in~\eqref{eqn:ofstrcit2}, $C_2(\alpha)=H_{\sf B}(\alpha*\epsilon)-H_{\sf B}(\epsilon)$  and $\alpha^*\in(0,1/2)$ satisfying $R=C_2(\alpha)+\frac{E_0(1/2,\alpha*\epsilon)}{3}$. 
}
\end{thm}
\begin{IEEEproof}
Appendix~\ref{app:sup}
\end{IEEEproof}

\begin{figure}[t]
\centerline{\includegraphics[scale=0.6]{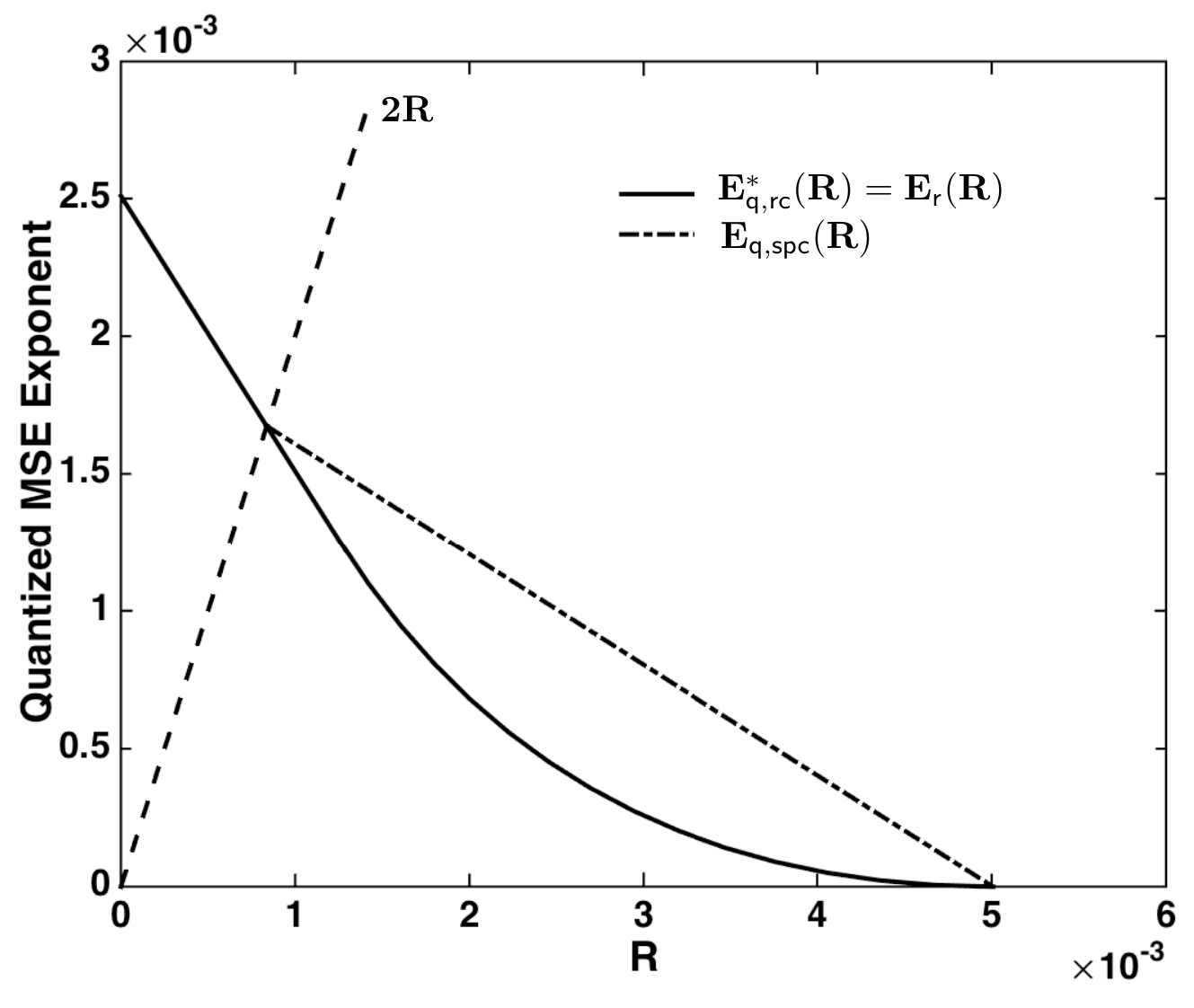}}
\caption{
A plot of $E^*_{\sf q, rc}(R)=E_{\sf r}(R)$, $E_{\sf q, spc}(R)$, and $2R$ for a BSC($\epsilon$) with $\epsilon=0.45$ where  $E^*_{\sf q, rc}(R)$ is the best achievable quantized-MSE exponent with random block coding and $E_{\sf q,spc}(R)$ is a lower bound on the best achievable quantized-MSE exponent $E^*_{\sf q,spc}(R)$ with superposition coding.
For any $R\in(E_0(1/2,\epsilon)/3,/C)$, there exists a gain in the achievable quantized-MSE exponent from superposition coding compared to that of random block coding.  
}
\label{fig:cq}
\end{figure}

In Fig~\ref{fig:cq}, we provide a plot of $E_{\sf q, spc}(R)$ in~\eqref{eqn:fact62} (dash-dot line), which is a lower bound on the best achievable quantized-MSE exponent $E^*_{\sf q, spc}(R)$ with superposition coding, and $E^*_{\sf q, rc}(R)$ in~\eqref{eqn:fact42} (solid line), which is the best achievable quantized-MSE exponent with random block coding, with the line $2R$ (dashed line). Here we consider a BSC($\epsilon$) with $\epsilon=0.45$. We can observe that the achievable quantized-MSE exponent $E_{\sf q, spc}(R)$ with superposition coding is strictly larger than the best achievable quantized-MSE exponent $E^*_{\sf q, rc}(R)$ with random block coding, at every rate $R\in (E_0(1/2,\epsilon)/3,C)$, as stated in Theorem~\ref{thm:sup}. 

We next consider the achievable MSE $\E[|X-\hat{X}_{N,\sf finite}|^2]$ with the superposition coding and demonstrate a gain in the MSE exponent in the high-rate regimes as compared to that of the random block coding. As shown in~\eqref{eqn:diff_q_MSE} and~\eqref{eqn:diff_q_MSE_exp} the MSE can be written as a sum of the quantized MSE $\E[c_{\sf q}(X,\hat{X}_{N,\sf finite})]\doteq e^{-NE^*_{\sf q, policy}(R)}$ and the estimation error from the finite-resolution estimator $\hat{X}_{N,\sf finite}$ as
\beq\label{eqn:1}
\begin{split}
\E[|X-\hat{X}_{N,\sf finite}|^2]&\doteq e^{-NE^*_{\sf q, policy}(R)}+e^{-N2R}\doteq e^{-N\min\{E^*_{\sf q, policy}(R),2R\}}.
\end{split}
\eeq
As shown in~\eqref{eqn:diff_q_MSE_exp}, the MSE exponent $E^*_{\sf MSE, policy}(R)$ and the quantized-MSE exponent $E^*_{\sf q, policy}(R)$ at a fixed rate $R>0$ are related as
\beq\label{eqn:comp_Es}
E^*_{\sf MSE, policy}(R)=\min\{E^*_{\sf q, policy}(R),2R\}.
\eeq
When $E^*_{\sf q, policy}(R)>2R$, the MSE exponent at a fixed rate $R$ is limited by the quantization error from the finite-resolution estimator of rate $R$. When  $E^*_{\sf q,policy}(R)\leq 2R$, on the other hand, the MSE exponent is governed by the quantized-MSE exponent, which depends on the error exponents of decoding-error probabilities of the two partial messages $(M_1,M_2)$ of rates $(R_1,R_2)$ where $R_1+R_2=R$. 

For random block coding, since the quantized-MSE exponent $E^*_{\sf q, rc}(R)$ is equal to $E_{\sf r}(R)$, the MSE exponent $E^*_{\sf MSE, rc}(R)$ of random block coding equals
\beq
E^*_{\sf MSE, rc}(R)=\min\{E_{\sf r}(R),2R\}.
\eeq
For a very noisy BSC($\epsilon$), the decoding-error exponent $E_{\sf r}(R)$ of random block coding can be approximated as~\eqref{eqn:randombck1}.
By using this approximation, we can show that, where $\epsilon\in[0.5-\delta,0.5]$ for a sufficiently small $\delta>0$,
\beq\label{eqn:fact66}
E^*_{\sf MSE, rc}(R)=\begin{cases}
2R,&0\leq R\leq E_0(1/2,\epsilon)/3,\\
E_{\sf r}(R), & E_0(1/2,\epsilon)/3< R\leq C.
\end{cases}
\eeq
In the low-rate regime of $0\leq R\leq E_0(1/2,\epsilon)/3$, the MSE exponent $E^*_{\sf MSE, rc}(R)$ of the random block coding is dominated by the estimation error from the finite-resolution estimator of rate $R$. On the other hand, in the high rate regime of $ E_0(1/2,\epsilon)/3< R\leq C$, the MSE exponent $E^*_{\sf MSE, rc}(R)$  is dominated by the quantized-MSE exponent $E^*_{\sf q, rc}(R)$, which is equal to $E_{\sf r}(R)$.

We next consider the MSE exponent $E^*_{\sf MSE, spc}(R)$ of superposition coding, which is equal to
\beq
E^*_{\sf MSE, spc}(R)=\min\{E^*_{\sf q, spc}(R),2R\}.
\eeq
In Theorem~\ref{thm:sup}, we demonstrated that the quantized-MSE exponent $E^*_{\sf q, spc}(R)$ of superposition coding is strictly larger than that of random block coding, i.e., $E^*_{\sf q, spc}(R)>E^*_{\sf q, rc}(R)=E_{\sf r}(R)$ at any rate $R\in(E_0(1/2,\epsilon)/3,C)$ for a very noisy BSC($\epsilon$). Combining this result with the fact that $E^*_{\sf MSE, rc}(R)=E_{\sf r}(R)>2R$ in this regime of $R\in(E_0(1/2,\epsilon)/3,C)$,
we can conclude that the MSE exponent $E^*_{\sf MSE, spc}(R)$ with superposition coding is strictly larger than the MSE exponent $E^*_{\sf MSE, rc}(R)$ of random block coding in $R\in(E_0(1/2,\epsilon)/3,C)$ for a very noisy BSC($\epsilon$).
\begin{cor}\label{cor}
{\it For a very noisy BSC($\epsilon$) with $\epsilon= 0.5-\delta$ for a sufficiently small $\delta>0$, the MSE exponent  $E^*_{\sf MSE,spc}(R)$ with  superposition coding  is strictly larger than that of random block coding $E^*_{\sf MSE, rc}(R)$ at any rate $R\in(E_0(1/2,\epsilon)/3,C)$, i.e,
\beq\label{MSE:spc:rc}
E^*_{\sf MSE, spc}(R)>E^*_{\sf MSE, rc}(R)=E_{\sf r}(R).
\eeq
}
\end{cor}
The non-adaptive block querying based on superposition coding thus achieves a strictly lager MSE exponent than that of random block coding, when the querying resolution of the block querying strategy scales as $k=NR/\ln 2$ bits for any rate $R\in (E_0(1/2,\epsilon)/3,C)$ over a very noisy BSC($\epsilon$).

In Fig~\ref{fig:cq}, we can see that the gain in the quantized-MSE exponent from superposition coding in $R\in(E_0(1/2,\epsilon)/3,C)$ also results in a gain in the MSE exponent in this high rate regime, since the MSE exponent $E^*_{\sf MSE, spc}(R)$ of superposition coding is proven to be at least  larger than $E_{\sf r}(R)$  in this regime, which is equal to the MSE exponent $E^*_{\sf MSE, rc}(R)$ of random block coding, as stated in Corollary~\ref{cor}. 

In this section, we focused our discussion on very noisy BSC($\epsilon$)s and proved gains in the achievable convergence rates of estimation errors from superposition coding  by using approximations of error exponents of decoding-error probabilities in the high-noise regime of $\epsilon\in(1/2-\delta,1/2)$ for a sufficiently small $\delta>0$. For other noise regimes, on the other hand, such a nice approximation of error exponents of decoding-error probabilities does not exist, and it is hard to compare the quantized-MSE exponent of superposition coding and that of random block coding. Instead, in the next section, we show empirical performances of the querying policies in mild-noise regimes  by comparing the quantized MSE of superposition coding and that of random block coding. 

\subsection{Simulations: Performance of Superposition Coding vs. Random Block Coding}
\begin{figure}[t]
\centerline{\includegraphics[scale=0.5]{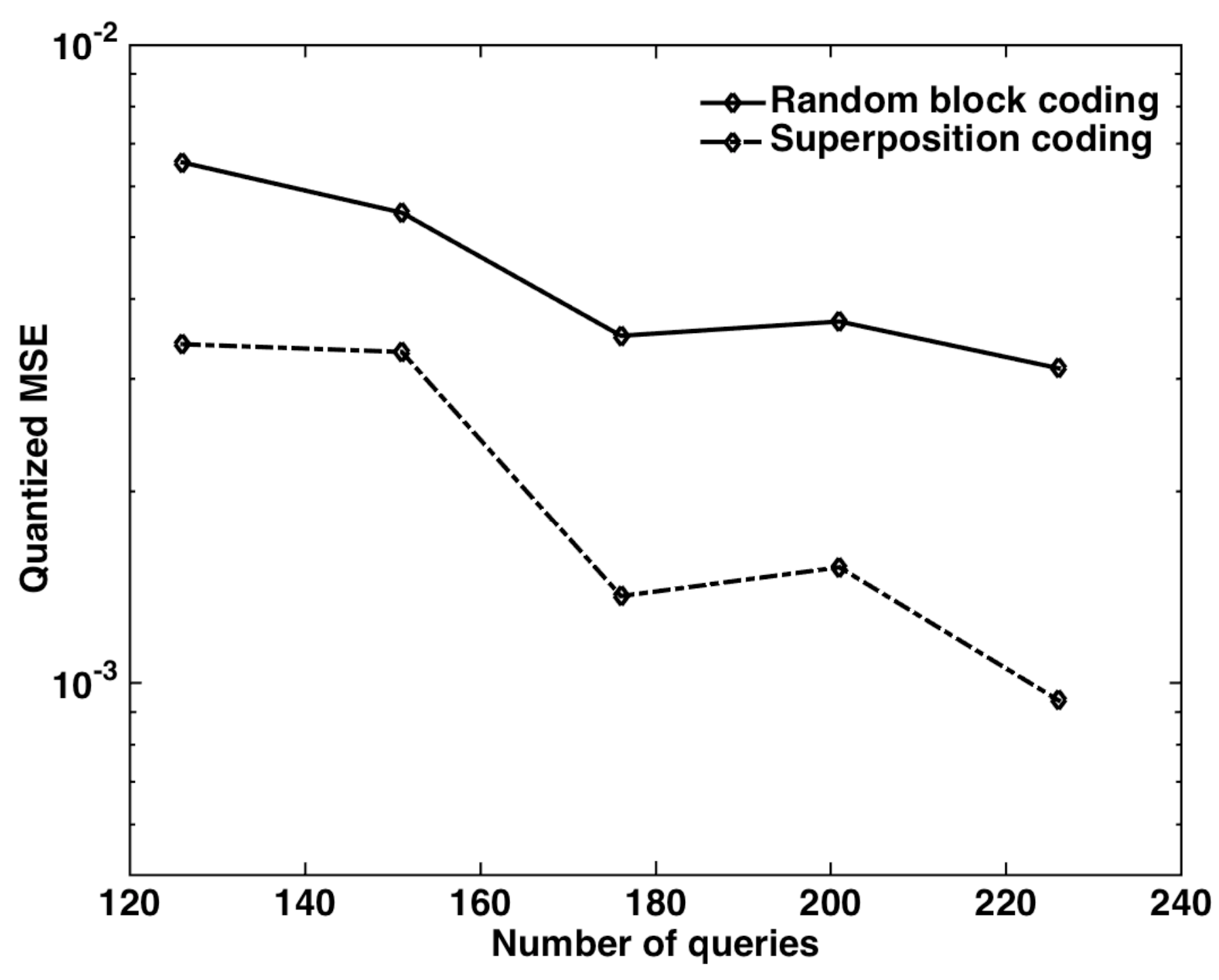}}
\caption{
Monte Carlo simulation for quantized-MSE $\E[c_{\sf q}(X,\hat{X}_{N,\sf finite})]$ of the querying policies based on superposition coding (dash-dot line) and of random block coding (solid line) as a function of the number of queries, where the rates $(R_1,R_2)$ of the partial messages $(M_1,M_2)$ are fixed as $(R_1,R_2)=(0.5(C-R_2),0.9C_2(\alpha))$ for capacity $C$ of the BSC($\epsilon$) and for the maximum achievable rate $C_2(\alpha)$ of $M_2$. The crossover probability $\epsilon$ of the BSC($\epsilon$) and the distribution parameter $\alpha$ of the superposition coding are set to be $\epsilon=0.3$ and $\alpha=0.1$, respectively. The markers in each line indicate the simulation points in terms of ($k_1,k_2$), the numbers of MSBs and of LSBs that are queried during the respective number $N$ of queries, for $N$ satisfying $N=(k_1\ln2)/R_1=(k_2\ln2)/R_2$. In this simulation, we checked five pairs of $(k_1,k_2)$ including $(5,4)$, $(6,5)$, $(6,6)$, $(7,7)$, and $(8,8)$ at the fixed rate pair $(R_1,R_2)$ with the increasing number of queries.
The number of Monte Carlo trials at each simulation point is equal to 3000.}
\label{fig:MC_RCvsSPC}
\end{figure}

In this section, we compare the performance of two querying policies, one based on superposition coding, which provides two different levels of error protection for MSBs vs. LSBs in the dyadic expansion of the target variable, and the other based on random block coding, which provides equal error protection to all the information bits in the dyadic expansion of the target variable.

Fig.~\ref{fig:MC_RCvsSPC} shows the empirical performance of these two non-adaptive block querying policies by comparing the quantized MSE $\E[c_q(X,\hat{X}_{N,\sf finite})]$ of superposition coding (dash-dot line) and that of random block coding (sold line) for a BSC($\epsilon$) with $\epsilon=0.3$, where the distribution parameter $\alpha$ of the superposition coding equals $\alpha=0.1$ and the rates of the two partial messages $M_1$ (MSBs) and $M_2$ (LSBs) are fixed as $(R_1,R_2)=(0.5(C-R_2),0.9C_2(\alpha))$ for $C=H_{\sf B}(1/2)-H_{\sf B}(\epsilon)$ and $C_2(\alpha)=H_{\sf B}(\alpha*\epsilon)-H_{\sf B}(\epsilon)$. 
We observe that the estimation error from the superposition coding is about one third of that of random block coding at the number of queries only about 200.
Therefore, even though Theorem~\ref{thm:sup} states the performance gain from the UEP superposition coding in the asymptotic regime for very noisy BSCs, empirical simulations show that the idea of designing a non-adaptive querying policy by using the UEP superposition coding provides performance gains in estimating the target variable even in non-asymptotic regimes of the number of queries and for wider range of noise levels of BSCs. 

\section{Comparison Between Performances Of the Four Different Querying Policies}\label{sec:allcompare}

In this section, we summarize and compare the four different querying policies discussed in this paper, including the adaptive bisection policy (Section~\ref{sec:sub1}), the non-adaptive UEP repetition policy (Section~\ref{sec:sub2}), the non-adaptive block querying based on random block coding (Section~\ref{sec:sub3}), and the non-adaptive block querying based on UEP superposition coding (Section~\ref{sec:super}).
Table~\ref{table:1} summarizes the MSE convergence rates and features of the four policies. Only the bisection policy uses past answers from the oracle to design the next query, while the other three policies determine a set of queries non-adaptively. Among the three non-adaptive block-querying policies, repetition policy and the policy based on superposition coding provide unequal error protection for MSBs vs. LSBs in the dyadic expansion of the target variable, while the block-querying policy based on random block coding provides equal error protection for every information bit. Repetition policy achieves MSE decreasing exponentially only in $\sqrt{N}$, while the other two non-adaptive block querying policies as well as the bisection policy achieve the linear in $N$ exponential rate of decrease. This is because the optimal repetition policy can extract only $k=O(\sqrt{N})$ information bits reliably by $N$ number of queries.  Non-adaptive block-querying policies based on either random block coding or superposition coding, on the other hand, can extract up to $k<NR$ bits for any positive rate $0<R<C$ by $N$ number of queries over a binary symmetric channel of capacity $C$. 
 Furthermore, superposition coding achieves a better MSE exponent than that of random block coding by providing unequal error protection for information bits.
\begin{table*}[t]
\caption{Comparison of four different querying policies}\label{table:1}
\begin{center}
    \begin{tabular}{| l | l | l | }
    \hline
   {\bf Policy} & {\bf MSE convergence rate} & {\bf Features}\\ \hline
   Bisection policy & $e^{-c_1N}$, $c_1>0$ & Adaptive\\ \hline
   Repetition policy & $e^{-c_2\sqrt{N}}$, $c_2>0$ & Non-adaptive, unequal error protection, no coding gain\\ \hline
   Random block coding & $e^{-c_3 N}$, $c_3>0$ & Non-adaptive, equal error protection, coding gain \\ \hline
   Superposition coding & $e^{-c_4 N}$, $c_4\geq c_3>0$ & Non-adaptive, unequal error protection, coding gain\\
     \hline
    \end{tabular} 
  \end{center}
\end{table*}

We next compare the achievable quantized-MSE exponent at a fixed rate $R>0$ for the three non-adaptive querying policies. 
\begin{itemize}
\item Non-adaptive UEP repetition querying: $E^*_{\sf q, repetition}(R)=0$ for every rate $R>0$.
\item Non-adaptive block querying based on random block coding: $E^*_{\sf q, rc}(R)=E_{\sf r}(R)$ for $E_{\sf r}(R)$ in~\eqref{eqn:randombck1}, i.e.,
\beq\label{eqn:randombck11}
E_{\sf r}(R)=\begin{cases}
E_0(1/2,\epsilon)-R, &  0\leq R< R_{\sf crit}(\epsilon),\\
D_{\sf B}(\gamma_{\sf GV}(R)\|\epsilon), & R_{\sf crit}(\epsilon)\leq R\leq C.
\end{cases}
\eeq
\item Non-adaptive block querying based on superposition coding: $E^*_{\sf q, spc}(R)\geq \max\{E_{\sf r}(R),E_{\sf q,spc}(R)\}$ where 
\beq\label{eqn:ofstrcit3}
\begin{split}
&E_{\sf q,spc}(R):=\max_{\alpha\in(0,1/2)}\min\{E_{\sf MSBs,SC}(R-C_2(\alpha),\alpha),\\
&\qquad\qquad \qquad\qquad \qquad\qquad 2(R-C_2(\alpha))\},
\end{split}
\eeq
for $E_{\sf MSBs,SC}(R_1,\alpha)$ in~\eqref{eqn:ofstrcit2}.
\end{itemize}

When we compare the three exponents $E^*_{\sf q,repetition}(R)$, $E^*_{\sf q, rc}(R)$, and $E^*_{\sf q, spc}(R)$ of the non-adaptive policies, we can show that
\beq
E^*_{\sf q, spc}(R)\geq E^*_{\sf q, rc}(R)\geq E^*_{\sf q, repetition}(R),
\eeq
which implies that the non-adaptive block querying based on superposition coding can achieve better error rates of convergence for the quantized MSE than do the other two non-adaptive policies.
The first inequality between $E^*_{\sf q, spc}(R)$ and $E^*_{\sf q, rc}(R)$ follows from the fact that random block coding is a special case of superposition coding where the parameter $\alpha$ equals 1/2.
In Theorem~\ref{thm:sup}, we also demonstrated  a strictly positive gain $E^*_{\sf q, spc}(R)> E^*_{\sf q, rc}(R)$ in the high rate regimes of $R$ for a very noisy BSC($\epsilon$) from the observation that the lower bound $E_{\sf q,spc}(R)$ in~\eqref{eqn:ofstrcit3} satisfies $E^*_{\sf q, spc}(R)>E_{\sf q,spc}(R)>E_{\sf r}(R)$.

\begin{figure}[t]
\centerline{\includegraphics[scale=0.5]{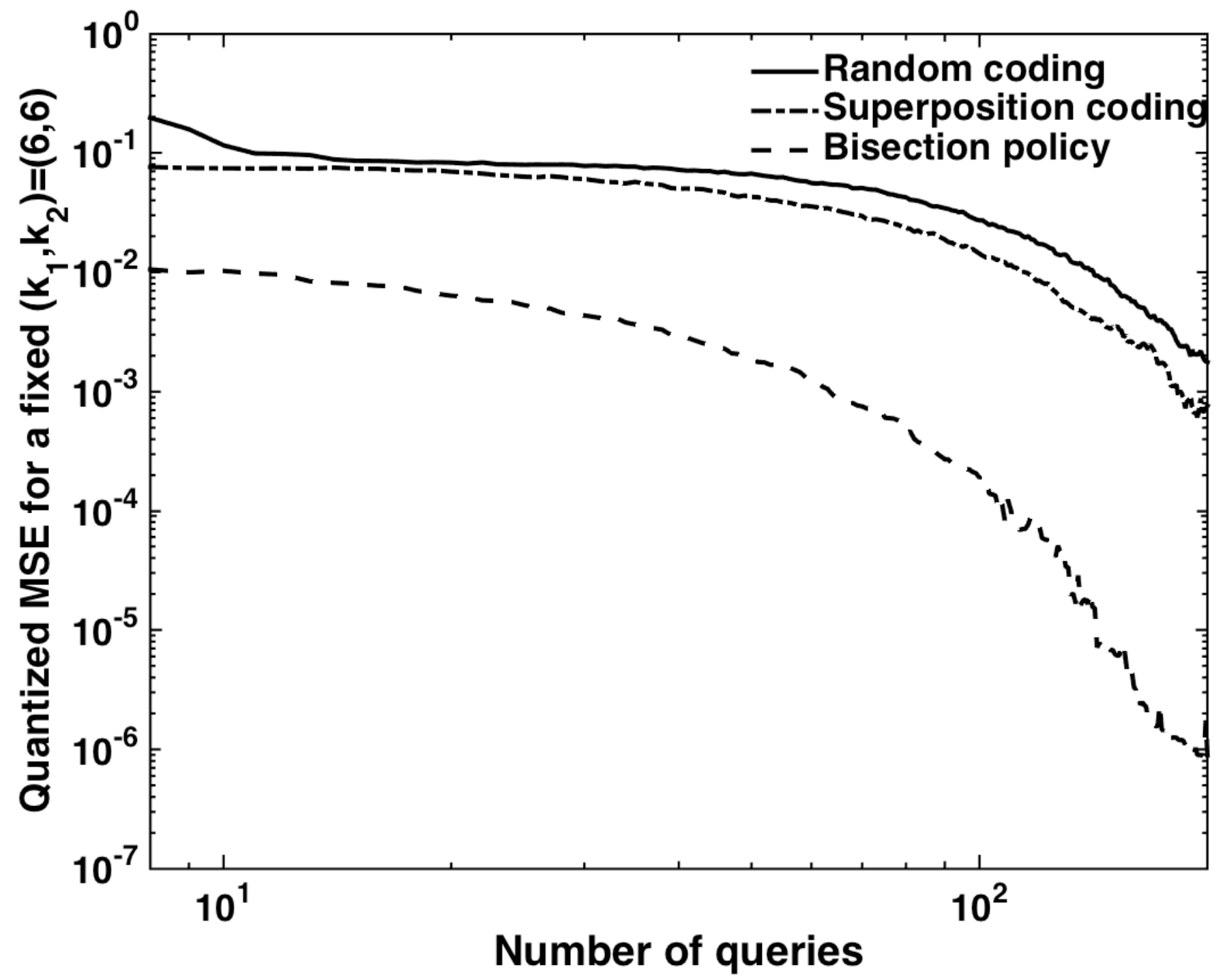}}
\caption{
Monte Carlo simulation for quantized-MSE performance of the querying policies based on random block coding (solid line), superposition coding (dash-dot line), and bisection policy (dashed line) as a function of the number of queries, where the pair of resolution bits $(k_1,k_2)$ for the MSBs and LSBs of the value of the target variable is fixed as $(k_1,k_2)=(6,6)$ for every $N$. The crossover probability $\epsilon$ of the BSC($\epsilon$) and the distribution parameter $\alpha$ of the superposition coding are set to $\epsilon=0.3$ and $\alpha=0.1$, respectively.
The number of Monte Carlo trials is equal to 3000.}
\label{fig:MC_fixedk}
\end{figure}

We next compare the performance between the adaptive bisection policy (BZ algorithm) and two non-adaptive querying policies, one based on random block coding and the other based on superposition coding.
In Lemma~\ref{lem:BZ_ref} of Section~\ref{sec:sub1}, we stated a lower bound on the quantized-MSE exponent for the adaptive BZ algorithm.  But we cannot use this lower bound in comparing the performance between the adaptive BZ algorithm and the two non-adaptive policies, since the tightness of this lower bound was not proven. 
Instead, Monte Carlo simulations are provided to compare the empirical performance of these three policies.

Fig.~\ref{fig:MC_fixedk} shows a plot of the empirical quantized MSE $\E[c_q(X,\hat{X}_{N,\sf finite})]$ of the two non-adaptive querying policies, random block coding (solid line) and superposition coding (dash-dot line), as well as that of the BZ algorithm (dashed line) as a function of the number $N$ of queries, where the pair of resolution bits $(k_1,k_2)$ for the MSBs and LSBs of the value of the target variable is fixed as $(k_1,k_2)=(6,6)$ for every $N$. It is observed that the bisection policy achieves the quantized MSE,  one to three orders of magnitude smaller than those of the two non-adaptive policies with number of queries less than 200. The non-adaptive policy based on superposition coding achieves gain in the quantized MSE performance over that of the random block coding, but falls short of achieving as fast convergence rate as that of the adaptive bisection policy in the non-asymptotic regime of number of queries.

\section{Future Directions}\label{sec:con}
In this paper, the problem of optimal query design was considered in the context of noisy 20 questions game with the goal of  estimating the value of a continuous target variable. We proposed a new non-adaptive block-querying policy based on superposition coding that could not only extract $k=NR/\ln2$ information bits ($0<R\leq C$) reliably over $N$ uses of a BSC($\epsilon$) of capacity $C$ but also provide two levels of unequal error protection (UEP) for the information bits.
Different from the UEP repetition querying policy considered in~\cite{jedynak2015} where the best achievable quantized MSE decreases exponentially only in $\sqrt{N}$ where $N$ is the number of queries, our non-adaptive querying policy based on superposition coding achieves linear in $N$ exponential rate of decrease, matching the rate of the bisection-based adaptive 20 questions scheme. Moreover, the achievable MSE exponent is larger than that of random block coding, which provides equal error protection for every information bit.

There are several open directions worthy of further study related to our work. 
First, the idea of designing a non-adaptive querying policy by using the UEP superposition coding can be applied to many other applications of data acquisition, possibly with diverse channel models, source (target) distributions, and cost functions. In applying the idea of UEP querying to general models of data acquisition, the important question is how to correctly assess the value (significance) of information bits that we try to extract by querying. For example, when a continuos random source, which we try to estimate, is not uniformly distributed over $[0,1]$, we need to first find the optimal quantization levels and the corresponding reproduction points with finite number of bits. Depending on the quantization levels and cost function, the significance of those information  bits might be varying. We need to measure the correct value of those bits in estimating the target variable and design a UEP querying policy to provide different levels of error protection depending on the value of those information bits.

Second, the proposed UEP querying policy based on superposition coding can be generalized to provide more than two levels of error protection.  This generalization may also improve the achievable estimation error since the resulting UEP querying policy would be able to use a fixed querying resource more efficiently by providing finer levels of error protection for information bits of different significance, compared to the two-level case we considered in this paper. Showing the improved performance, however, might require more complicated analysis on the error exponents of each of the partial messages. This generalization might also require development of more efficient decoding algorithms for the superposition codes. 
In our work, we mainly used successive-cancellation decoding rule, which is a sub-optimal decoding rule that successively decodes the two partial messages, public message (MSBs) and private message (LSBs).
Compared to the optimal maximum-likelihood decoding for each of the partial messages, the successive-cancellation decoding is computationally more efficient and also easier to analyze. 
But when the number of levels of error protection and the corresponding number of partial messages are increased, the performance of the successive-cancellation decoding might become worse than the two-level case, since the decoding error of the previous stage might keep propagating to all the later stages. Therefore, we need to develop another type of decoding rule to overcome this kind of challenge. 

\appendices

\section{Proof of Proposition~\ref{thm:opt_bis1}: Adaptive Bisection Policy}\label{sec:adaptive}

Successive-entropy-minimization policies choose the most informative querying region $Q_i$, which asks one bit of information about the target variable $X$ at each round, by satisfying 
\beq\label{eqn:1bit_cond_4app}
\begin{split}
&\Pr(X\in Q_i|Y_1^{i-1}=y_1^{i-1})=\Pr(X\notin Q_i|Y_1^{i-1}=y_1^{i-1})=1/2
\end{split}
\eeq
for answers $y_1^{i-1}\in\{0,1\}^{i-1}$ to previous queries. 
For a continuous random variable $X\sim p(x|y_1^{i-1})$, there exist diverse ways to design such a querying region $Q_i$ satisfying the condition~\eqref{eqn:1bit_cond_4app}. 

We quantify the value of the resulting observation $Y_i$ by the predicted variance reduction, defined as
\beq\label{eqn:var_gap_app}
{\sf Var}(X|Y_1^{i-1}=y_1^{i-1})-\E[{\sf Var}(X|Y_i, Y_1^{i-1}=y_1^{i-1})].
\eeq
Note that $\E[{\sf Var}(X|Y_i, Y_1^{i-1}=y_1^{i-1})]$ depends on the choice of $Q_i$ since the two possible posterior distributions of $X$, $p(x|y_i=0, y_1^{i-1})$ and $p(x|y_i=1, y_1^{i-1})$, after the $i$-th querying, are functions of the choice of $Q_i$.
The objective is to find the querying region $Q_i$ that not only satisfies~\eqref{eqn:1bit_cond_4app} but also maximizes the predicted variance reduction~\eqref{eqn:var_gap_app}. We aim to solve
\beq\label{eqn:var_gap_app1}
\begin{split}
&\max_{Q_i\in\mathcal{A}} \left({\sf Var}(X|Y_1^{i-1}=y_1^{i-1})-\E[{\sf Var}(X|Y_i, Y_1^{i-1}=y_1^{i-1})]\right).
\end{split}
\eeq
where $\mathcal{A}=\{Q: \Pr(X\in Q|Y_1^{i-1}=y_1^{i-1})=\Pr(X\notin Q|Y_1^{i-1}=y_1^{i-1})=1/2 \}$.
Proposition~\ref{thm:opt_bis1} states that the optimal $Q_i$ that is the solution of the optimization~\eqref{eqn:var_gap_app1} is the querying region that corresponds to the bisection policy, i.e., the optimal $Q_i$ is right region of the median of $p(x|y_1^{i-1})$. 



For adaptive sequential querying, given answers $y_1^{i-1}$ to the previous queries $(Q_1,\dots, Q_{i-1})$ we denote the updated posterior distribution $p(x|y_1^{i-1})$ of $X$ by $q(x):=p(x|y_1^{i-1})$. 
Define a one-step encoding map $d:[0,1]\to\{0,1\}$ that maps the value of $x\in[0,1]$ to a binary bit $Z=d(x)$. This encoding map $d(\cdot)$ is defined as the indicator function of the event $X\in Q$, i.e., $d(x)=1$ for every $x\in Q$ and $d(x)=0$ for every $x\notin Q$.
We derive the optimal one-step encoding map $d(\cdot)$ that maximally reduces the  conditional variance of $X\sim q(x)$ given the noisy answer $Y=Z\oplus N$ where $N\sim \text{Bernoulli}(\epsilon)$.

The conditional variance of $X$ given $Y$ can be written as 
\beq
\begin{split}
&\E[{\sf{Var}}(X|Y)]\\
&=\Pr(Y=0){\sf{Var}}(X|Y=0)+\Pr(Y=1){\sf{Var}}(X|Y=1)\\
&=\Pr(Y=0)\left(\E[X^2|Y=0]-\left(\E[X|Y=0]\right)^2\right) +\Pr(Y=1)\left(\E[X^2|Y=1]-\left(\E[X|Y=1]\right)^2\right)\\
&= \E[X^2]-\left(\Pr(Y=0)\left(\E[X|Y=0]\right)^2 +  \Pr(Y=1) \left(\E[X|Y=1]\right)^2\right).
\end{split}
\eeq
Since $\E[X^2]$ does not depend on the encoding map $d(\cdot)$, to minimize $\E[{\sf{Var}}(X|Y)]$ the  encoding map $d(\cdot)$ should maximize
\beq\label{eqn:argmaxd}
\begin{split}
&G_d:=\Pr(Y=0)\left(\E[X|Y=0]\right)^2+  \Pr(Y=1) \left(\E[X|Y=1]\right)^2.
\end{split}
\eeq
After applying Bayes' rule, $G_d$ in~\eqref{eqn:argmaxd} can be expressed in terms of the probabilities $\{\Pr(Z=0),\Pr(Z=1)\}$ and the conditional expectations  $\{\E[X|Z=0]=\E[X|X\in Q],\E[X|Z=1]=\E[X|X\notin Q]\}$ as
\beq
\begin{split}\label{eqn:argmaxd2}
G_d=& \frac{1}{\Pr(Y=0)}\left( (1-\epsilon)\cdot \Pr(Z=0)\E[X|Z=0] +\epsilon \cdot \Pr(Z=1)\E[X|Z=1] \right)^2 \\
& +\frac{1}{\Pr(Y=1)} \left( \epsilon\cdot \Pr(Z=0)\E[X|Z=0]+(1-\epsilon) \cdot \Pr(Z=1)\E[X|Z=1] \right)^2.
\end{split}
\eeq
Note that both of $\{\Pr(Z=0),\Pr(Z=1)\}$ and $\{\E[X|Z=0],\E[X|Z=1]\}$ depend on the encoding map $d(\cdot)$ as
\begin{align}
&\Pr(Z=0)=\Pr(X\notin Q_i)=\int_{d(x)=0} q(x) dx,\\
&\Pr(Z=1)=\Pr(X\in Q_i)=\int_{d(x)=1} q(x) dx,\\
&\E[X|Z=a]=\frac{1}{\Pr(Z=a)}\int_{d(x)=a} x q(x)  dx,\;\;\text{for} \;\; a=0,1.
\end{align}
When $m$ denotes the mean of $X\sim q(x)$, $\{\Pr(Z=0),\Pr(Z=1)\}$ and $\{\E[X|Z=0],\E[X|Z=1]\}$ should satisfy
\beq\label{eqn:pdconst}
\begin{split}
&m=\E[X]=\Pr(Z=0)\E[X|Z=0]+\Pr(Z=1)\E[X|Z=1].
\end{split}
\eeq
We show that among encoding maps $d(\cdot)$ with a fixed $\{\Pr(Z=0),\Pr(Z=1)\}$ the encoding map that maximizes $G_d$ in~\eqref{eqn:argmaxd2}, which thus minimizes $\E[{\sf Var}(X|Y)]$, under the constraint of~\eqref{eqn:pdconst} should be a step function.

\begin{lem}\label{lem:onestep_var} {\it Among encoding maps $d:[0,1]\to\{0,1\}$ with a fixed $\{\Pr(Z=0),\Pr(Z=1)\}$, the optimal encoding map that maximizes $G_d$ in~\eqref{eqn:argmaxd} under the constraint of~\eqref{eqn:pdconst} should maximize $|\E[X|Z=0]-m|$. For a fixed $\Pr(Z=0)$, in order to maximize $|\E[X|Z=0]-m|$ the optimal encoding map $d:[0,1]\to\{0,1\}$ should be either 
\beq\label{eqn:step11}
d(x)=
\begin{cases}
0 & x\leq t_1\\
1& x> t_1
\end{cases}
\eeq
for the threshold $t_1$ such that $\int_0^{t_1} q(x) dx=\Pr(Z=0)$, or
\beq\label{eqn:step22}
d(x)=
\begin{cases}
1 & x\leq t_2\\
0 & x> t_2
\end{cases}
\eeq
for the threshold $t_2$ such that $\int_{t_2}^1 q(x) dx=\Pr(Z=0)$.
}
\end{lem}
\begin{IEEEproof}
Let us define $\beta:=\Pr(Z=0)$ and $A_\beta:=\E[X|Z=0]$ for a fixed $0\leq \beta\leq 1$. From the constraints in~\eqref{eqn:pdconst}, it becomes  $\Pr(Z=1)\E[X|Z=1]=m-\beta A_\beta$. By using these parameters, we can rewrite $G_d$ in~\eqref{eqn:argmaxd2} as
\beq
\begin{split}
G_d=&\frac{\left((1-\epsilon)\beta A_\beta+\epsilon(m-\beta A_\beta)\right)^2}{\epsilon+(1-2\epsilon)\beta}+\frac{\left(\epsilon\beta A_\beta+(1-\epsilon)(m-\beta A_\beta)\right)^2}{(1-\epsilon)-(1-2\epsilon)\beta}\\
\end{split}
\eeq
By rearranging terms in the numerator and denominator of $G_d$, we can simply $G_d$ as
\beq\label{eqn:fd_alpha}
\begin{split}
G_d 
=&\frac{(1-2\epsilon)^2\beta^2(A_\beta-m)^2}{\epsilon(1-\epsilon)+(1-2\epsilon)^2\beta(1-\beta)}+m^2.
\end{split}
\eeq
Therefore, for a fixed $(\beta,\epsilon,m)$, $G_d$ is maximized when $A_\beta$ is as far as possible from the mean $m=\E[X]$ of $X\sim q(x)$. 
For a fixed $\beta=\Pr(Z=0)$,  the optimal encoding map $d:[0,1]\to\{0,1\}$ that maximizes $|A_{\beta}-m|=|\E[X|Z=0]-m|$ should be a step function of either~\eqref{eqn:step11} or~\eqref{eqn:step22}.
\end{IEEEproof}

For successive-entropy-minimization strategies that query one bit of information about the target variable $X\sim q(x)$ at each round, the probabilities of the event $\{Z=0\}$ and of $\{Z=1\}$ are balanced as
\beq\label{eqn:Zcond}
\Pr(Z=0)=\Pr(Z=1)=1/2.
\eeq
The thresholds of the two step functions~\eqref{eqn:step11} and~\eqref{eqn:step22} for this case become the same as the median of the distribution. 
Lemma~\ref{lem:onestep_var} thus implies that among policies satisfying~\eqref{eqn:Zcond}, the adaptive bisection policy is the optimal one-step policy that minimizes the conditional variance of $X$ given a noisy answer $Y$. 

\section{Proof of Lemma~\ref{lem:rep_biterror}: Bit-Error Probability with Repetition Coding}\label{app:lem:rep_biterror}
In Lemma~\ref{lem:rep_biterror}, we show that when a binary bit $B_i\sim\text{Bernoulli}(1/2)$ is repeatedly transmitted through a BSC($\epsilon$) by $N_i$ times, the decoding-error probability of $B_i$ with the majority voting algorithm is bounded below and above as
\beq\label{eqn:App_B_whatprove}
\frac{e^{-1/(3N_i)}}{\sqrt{{2\pi N_i}}}e^{-N_i D_{\sf B}\left(1/2\|\epsilon\right)}\leq \Pr(\hat{B}_i\neq B_i)\leq e^{-N_i D_{\sf B}(1/2\|\epsilon)}.
\eeq
We first prove the upper bound. For bit $B_i$ repeatedly queried $N_i$ times, denote the corresponding channel outputs  by $(Y_1,\dots, Y_{N_i})\in\{0,1\}^{N_i}$ and define $Y_j'=2Y_j-1\in\{-1,1\}$. The majority voting claims an estimate $\hat{B}_i=1$ when $\sum_{j=1}^{N_i} Y_j'> 0$, and $\hat{B}_i=0$ when $\sum_{j=1}^{N_i} Y_j'\leq 0$. Since the channel is symmetric and $B_i\sim\text{Bernoulli}(1/2)$, the bit error probability can be bounded above as
\beq
\begin{split}\label{eqn:bit_err_app}
&\Pr(\hat{B}_i\neq B_i)\\
&=\frac{1}{2}\Pr\left(\sum_{j=1}^{N_i} Y_j'>0\Bigg|B_i=0\right)+\frac{1}{2}\Pr\left(\sum_{j=1}^{N_i} Y_j'\leq 0\Bigg|B_i=1\right)\\
&\leq \Pr\left(\sum_{j=1}^{N_i} Y_j'\leq 0\Bigg|B_i=1\right)\\
&=\Pr\left( e^{-\lambda \sum_{j=1}^{N_i} Y_j'}\geq 1\Big|B_i=1\right), \text{ for } \lambda>0\\
&\leq\E\left[e^{-\lambda \sum_{j=1}^{N_i} Y_j'}\Big|B_i=1\right],
\end{split}
\eeq
where the last inequality is from the Markov's inequality.

Using the conditional independence of $Y_1',\dots,Y_{N_i}'$ given $B_i=1$ and the fact that $\Pr(Y_j'=1|B_i=1)=1-\epsilon$ and  $\Pr(Y_j'=-1|B_i=1)=\epsilon$,
\beq
\begin{split}
\E\left[e^{-\lambda \sum_{j=1}^{N_i} Y_j'}\Big|B_i=1\right]=&\prod_{j=1}^{N_i} \E\left[e^{-\lambda  Y_j'}\Big|B_i=1\right]\\
=& \prod_{j=1}^{N_i} \left((1-\epsilon)e^{-\lambda}+\epsilon \cdot e^{\lambda}\right).
\end{split}
\eeq
When $\lambda=\ln\sqrt{\frac{1-\epsilon}{\epsilon}}$, the term $ \left((1-\epsilon)e^{-\lambda}+\epsilon \cdot e^{\lambda}\right)$ is minimized as $2\sqrt{\epsilon(1-\epsilon)}=e^{-D_{\sf B}(1/2\|\epsilon)}$. Therefore, the bit-error probability can be bounded above as
\beq
\begin{split}
\Pr(\hat{B}_i\neq B_i)&\leq \min_{\lambda} \left(\prod_{j=1}^{N_i} \left((1-\epsilon)e^{-\lambda}+\epsilon \cdot e^{\lambda}\right)\right)\\
&=e^{-N_iD_{\sf B}(1/2\|\epsilon)}.
\end{split}
\eeq
We next prove the lower bound in~\eqref{eqn:App_B_whatprove}. The bit error probability can be bounded below as
\beq\label{eqn:Bilower}
\Pr(\hat{B}_i\neq B_i)\geq \frac{1}{2}\Pr\left( \sum_{j=1}^{N_i}Y_j'=0\Bigg|B_i=1\right),
\eeq
where $\Pr\left( \sum_{j=1}^{N_i}Y_j'=0\big|B_i=1\right)$ is the probability of the event that a half of the transmitted bits are flipped by the channel noise, which is distributed by Bernoulli$(\epsilon)$ distribution.
This probability is bounded below as
\beq
\begin{split}
&\Pr\left( \sum_{j=1}^{N_i}Y_j'=0\Bigg|B_i=1\right)\\
&={ N_i \choose N_i/2}\epsilon^{N_i/2}(1-\epsilon)^{N_i/2}\\
&\geq  \sqrt{\frac{2}{\pi N_i}}e^{-1/(3N_i)}2^{N_i}\epsilon^{N_i/2}(1-\epsilon)^{N_i/2}\\
&=\sqrt{\frac{2}{\pi N_i}}e^{-1/(3N_i)}e^{-N_i D_{\sf B}\left(1/2\|\epsilon\right)}
\end{split}
\eeq
where the middle inequality is from the Stirling bound. By plugging this lower bound into~\eqref{eqn:Bilower}, we obtain
\beq
\Pr(\hat{B}_i\neq B_i)\geq \frac{e^{-1/(3N_i)}}{\sqrt{{2\pi N_i}}}e^{-N_i D_{\sf B}\left(1/2\|\epsilon\right)}.
\eeq

\section{Proof of Lemma~\ref{prop:RC_exp}: quantized-MSE exponent with Random Block Coding}\label{app:prop:RC_exp}
In Lemma~\ref{prop:RC_exp}, we show that  the best achievable  quantized-MSE exponent with the random block codes of rate $R$ is
\beq
\begin{split}\label{eqn:exp_RC1_app}
E^*_{\sf q, rc}(R)=E_{\sf r}(R)
\end{split}
\eeq
for $E_{\sf r}(R)$ in~\eqref{eqn:randombck1}.
 The achievability of~\eqref{eqn:exp_RC1_app} was shown by the bound~\eqref{eqn:thm_rc2}.
 In this section we prove  the converse, i.e., the quantized-MSE exponent with the random block codes of rate $R$ cannot be better than $E_{\sf r}(R)$.
We prove this by providing a lower bound on the quantized MSE.

Consider the quantized MSE expanded in terms of the conditional bit-error probabilities.
\beq
\begin{split}
&\E[c_{\sf q}(X,\hat{X}_{N,\sf finite})]\\
&= \sum_{i=1}^{k} \left(\Pr(\hat{B}_i\neq {B}_i,\hat{B}_1^{i-1}=B_1^{i-1})\E[c_{\sf q}(X,\hat{X}_{N,\sf finite}) \big| \hat{B}_i\neq {B}_i,\hat{B}_1^{i-1}=B_1^{i-1}]\right).
\end{split}
\eeq
Since $\Pr(\hat{M}= M)\to 1$ as $N\to\infty$ for the random block codes of rate $ R\in(0, C)$, we know that $\Pr(\hat{B}_1^{i}= B_1^{i})\doteq 1$ for any $i=1,\dots, k$.
 Therefore, we can write $\E[c_{\sf q}(X,\hat{X}_{N,\sf finite})]$ as
\beq
\begin{split}\label{eqn:cqexpansion}
&\E[c_{\sf q}(X,\hat{X}_{N,\sf finite})]\\
&\doteq\sum_{i=1}^{k} \left(\Pr(\hat{B}_i\neq {B}_i|\hat{B}_1^{i-1}=B_1^{i-1})\E[c_{\sf q}(X,\hat{X}_{N,\sf finite}) \big| \hat{B}_i\neq {B}_i,\hat{B}_1^{i-1}=B_1^{i-1}]\right).
\end{split}
\eeq
We then show that the first term of the summation in the right-hand side is 
\beq\label{eqn:first_sum1}
\Pr(\hat{B}_1\neq B_1)\E[c_{\sf q}(X,\hat{X}_{N,\sf finite}) |\hat{B}_1\neq B_1]\dot{\geq}e^{-NE_{\sf r}(R)},
\eeq
for the random coding error exponent $E_{\sf r}(R)$.

Note that
\beq
\Pr(\hat{B}_1\neq B_1)\leq \Pr(\hat{M}\neq M)\leq \sum_{i=1}^k \Pr(\hat{B}_i\neq B_i).
\eeq
The average bit error probability $\Pr(\hat{B}_i\neq B_i)$ for the random block codes is the same for every $i\in\{1,\dots,k\}$ from the symmetry of encoding process for the information bits $\{B_i\}$.
Since $k=NR/\ln2$ increases linearly in $N$, the exponent of $\Pr(\hat{B}_1\neq B_1)$ is the same as that of $\Pr(\hat{M}\neq M)\doteq e^{-NE_{\sf r}(R)}$, i.e.,
$
\Pr(\hat{B}_1\neq B_1)\doteq e^{-NE_{\sf r}(R)}.
$
Moreover,
\beq
\E[c_{\sf q}(X,\hat{X}_{N,\sf finite})|\hat{B}_1\neq B_1]\doteq 1.
\eeq
This can be shown by calculating a lower bound on $\E[c_{\sf q}(X,\hat{X}_{N,\sf finite})|\hat{B}_1\neq B_1]$.
When $\hat{B}_1=1$ and $B_1=0$, the best $\hat{X}_N$ that minimizes the conditional expectation is $\hat{X}_N=1/2$.
Conditioned on $B_1=0$, $X$ is uniformly distributed over $[0,1/2]$, and thus
\beq
\begin{split}
&\E[c_{\sf q}(X,\hat{X}_N)|\hat{B}_1=1, B_1=0]\geq \int_{0}^{1/2} 2 (x-1/2)^2 dx=1/12.
\end{split}
\eeq
The same bound also holds when $\hat{B}_1=0$ and $B_1=1$.
From this lower bound on $\E[c_{\sf q}(X,\hat{X}_N)|\hat{B}_1=1, B_1=0]$ and the bit error probability $\Pr(\hat{B}_1\neq B_1)\doteq  e^{-N E_{\sf r}(R)}$, the lower bound in~\eqref{eqn:first_sum1} can be proven. 
From the bound~\eqref{eqn:first_sum1} and the expansion on $\E[c_{\sf q}(X,\hat{X}_{N,\sf finite})]$ in~\eqref{eqn:cqexpansion}, we can conclude that
\beq
\E[c_{\sf q}(X,\hat{X}_{N,\sf finite})]\dot{\geq}e^{-NE_{\sf r}(R)}.
\eeq

\section{Proofs of Lemma~\ref{lem:sup_exp_general}: decoding-error exponents of Partial Messages Encoded by Superposition Coding}\label{sec:pf_sup_exp}

In Lemma~\ref{lem:sup_exp_general}, we show that superposition coding provides a better, or at least as good, error protection for the more important partial message $M_1$ (MSBs) than that of the random block coding. We provide two lower bounds $E_{\sf MSBs, JML}(R_1,R_2)$ and $E_{\sf MSBs,SC}(R_1,\alpha)$ on the best achievable error exponent $E^*_{\sf MSBs}(R_1,R_2,\alpha)$ and prove that these exponents are larger than the best achievable exponent $E_{\sf r}(R_1+R_2)$ of random block coding.

The first lower bound $E_{\sf MSBs, JML}(R_1,R_2)$ is defined as the best achievable error exponent for $\Pr(\hat{M}_1\neq M_1)$ with joint-maximum-likelihood (JML) decoding rule for superposition coding.
We show that $E_{\sf MSBs, JML}(R_1,R_2)\geq E_{\sf r}(R_1+R_2)$ for every $(R_1,R_2,\alpha)$ in the following lemma. 
\begin{lem}\label{lem:JML_errbd}
{\it For a given $(R_1,R_2,\alpha)$, the best achievable error exponent $E_{\sf MSBs, JML}(R_1,R_2)$ with joint-ML decoding rule for superposition coding is bounded below as
\beq
E_{\sf MSBs, JML}(R_1,R_2)\geq E^{\sf LB}_{\sf MSBs, JML}(R_1, R_2)
\eeq
where 
\beq\label{eqn:Em1JML_match}
\begin{split}
&E^{\sf LB}_{\sf MSBs, JML}(R_1, R_2)=
\begin{cases}
E_0(1/2,\epsilon)-R_2-R_1, & R_1\leq \max\{0,R_{\sf crit}(\epsilon)-R_2\},\\
D_{\sf B}(\gamma_{\sf GV}(R_1+R_2)\|\epsilon), &\max\{0,R_{\sf crit}(\epsilon)-R_2\}<R_1\leq H_{\sf B}(1/2)-H_{\sf B}(\epsilon)-R_2.
\end{cases}
\end{split}
\eeq
}
\end{lem}
\begin{IEEEproof}
Appendix~\ref{app:JML_errbd}
\end{IEEEproof}
Note that  for any given $(R_1,R_2,\alpha)$ the achievable exponent $E^{\sf LB}_{\sf MSBs, JML}(R_1, R_2)$ is equal to $E_{\sf r}(R_1+R_2)$ in~\eqref{eqn:randombck1}.
Since the joint-maximum-likelihood decoding is a sub-optimal decoding rule, the fact that $E_{\sf MSBs, JML}(R_1, R_2)\geq E^{\sf LB}_{\sf MSBs, JML}(R_1, R_2)=E_{\sf r}(R_1+R_2)$ implies that the superposition coding provides a better, or at least as good, error protection for the partial message $M_1$ than that of the random block coding for every $(R_1,R_2)$, regardless of the choice of $\alpha\in(0,1/2)$.

We next prove a strictly positive gain in the error exponent $E^*_{\sf MSBs}(R_1,R_2,\alpha)$ of superposition coding, by providing another lower bound $E_{\sf MSBs, SC}(R_1,\alpha)$ on $E^*_{\sf MSBs}(R_1,R_2,\alpha)$. The exponent  $E_{\sf MSBs, SC}(R_1,\alpha)$ is the best achievable error exponent with successive-cancellation (SC) decoding. 
As shown in Eq.~\eqref{eqn:ofstrcit2}, the exponent can be written as 
\beq\label{eqn:EMBSsapp}
\begin{split}
&E_{\sf MSBs, \sf SC}(R_1, \alpha)=
\begin{cases}
E_0(1/2,\alpha*\epsilon)-R_1, &0\leq R_1\leq R_{\sf crit}(\alpha*\epsilon),\\
D_{\sf B}(\gamma_{\sf GV}(R_1)\|\alpha*\epsilon), &R_{\sf crit}(\alpha*\epsilon)<R_1\leq C-C_2(\alpha),
\end{cases}
\end{split}
\eeq
where $E_0(a,b)=-\ln(1-2a(1-a)(\sqrt{b}-\sqrt{1-b})^2)$ and thus $E_0(1/2,\alpha*\epsilon)=-\ln(1/2+\sqrt{(\alpha*\epsilon)(1-(\alpha*\epsilon))})$, $C=H_{\sf B}(1/2)-H_{\sf B}(\epsilon)$, $C_2(\alpha)=H_{\sf B}(\alpha*\epsilon)-H_{\sf B}(\epsilon)$, $R_{\sf crit}(\alpha*\epsilon)=D_{\sf B}(\gamma_{\sf crit}(\alpha*\epsilon)\|1/2)$ and $\gamma_{\sf crit}(\alpha*\epsilon)=\frac{\sqrt{\alpha*\epsilon}}{\sqrt{\alpha*\epsilon}+\sqrt{1-\alpha*\epsilon}}$.

\begin{figure}[t]
\centerline{\includegraphics[scale=0.6]{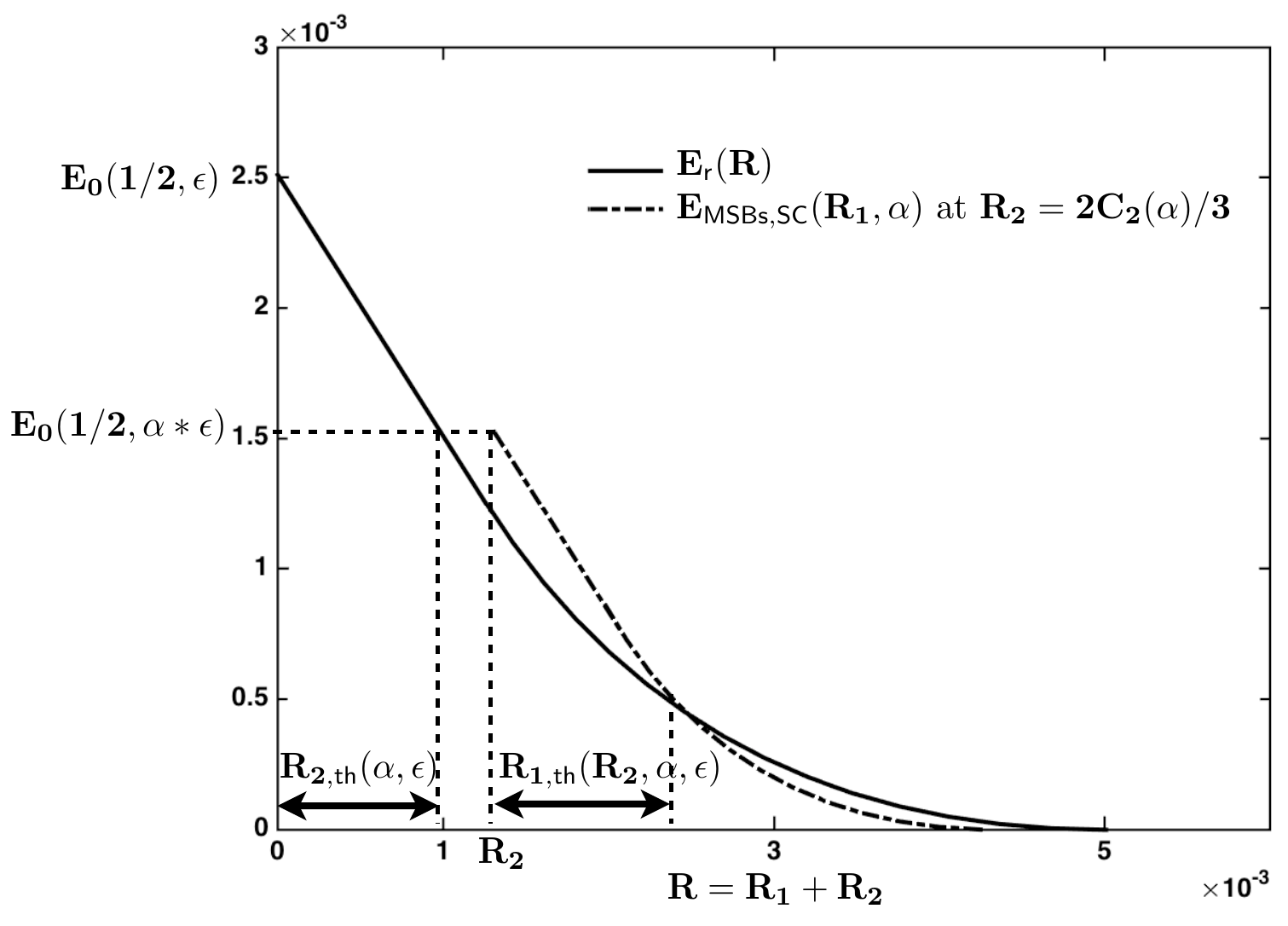}}
\caption{
A plot of error exponents $E_{\sf r}(R_1+R_2)$ (solid line) and $E_{\sf MSBs,SC}(R_1,\alpha)$ (dash-dot line) for the more important partial message $M_1$ (MSBs) with random block coding and with superposition coding, respectively, where $\epsilon=0.45$, $\alpha=0.11$, and $R_2=2C_2(\alpha)/3$. The error exponent $E_{\sf MSBs,SC}(R_1,\alpha)$ of superposition coding is strictly larger than that of the random block coding $E_{\sf r}(R_1,R_2)$ over $R_1\in(0,R_{1,\sf th}(R_2,\alpha,\epsilon))$ for a fixed $R_2=2C_2(\alpha)/3>R_{2,\sf th}(\alpha,\epsilon)$.
}
\label{fig:gain_proof}
\end{figure}

By comparing $E_{\sf MSBs, SC}(R_1, \alpha)$ in~\eqref{eqn:EMBSsapp} with the error exponent $E_{\sf r}(R_1+R_2)$ in~\eqref{eqn:randombck1} of random block coding, we show that for a  sufficiently large $R_2$ and sufficiently small $R_1$ superposition coding provides a strictly better error protection for the partial message $M_1$ than does random block coding. More precisely, for a given $\alpha\in(0,1/2)$ and a fixed  $\epsilon\in(0,1/2)$, we find thresholds $R_{2,\sf th}(\alpha,\epsilon)$ on $R_2$ and $R_{1,\sf th}(R_2,\alpha,\epsilon)$ on $R_1$ such that when $R_{2,\sf th}(\alpha,\epsilon)<R_2<C_2(\alpha)$ and $0\leq R_1<R_{1,\sf th}(R_2,\alpha,\epsilon)$, it is guaranteed that
\beq
E_{\sf MSBs, SC}(R_1,\alpha)>E_{\sf r}(R_1+R_2).
\eeq
The threshold on $R_2$, denoted $R_{2,\sf th}(\alpha,\epsilon)$, is defined as the rate $R$ at which the random-coding error exponent $E_{\sf r}(R)$ equals the error exponent $E_{\sf MSBs, SC}(0,\alpha)=E_0(1/2,\alpha*\epsilon)$ of superposition coding at $R_1=0$.
For a fixed rate $R_2>R_{2,\sf th}(\alpha,\epsilon)$, the threshold on $R_1$ denoted $R_{1,\sf th}(R_2,\alpha,\epsilon)$, is defined as the minimum rate $R_1$ where the error exponent $E_{\sf MSBs, SC}(R_1,\alpha)$ of superposition coding equals the random-coding error exponent $E_{\sf r}(R_1+R_2)$.
There always exists such a threshold $R_{1,\sf th}(R_2,\alpha,\epsilon)$ in the range $[0,C-R_2]$ since both $E_{\sf MSBs, SC}(R_1,\alpha)$ and $E_{\sf r}(R_1+R_2)$ keeps decreasing as $R_1$ increases and $E_{\sf MSBs, SC}(R_1,\alpha)>E_{\sf r}(R_1+R_2)$ at $R_1=0$ and $E_{\sf MSBs, SC}(R_1,\alpha)=E_{\sf r}(R_1+R_2)$ as $R_1$ approaches to $C-R_2$.
In Fig.~\ref{fig:gain_proof}, we plot $E_{\sf MSBs, SC}(R_1,\alpha)$ (dash-dot line) and $E_{\sf r}(R_1+R_2)$ (solid line) for $\alpha=0.11$, $\epsilon=0.45$ and $R_2=2C_2(\alpha)/3>R_{2,\sf th}(\alpha,\epsilon)$. From the plot, we can observe a gain in the error exponent of $\Pr(\hat{M}_1\neq M_1)$ from superposition coding for every $0\leq R_1\leq R_{1,\sf th}(R_2,\alpha,\epsilon)$ at a fixed $R_2=2C_2(\alpha)/3>R_{2,\sf th}(\alpha,\epsilon)$.
From the plot, we can also see how the thresholds $R_{2,\sf th}(\alpha,\epsilon)$ and $R_{1,\sf th}(R_2,\alpha,\epsilon)$ are determined for this particular example.

Lastly, we state a lower bound on the error exponent  $E_{\sf LSBs, SC}(R_2,\alpha)$ of $\Pr(\hat{M}_2\neq M_2|\hat{M}_1=M_1)$ with successive-cancellation decoding rule for superposition coding. This result was provided in~\cite{gallager1974capacity}:
\beq\label{eqn:E0M2_G}
E_{\sf LSBs, SC}(R_2,\alpha):=\max_{0\leq\rho \leq 1}\left[F_{0}(\rho,\alpha)-\rho R_2\right]
\eeq
where
\beq\label{eqn:E0M2_G_exp}
\begin{split}
&F_{0}(\rho,\alpha)=-\ln\left\{\sum_{y\in\{0,1\}}\left[\sum_{v\in\{0,1\}}p_V(v)p_{Y|V}(y|v)^{\frac{1}{1+\rho}}\right]^{1+\rho}\right\}
\end{split}
\eeq
for $p_V(v)$ being Bernoulli($\alpha$) distribution and $p_{Y|V}(y|v)$ being the transition probability of BSC($\epsilon$).
We can show that $E_{\sf LSBs, SC}(R_2,\alpha)>0$ for $R_2<C_2(\alpha)$ where $C_2(\alpha)$  is equal to 
\beq
C_2(\alpha)=\frac{\partial F_{0}(\rho,\alpha)}{\partial\rho}\Big|_{\rho=0}=H_{\sf B}(\alpha*\epsilon)-H_{\sf B}(\epsilon).
\eeq
Note that as $\alpha$ decreases from 1/2 to 0, the maximum rate $C_2(\alpha)$ of the less important partial message $M_2$ (LSBs) keeps decreasing.


\section{Proof of Lemma~\ref{lem:strict1}: Gain in the decoding-error exponent of MSBs from Superposition Coding for a Very Noisy BSC($\epsilon$)}\label{app:cor:strict}
In Lemma~\ref{lem:strict1}, we show that for a very noisy BSC($\epsilon$) with $\epsilon=1/2-\delta$ for a sufficiently small $\delta>0$, superposition coding achieves a strictly positive gain in the error exponent of $\Pr(\hat{M}_1\neq M_1)$ for every rate $R_1\in(0,C-C_2(\alpha))$ of the more important message $M_1$ (MSBs) compared to that of random block coding, when the rate $R_2$ of the less important message $M_2$ (LSBs) equals the maximum rate $C_2(\alpha)=H_{\sf B}(\alpha*\epsilon)-H_{\sf B}(\epsilon)$ for a fixed parameter $\alpha\in(0,1/2)$ of superposition coding. 

When we fix $R_2=C_2(\alpha)$,  the best achievable $\Pr(\hat{M}_1\neq M_1)$ of random block coding at a rate $R_1\in[0,C-C_2(\alpha)]$ is 
\beq
\Pr(\hat{M}_1\neq M_1)\doteq  e^{-NE_{\sf r}(R_1+R_2)}
\eeq
where
\beq
\begin{split}
&E_{\sf r}(R_1+R_2)=
\begin{cases}
E_0(1/2,\epsilon)-C_2(\alpha)-R_1, & 0\leq R_1\leq \max\{0,R_{\sf crit}(\epsilon)-C_2(\alpha)\},\\
D_{\sf B}(\gamma_{\sf GV}(R_1+C_2(\alpha))\|\epsilon),&\max\{0,R_{\sf crit}(\epsilon)-C_2(\alpha)\}<R_1\leq C-C_2(\alpha),
\end{cases}
\end{split}
\eeq
for $E_0(1/2,\epsilon)=\ln2-\ln(1+2\sqrt{\epsilon(1-\epsilon)})$, $C=H_{\sf B}(1/2)-H_{\sf B}(\epsilon)$, $C_2(\alpha)=H_{\sf B}(\alpha*\epsilon)-H_{\sf B}(\epsilon)$ and $R_{\sf crit}(\epsilon)=D_{\sf B}(\gamma_{\sf crit}(\epsilon)\|1/2)$ where $\gamma_{\sf crit}(\epsilon)=\frac{\sqrt{\epsilon}}{\sqrt{\epsilon}+\sqrt{1-\epsilon}}$.

With superposition coding and successive cancellation (SC) decoding, we can achieve 
\beq
\Pr(\hat{M}_1\neq M_1)\dot{\leq} e^{-NE_{\sf MSBs, SC}(R_1,\alpha)}
\eeq
where
\beq
\begin{split}
&E_{\sf MSBs, SC}(R_1,\alpha)=\begin{cases}
E_0(1/2,\alpha*\epsilon)-R_1, &0\leq R_1\leq R_{\sf crit}(\alpha*\epsilon),\\
D_{\sf B}(\gamma_{\sf GV}(R_1)\|\alpha*\epsilon), & R_{\sf crit}(\alpha*\epsilon)<R_1\leq C-C_2(\alpha), 
\end{cases}
\end{split}
\eeq
for the Gilbert-Varshamov distance $\gamma_{\sf G}(R)\in[0,1/2]$ that is defined as $D_{\sf B}(\gamma_{\sf GV}(R)\|1/2)=R$.

To prove that $E_{\sf MSBs, SC}(R_1,\alpha)> E_{\sf r}(R_1+R_2)$ for every $R_1\in[0,C-C_2(\alpha))$ at $R_2=C_2(\alpha)$, we need to demonstrate the following three statements for $\epsilon\approx 0.5$ at every $\alpha\in(0,1/2)$,
\begin{enumerate}\label{enu:three}
\item $E_0(1/2,\alpha*\epsilon)>E_0(1/2,\epsilon)-C_2(\alpha)$.
\item $E_0(1/2,\alpha*\epsilon)-R_1>D_{\sf B}(\gamma_{\sf GV}(R_1+C_2(\alpha))\|\epsilon)$ for $0\leq R_1\leq R_{\sf crit}(\alpha*\epsilon)$ when $R_{\sf crit}(\epsilon)<C_2(\alpha)$.
\item  $D_{\sf B}(\gamma_{\sf GV}(R_1)\|\alpha*\epsilon)>D_{\sf B}(\gamma_{\sf GV}(R_1+C_2(\alpha))\|\epsilon)$ for $R_{\sf crit}(\alpha*\epsilon)<R_1<C-C_2(\alpha)$.
\end{enumerate}
Once these three statements are proven, it is sufficient to show that $E_{\sf MSBs, SC}(R_1,\alpha)> E_{\sf r}(R_1+R_2)$ for every $0\leq R_1\leq C-C_2(\alpha)$ when $R_2=C_2(\alpha)$.

The first statement 1) $E_0(1/2,\alpha*\epsilon)>E_0(1/2,\epsilon)-C_2(\alpha)$ is equivalent to
\beq
\begin{split}
&H_{\sf B}(\alpha*\epsilon)-\ln(\sqrt{\alpha*\epsilon}+\sqrt{1-\alpha*\epsilon})^2>H_{\sf B}(\epsilon)-\ln(\sqrt{\epsilon}+\sqrt{1-\epsilon})^2.
\end{split}
\eeq
When we define $f(x)=H_{\sf B}(x)-\ln(\sqrt{x}+\sqrt{1-x})^2$, the above inequality is equivalent to $f(\alpha*\epsilon)-f(\epsilon)>0$. Note that $0\leq \epsilon<\alpha*\epsilon\leq 1/2$ for every $\alpha\in(0,1/2)$.
Moreover, the derivative of $f(x)$ is positive in the regime of $0.05\leq x\leq 1/2$. Therefore, for a BSC($\epsilon$) with $\epsilon\geq 0.05$, the statement 1) $E_0(1/2,\alpha*\epsilon)>E_0(1/2,\epsilon)-C_2(\alpha)$ is true.

We next prove the statement 3) $D_{\sf B}(\gamma_{\sf GV}(R_1)\|\alpha*\epsilon)>D_{\sf B}(\gamma_{\sf GV}(R_1+C_2(\alpha))\|\epsilon)$ for $R_{\sf crit}(\alpha*\epsilon)<R_1<C-C_2(\alpha)=H_{\sf B}(1/2)-H_{\sf B}(\alpha*\epsilon)$, which will also be used to prove the statement 2) later.
First, note that at $R_1=H_{\sf B}(1/2)-H_{\sf B}(\alpha*\epsilon)$, $D_{\sf B}(\gamma_{\sf GV}(R_1)\|\alpha*\epsilon)=D_{\sf B}(\gamma_{\sf GV}(R_1+C_2(\alpha))\|\epsilon)=0$ from the definition of $\gamma_{\sf GV}(R)$.
We will prove that $D_{\sf B}(\gamma_{\sf GV}(R_1)\|\alpha*\epsilon)-D_{\sf B}(\gamma_{\sf GV}(R_1+C_2(\alpha))\|\epsilon)$ strictly decreases in $R_1\in(R_{\sf crit}(\alpha*\epsilon),H_{\sf B}(1/2)-H_{\sf B}(\alpha*\epsilon)]$.
Since $D_{\sf B}(\gamma_{\sf GV}(R_1)\|\alpha*\epsilon)=D_{\sf B}(\gamma_{\sf GV}(R_1+C_2(\alpha)\|\epsilon)=0$ at $R_1=H_{\sf B}(1/2)-H_{\sf B}(\alpha*\epsilon)$, the fact that the difference between the two divergences strictly decreases implies the statemenet 3). To show that $D_{\sf B}(\gamma_{\sf GV}(R_1)\|\alpha*\epsilon)-D_{\sf B}(\gamma_{\sf GV}(R_1+C_2(\alpha))\|\epsilon)$ keeps decreasing as $R_1$ increases, we will show that
\beq\label{eqn:derdiv1}
\frac{\partial}{\partial R_1}D_{\sf B}(\gamma_{\sf GV}(R_1)\|\alpha*\epsilon)<\frac{\partial}{\partial R_1}D_{\sf B}(\gamma_{\sf GV}(R_1+C_2(\alpha))\|\epsilon).
\eeq
From the definition of $\gamma_{\sf GV}(R)$, it satisfies $\ln2+\gamma_{\sf GV}(R)\ln\gamma_{\sf GV}(R)+(1-\gamma_{\sf GV}(R))\ln (1-\gamma_{\sf GV}(R))=R$. By differentiating both sides by $\gamma_{\sf GV}(R)$ and re-arranging the terms, we get
\beq\label{eqn:der_gammagv}
\frac{\partial \gamma_{\sf GV}(R)}{\partial R}=-\frac{1}{\ln\frac{1-\gamma_{\sf GV}(R)}{\gamma_{\sf GV}(R)}}.
\eeq
From
$
\frac{\partial}{\partial x} D_{\sf B}(x\|y)=\ln\left(\frac{x}{1-x}\frac{1-y}{y}\right)
$ and~\eqref{eqn:der_gammagv},
we have
\beq
\begin{split}
&\frac{\partial}{\partial R_1}D_{\sf B}(\gamma_{\sf GV}(R_1)\|\alpha*\epsilon)=1-\frac{\ln\frac{1-\alpha*\epsilon}{\alpha*\epsilon}}{\ln\frac{1-\gamma_{\sf GV}(R_1)}{\gamma_{\sf GV}(R_1)}},\\
&\frac{\partial}{\partial R_1}D_{\sf B}(\gamma_{\sf GV}(R_1+C_2(\alpha))\|\epsilon)=1-\frac{\ln\frac{1-\epsilon}{\epsilon}}{\ln\frac{1-\gamma_{\sf GV}(R_1+C_2(\alpha))}{\gamma_{\sf GV}(R_1+C_2(\alpha))}}.
\end{split}
\eeq
Therefore, showing~\eqref{eqn:derdiv1} is equivalent to showing
\beq\label{eqn:derdiv3}
\frac{\ln\frac{1-\epsilon}{\epsilon}}{\ln\frac{1-\alpha*\epsilon}{\alpha*\epsilon}}<\frac{\ln\frac{1-\gamma_{\sf GV}(R_1+C_2(\alpha))}{\gamma_{\sf GV}(R_1+C_2(\alpha))}}{\ln\frac{1-\gamma_{\sf GV}(R_1)}{\gamma_{\sf GV}(R_1)}}.
\eeq
To prove this inequality, we will first show that 
\beq\label{eqn:derdiv2}
\gamma_{\sf GV}(R_1)-\gamma_{\sf GV}(R_1+C_2(\alpha))\geq \alpha*\epsilon-\epsilon
\eeq
Note that $\alpha*\epsilon=\gamma_{\sf GV}(H_{\sf B}(1/2)-H_{\sf B}(\alpha*\epsilon))$ and $\epsilon=\gamma_{\sf GV}(H_{\sf B}(1/2)-H_{\sf B}(\epsilon))$.
Therefore,~\eqref{eqn:derdiv2} can be written as
\beq\label{eqn:derdiv2_ext}
\begin{split}
&\gamma_{\sf GV}(R_1)-\gamma_{\sf GV}(R_1+C_2(\alpha))\geq \gamma_{\sf GV}(H_{\sf B}(1/2)-H_{\sf B}(\alpha*\epsilon))-\gamma_{\sf GV}(H_{\sf B}(1/2)-H_{\sf B}(\epsilon)).
\end{split}
\eeq
Note that $\gamma_{\sf GV}(R)\in[0,1/2]$ is convex and decreasing in $R\geq 0$. 
Moreover, we know that $(H_{\sf B}(1/2)-H_{\sf B}(\epsilon))-(H_{\sf B}(1/2)-H_{\sf B}(\alpha*\epsilon))=C_2(\alpha)$. Since we consider the regime where $R_1\leq H_{\sf B}(1/2)-H_{\sf B}(\alpha*\epsilon)$, from the convexity of $\gamma_{\sf GV}(R)$, the inequality in~\eqref{eqn:derdiv2_ext} can be implied.
Again, since $\alpha*\epsilon-\epsilon=\gamma_{\sf GV}(H_{\sf B}(1/2)-H_{\sf B}(\alpha*\epsilon))-\gamma_{\sf GV}(H_{\sf B}(1/2)-H_{\sf B}(\epsilon))$, for $c:=\alpha*\epsilon-\epsilon$, we have $\gamma_{\sf GV}(R_1+C_2(\alpha))\leq  \gamma_{\sf GV}(R_1)-c\leq 1/2$ from~\eqref{eqn:derdiv2_ext}.

Since $\ln\frac{1-x}{x}$ is decreasing in $0\leq x\leq 1/2$ and $\gamma_{\sf GV}(R_1+C_2(\alpha))\leq  \gamma_{\sf GV}(R_1)-c\leq 1/2$, we have
\beq
\frac{\ln\frac{1-(\gamma_{\sf GV}(R_1)-c)}{(\gamma_{\sf GV}(R_1)-c)}}{\ln\frac{1-\gamma_{\sf GV}(R_1)}{\gamma_{\sf GV}(R_1)}}\leq \frac{\ln\frac{1-\gamma_{\sf GV}(R_1+C_2(\alpha))}{\gamma_{\sf GV}(R_1+C_2(\alpha))}}{\ln\frac{1-\gamma_{\sf GV}(R_1)}{\gamma_{\sf GV}(R_1)}}.
\eeq
Therefore, to prove~\eqref{eqn:derdiv3}, it is sufficient to show that
\beq\label{eqn:derdiv4}
\frac{\ln\frac{1-\epsilon}{\epsilon}}{\ln\frac{1-\alpha*\epsilon}{\alpha*\epsilon}}< \frac{\ln\frac{1-(\gamma_{\sf GV}(R_1)-c)}{(\gamma_{\sf GV}(R_1)-c)}}{\ln\frac{1-\gamma_{\sf GV}(R_1)}{\gamma_{\sf GV}(R_1)}}.
\eeq

Since $\alpha*\epsilon< \gamma_{\sf GV}(R_1)$ and $\epsilon=\alpha*\epsilon-c$, if we can prove that 
\beq\label{eqn:derdiv5}
\frac{\ln\frac{1-(x-c)}{(x-c)}}{\ln\frac{1-x}{x}}
\eeq
is increasing in $x\in[\alpha*\epsilon,1/2]$, the inequality in~\eqref{eqn:derdiv4} holds.
We will prove this by showing that the derivative of~\eqref{eqn:derdiv5} in $x$ is positive for a very noisy BSC($\epsilon$) with $\epsilon\approx1/2$.
The derivative of~\eqref{eqn:derdiv5} is positive iff
\beq\label{eqn:derdiv6}
\begin{split}
&\frac{-1}{(x-c)(1-(x-c))}\ln\frac{1-x}{x}+\frac{1}{x(1-x)}\ln\frac{1-(x-c)}{(x-c)}>0.
\end{split}
\eeq 
From  $\alpha*\epsilon=\epsilon+\alpha(1-2\epsilon)$,  when $\epsilon\approx 1/2$ it is implied that $c=\alpha*\epsilon-\epsilon\approx 0$ and $\alpha*\epsilon\approx 1/2$.
Therefore, in the regime of  $\alpha*\epsilon\leq x\leq 1/2$, $c/x\approx 0$ and $c/(1-x)\approx 0$.
In this regime, we can approximate the terms in the left-hand side of~\eqref{eqn:derdiv6} as
\begin{align}
&\ln\frac{1-(x-c)}{(x-c)}=\ln\frac{(1-x)\left(1+\frac{c}{1-x}\right)}{x\left(1-\frac{c}{x}\right)}=\ln\frac{1-x}{x}+\frac{c}{1-x}+\frac{c}{x}+O(c^2),\\
&\frac{-1}{(x-c)(1-(x-c))}=\frac{-1}{x(1-x)}\frac{1}{\left(1-\frac{c}{x}\right)\left(1+\frac{c}{1-x}\right)}=\frac{-1}{x(1-x)}\left(1+\frac{c(1-2x)}{x(1-x)}+O(c^2)\right).
\end{align}
By plugging these approximations, the left-hand side of~\eqref{eqn:derdiv6} is approximated as
\beq\label{eqn:derdiv7}
\frac{c}{x^2(1-x)^2}\left(1-(1-2x)\ln\frac{1-x}{x}\right)+O(c^2).
\eeq 
Since $0<(1-2x)\ln\frac{1-x}{x}\ll 1$ for $x= 1/2-\delta$ for an arbitrarily small $\delta>0$, it can be shown that~\eqref{eqn:derdiv7} is positive. This implies that~\eqref{eqn:derdiv6} is valid for the very noisy channel, and thus~\eqref{eqn:derdiv4} is true.
This concludes the proof for the statement 3) $D_{\sf B}(\gamma_{\sf GV}(R_1)\|\alpha*\epsilon)>D_{\sf B}(\gamma_{\sf GV}(R_1+C_2(\alpha))\|\epsilon)$ for $R_{\sf crit}(\alpha*\epsilon)<R_1<C-C_2(\alpha)=H_{\sf B}(1/2)-H_{\sf B}(\alpha*\epsilon)$.

Lastly, we prove the statement 2) $E_0(1/2,\alpha*\epsilon)-R_1>D_{\sf B}(\gamma_{\sf GV}(R_1+C_2(\alpha))\|\epsilon)$ for $0\leq R_1\leq R_{\sf crit}(\alpha*\epsilon)$ when $R_{\sf crit}(\epsilon)<C_2(\alpha)$. Statement 3) implies that at $R_1=R_{\sf crit}(\alpha*\epsilon)$, $E_0(1/2,\alpha*\epsilon)-R_1>D_{\sf B}(\gamma_{\sf GV}(R_1+C_2(\alpha)\|\epsilon)$, since $E_0(1/2,\alpha*\epsilon)-R_1=D_{\sf B}(\gamma_{\sf GV}(R_1)\|\alpha*\epsilon)$ at $R_1=R_{\sf crit}(\alpha*\epsilon)$. When $R_{\sf crit}(\epsilon)<C_2(\alpha)$, the derivative of $D_{\sf B}(\gamma_{\sf GV}(R_1+C_2(\alpha))$ in $R_1\in[0,R_{\sf crit}(\alpha*\epsilon)]$ is
\beq
-1< \frac{\partial}{\partial R_1}D_{\sf B}(\gamma_{\sf GV}(R_1+C_2(\alpha))\|\epsilon)\leq 0.
\eeq
On the other hand, the derivative of $(E_0(1/2,\alpha*\epsilon)-R_1)$ in $R_1$ is $\frac{\partial}{\partial R_1}(E_0(1/2,\alpha*\epsilon)-R_1)=-1$. Since $(E_0(1/2,\alpha*\epsilon)-R_1)$ decreases faster than $D_{\sf B}(\gamma_{\sf GV}(R_1+C_2(\alpha))\|\epsilon)$ in $R_1\in[0,R_{\sf crit}(\alpha*\epsilon)]$, while $(E_0(1/2,\alpha*\epsilon)-R_1)$ is still greater than $D_{\sf B}(\gamma_{\sf GV}(R_1+C_2(\alpha))\|\epsilon)$ at $R_1=R_{\sf crit}(\alpha*\epsilon) $, it is implied that
\beq
E_0(1/2,\alpha*\epsilon)-R_1> D_{\sf B}(\gamma_{\sf GV}(R_1+C_2(\alpha))\|\epsilon)
\eeq
in $R_1\in[0,R_{\sf crit}(\alpha*\epsilon)]$. 

We proved the three statements 1), 2) and 3),  and these three statements imply the Lemma~\ref{lem:strict1}.

\section{Proof of Lemma~\ref{lem:JML_errbd}: A Lower Bound on the decoding-error exponent of the More Important Partial Message with Joint-ML Decoding Rule for Superposition Coding}\label{app:JML_errbd}
In Lemma~\ref{lem:JML_errbd}, we show that the decoding-error exponent $E_{\sf MSBs, \sf JML}(R_1, R_2)$ of the more important partial message $M_1$ (MSBs) with joint-ML decoding rule for superposition coding is bounded below as
\beq
E_{\sf MSBs, \sf JML}(R_1, R_2)\geq E^{\sf LB}_{\sf MSBs, \sf JML}(R_1, R_2)
\eeq
where
\beq
\begin{split}
&E^{\sf LB}_{\sf MSBs, JML}(R_1, R_2)=
\begin{cases}
E_0(1/2,\epsilon)-R_2-R_1, & R_1\leq \max\{0,R_{\sf crit}(\epsilon)-R_2\},\\
D_{\sf B}(\gamma_{\sf GV}(R_1+R_2)\|\epsilon), &\max\{0,R_{\sf crit}(\epsilon)-R_2\}<R_1\leq H_{\sf B}(1/2)-H_{\sf B}(\epsilon)-R_2.
\end{cases}
\end{split}
\eeq
In this section, we prove this lemma.

Consider superposition codes composed of codewords $\{\bz^{(m_1,m_2)}\}$, $m_1\in\{0,\dots, e^{NR_1}-1\}$, $m_2\in\{0,\dots, e^{NR_2}-1\}$, where $\bz^{(m_1,m_2)}=\bu^{(m_1)}\oplus \bv^{(m_2)} $ and $\bu^{(m_1)}$ consists of $N$ i.i.d. symbols of Bernoulli(1/2) distribution and  $\bv^{(m_2)}$ of Bernoulli($\alpha$) distribution for $\alpha\in(0,1/2)$. 
Without loss of generality, we suppose that $\bz^{(0,0)}$ is the correct codeword, which is transmitted by $N$ uses of a BSC($\epsilon$), and analyze the decoding-error probability of $m_1=0$.
Given the received word $\by=\bz^{(0,0)}\oplus \bn$, the joint maximum likelihood decoding rule finds a unique codeword $\bz^{(\hat{m}_1,\hat{m}_2)}$ that is closest to $\by$.
When we denote the decoded message as $(\hat{m}_1,\hat{m}_2)$, the decoding error happens only when $\hat{m}_1\neq 0$, regardless of whether or not $\hat{m}_2=0$.

The decoding-error event of the partial message $M_1$, denoted $\mathcal{E_{\sf JML}}$, occurs if there exists a codeword $\bz^{(m_1,m_2)}$ with $m_1\neq 0$ whose distance from $\by$ is less than or equal to  the minimum of all distances between $\by$ and $\bz^{(m_1,m_2)}$ for $m_1=0$, i.e., when the minimum distance  between $\by$ and any incorrect codeword $\bz^{(m_1,m_2)}$ with $m_1\neq 0$ is $N\delta$ and  the minimum distance between $\by$ and any codeword $\bz^{(m_1,m_2)}$ with $m_1=0$ is $N\tau$,  the decoding error occurs for the event $\mathcal{E_{\sf JML}}=\{\delta\leq \tau\}$.

Because only the distances of codewords from $\by$ matter, we consider the ``output-centered analysis'' proposed in~\cite{forney2001exponential} where all codewords are translated by $\by$.
Let $\bw^{(m_1,m_2)}=\bz^{(m_1,m_2)}\oplus \by=\bz^{(m_1,m_2)}\oplus\bz^{(0,0)}\oplus\bn$ denote the translated codewords.
For the correct $(m_1,m_2)=(0,0)$, the translated codeword $\bw^{(0,0)}$ is equal to the channel noise word $\bn$ and is independent of $\by$. 
The set of translated codewords for $m_1=0$, $\bw^{(0,m_2)}=\bv^{(m_2)}\oplus \bv^{(0)}\oplus \bn$, are independent of $\by$ but dependent on $\bn$. Moreover, $\{\bw^{(0,m_2)}\}$, $m_2\in\{0,\dots, e^{NR_2}-1\}$ are mutually dependent. The rest of the translated codewords with $m_1\neq 0$, i.e., $\{\bw^{(m_1, m_2)}\}$ for $ m_1\in\{1,\dots, e^{NR_1}-1\}$ and $m_2\in\{0,\dots, e^{NR_2}-1\}$, are independent of $\by$, $\bn$, and  $\{\bw^{(0,m_2)}\}$ for $ m_2\in\{0,\dots, e^{NR_2}-1\}$. However, the codewords $\{\bw^{(m_1, m_2)}\}$, $m_2\in\{0,\dots, e^{NR_2}-1\}$,  for a fixed $m_1$  are mutually dependent. Lastly, all possible received words $\by$ are equiprobable: $p(\by)=2^{-N}$.
The probability distribution of the decoding system consisting of the translated codewords and a received word $\by$ is thus decomposed as
\beq\label{eqn:decompose}
\begin{split}
&p(\{\bw^{(0,m_2)}\},\by,\{\bw^{(m_1\neq 0,m_2)}\})=2^{-N}p(\{\bw^{(0,m_2)}\})p(\{\bw^{(m_1\neq 0,m_2)}\}).
\end{split}
\eeq
Therefore, we can think of the whole decoding system as the one consisting of two independent subsystems, one comprising the translated codewords with $m_1=0$ and the other comprising the translated codewords with $m_1\neq 0$. 
We analyze the lower bound on the decoding-error probability of $M_1$ with the joint-ML decoding rule by using the fact that $\{\bw^{(0,m_2)}\}$ and $\{\bw^{(m_1\neq 0,m_2)}\}$ are independent. 

The decoding-error event $\mathcal{E}_{\sf JML}$ occurs if the minimum weight $N\tau$ of  $\{\bw^{(0,m_2)}\}$ is greater than or equal to the minimum weight $N\delta$ of $\{\bw^{(m_1\neq0,m_2)}\}$, i.e., $\mathcal{E_{\sf JML}}=\{\delta\leq \tau\}$. Define an error event $\mathcal{E}_\gamma=\{\delta\leq \gamma\leq\tau\}$ for a fixed $\gamma\in \Gamma=\{\gamma:0\leq \gamma\leq 1, N\gamma\in \mathcal{N}_0\}$ for the set $\mathcal{N}_0$ of non-negative integers. The decoding-error probability is equal to $\Pr(\mathcal{E}_{\sf JML})=\Pr(\delta\leq \tau)=\sum_{\gamma\in\Gamma} \Pr(\mathcal{E}_\gamma)$.
We first analyze $\Pr(\mathcal{E}_\gamma)$ and then find the typical $\gamma$ that dominates the exponentially decreasing rate of $\Pr(\mathcal{E}_{\sf JML})$ in the asymptotic regime.

From the independency between  $\{\bw^{(0,m_2)}\}$ and $\{\bw^{(m_1\neq0,m_2)}\}$,
\beq
\Pr(\mathcal{E}_\gamma)=\Pr(\delta\leq \gamma)\Pr(\tau \geq\gamma).
\eeq
We first establish an upper bound on $\Pr(\delta\leq \gamma)$. Note that $N\delta=\min_{(m_1\neq0,m_2)}w_H\left(\bw^{(m_1,m_2)}\right)$ where $w_H(\cdot)$ is the Hamming weight of the sequence.
We need to analyze the distribution of $w_H\left(\bw^{(m_1\neq 0,m_2)}\right)$.
Note that every symbol of every $\bw^{(m_1\neq 0,m_2)}$ is equiprobable.
By the Chernoff exponent lemma, for $\gamma< 1/2$, the probability that the Hamming weight $w_H\left(\bw^{(m_1\neq0, m_2)}\right)$ of a given incorrect codeword $\bw^{(m_1\neq 0, m_2)}$ is less than or equal to $N\gamma$ is
\beq\label{eqn:output_g_E1}
\Pr\left(w_H(\bw^{(m_1\neq 0, m_2)})\leq N\gamma\right)\doteq e^{-ND_{\sf B}(\gamma\|1/2)}.
\eeq
Since there are $(e^{NR}-e^{NR_2})$ codewords with $m_1\neq 0$, by the union bound
\beq
\begin{split}
&\Pr(\delta\leq \gamma)=\Pr\left(\min_{(m_1\neq 0, m_2)}w_H(\bw^{m_1\neq 0, m_2})\leq N\gamma\right)\\
&\dot{\leq}
\begin{cases}
e^{-N(D_{\sf B}(\gamma\|1/2)-R)},& \gamma\leq \gamma_{\sf GV}(R),\\
1,& \gamma>\gamma_{\sf GV}(R).
\end{cases}
\end{split}
\eeq
We next analyze $\Pr(\tau\geq \gamma)$ where $\tau=\min_{(m_1=0,m_2)}w_H(\bw^{(m_1,m_2)})$.
The translated correct codeword $\bw^{(0,0)}$ is equal to $\bn$ and is distributed by
\beq
p(\bn)=\epsilon^{w_H(\bn)}(1-\epsilon)^{N-w_H(\bn)}.
\eeq
Therefore, for $\gamma> \epsilon$, by the Chernoff exponent lemma we have
\beq
\Pr(w_H(\bn)\geq N\gamma)\doteq e^{-ND_{\sf B}(\gamma\|\epsilon)}.
\eeq
The event $ \tau\geq \gamma$ occurs when the Hamming weight $w_H\left(\bw^{(m_1=0,m_2)}\right)$ of {\it every} codeword $\{\bw^{(m_1=0,m_2)},0\leq m_2\leq e^{NR_2}-1\}$ is greater than or equal to $N\gamma$.
Therefore, for $\epsilon<\gamma<1/2$,
\beq
\begin{split}\label{eqn:output_g_E2}
&\Pr(\tau\geq \gamma)=\Pr\left(\min_{(m_1=0, m_2)}w_H(\bw^{(m_1= 0, m_2)})\geq N\gamma\right)\\
&\leq \Pr(w_H(\bn)\geq N\gamma)\doteq e^{-ND_{\sf B}(\gamma\|\epsilon)}.
\end{split}
\eeq
This bound may not be exponentially tight for $0\leq \alpha<1/2$. 
However, when $\alpha=1/2$,  since $\{\bw^{(0,m_2\neq0)},\bn\}$ are independent to each other and every symbol of every $\bw^{(0,m_2\neq 0)}$ is independent and equiprobable, for $\epsilon<\gamma<1/2$
\beq
\Pr\left(w_H(\bw^{(0,m_2\neq0)})\geq N\gamma\right)\doteq 1.
\eeq
Therefore, for the case of $\alpha=1/2$, the upper bound in~\eqref{eqn:output_g_E2} becomes exponentially tight. 

From~\eqref{eqn:decompose},~\eqref{eqn:output_g_E1} and~\eqref{eqn:output_g_E2},
\beq
\Pr(\mathcal{E}_\gamma)\dot{\leq}
\begin{cases}
e^{-N(D_{\sf B}(\gamma\|\epsilon)+D_{\sf B}(\gamma\|1/2)-R)},&\epsilon< \gamma\leq \gamma_{\sf GV}(R),\\
e^{-ND_{\sf B}(\gamma\|\epsilon)},& \gamma> \gamma_{\sf GV}(R).
\end{cases}
\eeq

By using this result, we calculate the achievable error exponent of $\Pr(\mathcal{E}_{\sf JML})=\sum_{\gamma\in\Gamma}\Pr(\mathcal{E}_\gamma)$ by finding $\gamma$ that dominates the exponentially decreasing rate of $\Pr(\mathcal{E}_{\sf JML})$ as $N$ increases. The resulting $\Pr(\mathcal{E}_{\sf JML})$ is
\beq
\Pr(\mathcal{E}_{\sf JML})\dot{\leq}
\begin{cases}
e^{-N(E_0(1/2,\epsilon)-R)},&0\leq R<R_{\sf crit}(\epsilon),\\
e^{-ND_{\sf B}(\gamma_{\sf GV}(R)\|\epsilon)},&R_{\sf crit}(\epsilon)\leq R<C,
\end{cases}
\eeq
where $E_0(a,b)=-\ln(1-2a(1-a)(\sqrt{b}-\sqrt{1-b})^2)$ and thus $E_0(1/2,\epsilon)=\ln 2-\ln(1+2\sqrt{\epsilon(1-\epsilon})$. And, $R_{\sf crit}(\epsilon)=D_{\sf B}(\gamma_{\sf crit}(\epsilon)\|1/2)$ with $\gamma_{\sf crit}(\epsilon)=\frac{\sqrt{\epsilon}}{\sqrt{\epsilon}+\sqrt{1-\epsilon}}$, and $C=H_{\sf B}(1/2)-H_{\sf B}(\epsilon)$.

Therefore, the best achievable error exponent  $E_{\sf MSBs,\sf JML}(R_1,R_2)$ with joint ML decoding for superposition coding of rates $(R_1,R_2)$ is bounded below as $E_{\sf MSBs,\sf JML}(R_1,R_2)\geq E^{\sf LB}_{\sf MSBs,\sf JML}(R_1,R_2)$ where
\beq
\begin{split}
&E^{\sf LB}_{\sf MSBs,\sf JML}(R_1, R_2)=\\
&
\begin{cases}
E_0(1/2,\epsilon)-R_2-R_1, &R_1< \max\{0,R_{\sf crit}(\epsilon)-R_2\},\\
D_{\sf B}(\gamma_{\sf GV}(R_1+R_2)\|\epsilon), &\max\{0,R_{\sf crit}(\epsilon)-R_2\}\leq R_1<H_{\sf B}(1/2)-H_{\sf B}(\epsilon)-R_2.
\end{cases}
\end{split}
\eeq

\section{Proof of Theorem~\ref{thm:sup}: Gains in the quantized-MSE exponent from Superposition Coding}\label{app:sup}
 In Theorem~\ref{thm:sup},  we show that for a very noisy BSC($\epsilon$) with $\epsilon= 0.5-\delta$ for a sufficiently small $\delta>0$, the best achievable quantized-MSE exponent  $E^*_{\sf q,spc}(R)$ with superposition coding  is strictly larger than that of random block coding $E^*_{\sf q, rc}(R)$ for every rate $R\in(E_0(1/2,\epsilon)/3,C)$. Here we prove this theorem by finding a lower bound $E_{\sf q,spc}(R)$ on  $E^*_{\sf q,spc}(R)$ and showing that this lower bound is strictly larger than $E^*_{\sf q, rc}(R)$ at any rate $R\in(E_0(1/2,\epsilon)/3,C)$.
 
The quantized MSE $\E[c_{\sf q}(X,\hat{X}_{N,\sf finite})]$ can be bounded above in terms of decoding-error probabilities of the two partial messages $M_1$ (MSBs) and $M_2$ (LSBs) as
\beq
\begin{split}
&\E[c_{\sf q}(X,\hat{X}_{N,\sf finite})]\leq \Pr(\hat{M}_1\neq M_1)+\Pr(\hat{M}_2\neq M_2|\hat{M}_1=M_1)e^{-2NR_1}.
\end{split}
\eeq
In Lemma~\ref{lem:strict1}, we show that successive-cancellation decoding rule for superposition codes of distribution parameter $\alpha\in(0,1/2)$ achieves $\Pr(\hat{M}_1\neq M_1)$ such that
\beq
\Pr(\hat{M}_1\neq M_1)\dot{\leq}e^{-NE_{\sf MSBs, SC}(R_1,\alpha)}
\eeq where
 \beq\label{eqn:ofstrcit2_app}
 \begin{split}
&E_{\sf MSBs, \sf SC}(R_1, \alpha)=
\begin{cases}
E_0(1/2,\alpha*\epsilon)-R_1, &0\leq R_1\leq R_{\sf crit}(\alpha*\epsilon),\\
D_{\sf B}(\gamma_{\sf GV}(R_1)\|\alpha*\epsilon), &R_{\sf crit}(\alpha*\epsilon)<R_1\leq C-C_2(\alpha).
\end{cases}
\end{split}
\eeq
By using this bound and the bound on the conditional decoding-error probability of $M_2$ with the successive-cancellation decoding,
\beq
\Pr(\hat{M}_2\neq M_2|\hat{M}_1=M_1 )\dot{\leq} e^{-NE_{\sf LSBs, SC}(R_2,\alpha)},
\eeq
 the quantized MSE with the superposition coding can be bounded above as
\beq\label{eqn:star}\nonumber
\begin{split}
&\E[c_{\sf q}(X,\hat{X}_{N,\sf finite})]\\
&\leq e^{-N E_{\sf MSBs, SC}(R_1,\alpha)}+e^{-N E_{\sf LSBs,SC}(R_2,\alpha)}e^{-2NR_1}\\
&\doteq e^{-N\min\{E_{\sf MSBs, SC}(R_1,\alpha),E_{\sf LSBs, SC}(R_2,\alpha)+2R_1 \}}.
\end{split}
\eeq
Therefore, the best achievable quantized-MSE exponent $E^*_{\sf q, spc}(R)$ with the superposition coding of rate $R=R_1+R_2$ is bounded below as
\beq\label{eqn:61}
\begin{split}
&E^*_{\sf q, spc}(R)\geq\max_{\substack{\{(R_1,R_2,\alpha):\\ R_1+R_2=R, \\ \alpha\in(0,1/2)\}}}\min\{E^*_{\sf MSBs}(R_1,R_2,\alpha),E^*_{\sf LSBs}(R_2,\alpha)+2R_1\}.
\end{split}
\eeq


For a given $\alpha\in(0,1/2)$, when we choose the rate $R_2$ of the partial message $M_2$ (LSBs) equal to $C_2(\alpha)=H_{\sf B}(\alpha*\epsilon)-H_{\sf B}(\epsilon)$, which is the maximum possible rate of $M_2$ to guarantee $\Pr(\hat{M}_2\neq M_2|\hat{M}_1= M_1)\to 0$ as $N\to\infty$, the resulting error exponent $E_{\sf LSBs,SC}(R_2,\alpha)$ equals 0. This particular choice of $R_2=C_2(\alpha)$ provides a lower bound on $E^*_{\sf q,spc}(R)$ such that
\beq
\begin{split}
&E^*_{\sf q, spc}(R) \geq\max_{\alpha\in(0,1/2)}\min\{E_{\sf MSBs, SC}(R-C_2(\alpha),\alpha), 2(R-C_2(\alpha))\}.
\end{split}
\eeq

We next find $\alpha\in(0,1/2)$ that maximizes $\min\{E_{\sf MSBs, SC}(R-C_2(\alpha),\alpha), 2(R-C_2(\alpha))\}$. More specifically, we find $\alpha$ that makes 
\beq\label{eqn:thm2_cond}
E_{\sf MSBs, SC}(R-C_2(\alpha),\alpha)=2(R-C_2(\alpha)).
\eeq
For a very noisy BSC($\epsilon$) with $\epsilon\approx 1/2$, as shown in~\cite{gallager1968information} (pp. 147-149), the error exponent $E_{\sf MSBs, SC}(R_1,\alpha)$ in~\eqref{eqn:ofstrcit2} can be approximated as
\beq\label{eqn:randombck_approx11}
\begin{split}
&E_{\sf MSBs, SC }(R_1,\alpha)\approx\begin{cases}
\frac{C-C_2(\alpha)}{2}-R_1, & 0\leq  R_1< \frac{C-C_2(\alpha)}{4},\\
(\sqrt{C-C_2(\alpha)}-\sqrt{R_1})^2, & \frac{C-C_2(\alpha)}{4}\leq R_1\leq C-C_2(\alpha).
\end{cases}
\end{split}
\eeq
By using this approximation, we can see that when $\epsilon\approx 1/2$, $R_1=R-C_2(\alpha)$ satisfying~\eqref{eqn:thm2_cond} is in the interval $0\leq R_1\leq R_{\sf crit}(\alpha*\epsilon)\approx \frac{C-C_2(\alpha)}{4}$, since
at $R_1=R-C_2(\alpha)=E_0(1/2,\alpha*\epsilon)/3\approx \frac{C-C_2(\alpha)}{6}$, which is strictly smaller than $R_{\sf crit}(\alpha*\epsilon) \approx  \frac{C-C_2(\alpha)}{4}$, the condition~\eqref{eqn:thm2_cond} is satisfied. 
So, at $\alpha\in(0,1/2)$ satisfying 
\beq\label{eqn:foralpha}
R=C_2(\alpha)+\frac{E_0(1/2,\alpha*\epsilon)}{3},
\eeq
the condition~\eqref{eqn:thm2_cond} is met. Moreover, when $\epsilon\approx 1/2$, the derivative of $C_2(\alpha)+\frac{E_0(1/2,\alpha*\epsilon)}{3}$ with respect to $\alpha$ is positive in $\alpha\in(0,1/2)$.
Therefore, $C_2(\alpha)+\frac{E_0(1/2,\alpha*\epsilon)}{3}$ increases from $E_0(1/2,\epsilon)/3$ to $C=H_{\sf B}(1/2)-H_{\sf B}(\epsilon)$ as $\alpha$ increases from 0 to 1/2.
It means that for all $R\in(E_0(1/2,\epsilon)/3,C)$, there always exists a unique $\alpha^*\in(0,1/2)$ satisfying~\eqref{eqn:foralpha}. 


Lastly, as shown in Lemma~\ref{lem:strict1}, since $E_{\sf MSBs, SC}(R_1,\alpha)>E^*_{\sf q, rc}(R)=E_{\sf r}(R_1+R_2)$ for every $R_1\in[0,C-C_2(\alpha))$ when $R_2=C_2(\alpha)$, for our choice of $R_1^*(\alpha^*)=R-C_2(\alpha^*)\in[0,C-C_2(\alpha))$ and $R_2=C_2(\alpha^*)$ it can be shown that
\beq
E_{\sf MSBs, SC}(R-C_2(\alpha^*),\alpha^*)>E_{\sf r}(R).
\eeq
This completes the proof of the theorem.

\begin{figure}[t]
\centerline{\includegraphics[scale=0.6]{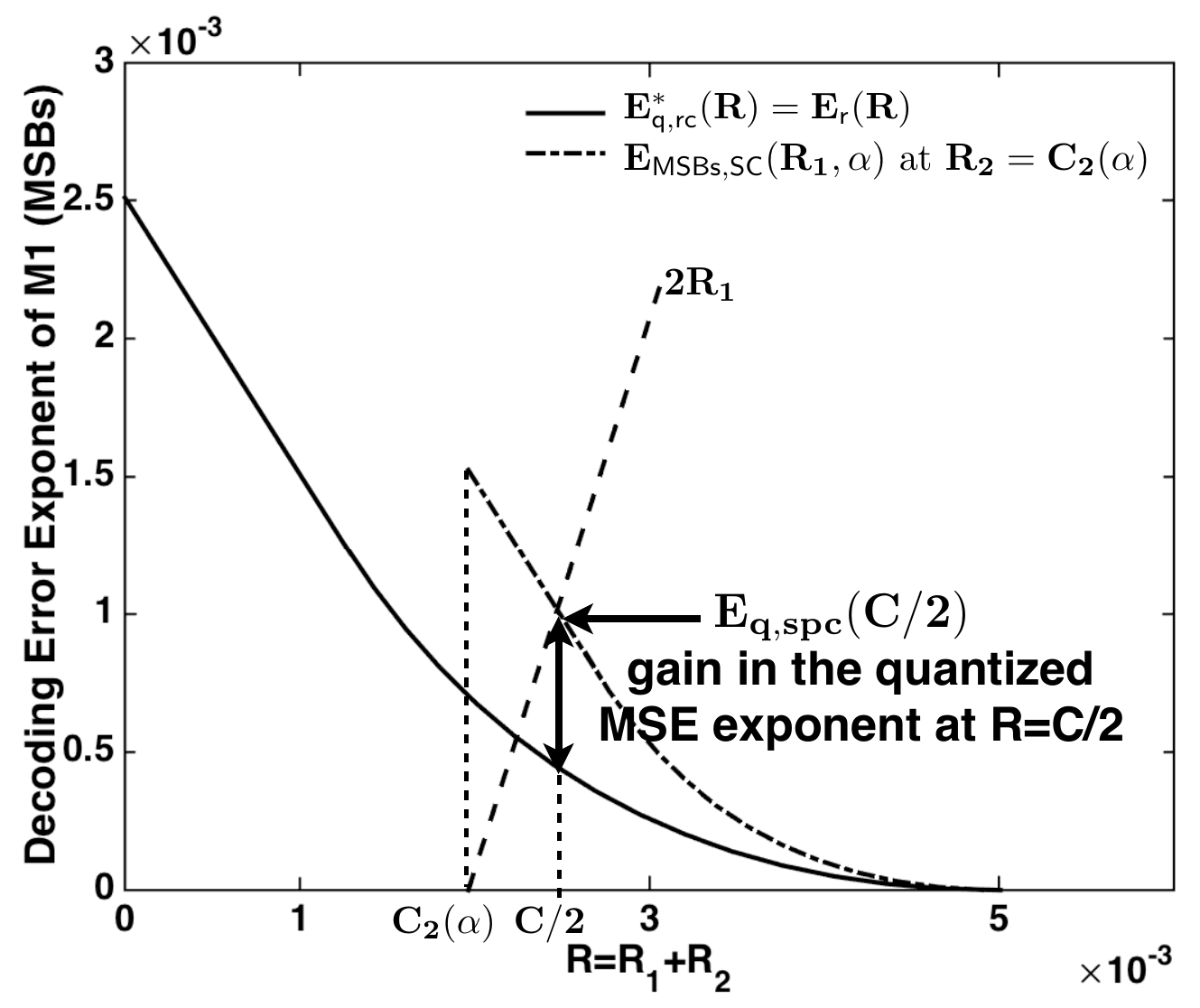}}
\caption{A plot of  
$E^*_{\sf q,rc}(R_1+R_2)=E_{\sf r}(R_1+R_2)$ (the best achievable quantized MSE with random block coding), 
$E_{\sf MSBs, SC}(R_1,\alpha)$ (the achievable decoding-error exponent of MSBs with superposition coding), and the line $2R_1$  at a fixed $R_2=C_2(\alpha)$ where $\epsilon=0.45$ and $\alpha=0.11$. 
When we choose $\alpha$ that satisfies the condition in~\eqref{eqn:thm2_cond}, i.e., $E_{\sf MSBs, SC}(R-C_2(\alpha),\alpha)=2(R-C_2(\alpha))$ at $R=C/2$,  the achievable  quantized-MSE exponent $E_{\sf q, spc}(C/2)$ with the superposition coding  equals the value of $E_{\sf MSBs, SC}(R_1,\alpha)$ at $R_1$ where $E_{\sf MSBs, SC}(R_1,\alpha)$ and $2R_1$ cross each other. The achievable quantized-MSE exponent $E_{\sf q, spc}(C/2)$ at $R=C/2$ is  strictly larger than the best achievable quantized-MSE exponent $E^*_{\sf q,rc}(C/2)=E_{\sf r}(C/2)$ with random block coding as shown in this figure.
}
\label{fig:cqgainR1mod}
\end{figure}

In Fig~\ref{fig:cqgainR1mod}, we illustrate the gain in the quantized-MSE exponent from superposition coding at $R=C/2$.  When we choose $\alpha$ that satisfies the condition in~\eqref{eqn:thm2_cond}, i.e., $E_{\sf MSBs, SC}(R-C_2(\alpha),\alpha)=2(R-C_2(\alpha))$ at $R=C/2$, the achievable  quantized-MSE exponent with the superposition coding, which is denoted $E_{\sf q, spc}(C/2)$, is equal to the value of $E_{\sf MSBs, SC}(R_1,\alpha)$ at $R_1$ where $E_{\sf MSBs, SC}(R_1,\alpha)$ and $2R_1$ cross each other. In this plot, we can check that the achievable quantized-MSE exponent $E_{\sf q, spc}(C/2)$ with superposition coding is  strictly larger than the best achievable quantized-MSE exponent $E^*_{\sf q, rc}(C/2)=E_{\sf r}(C/2)$ with random block coding.






\bibliographystyle{IEEEtran}

\end{document}